\pdfoutput=1
\documentclass[11pt]{article}
\usepackage[margin=1in]{geometry}
\usepackage[T1]{fontenc}
\usepackage{lmodern,microtype,amsmath,amssymb,amsthm,bm,booktabs,graphicx}
\usepackage{xcolor}

\usepackage[authoryear,round]{natbib}
\usepackage[ruled,vlined]{algorithm2e}
\usepackage{xurl}
\usepackage[colorlinks=true,linkcolor=blue,citecolor=blue,urlcolor=blue,hyperfootnotes=false]{hyperref}
\usepackage{setspace}
\usepackage{needspace}
\usepackage{adjustbox}
\numberwithin{equation}{section}
\newtheorem{theorem}{Theorem}[section]
\newtheorem{proposition}[theorem]{Proposition}

\newtheorem{corollary}[theorem]{Corollary}
\theoremstyle{definition}
\newtheorem{assumption}[theorem]{Assumption}

\newtheorem{example}[theorem]{Example}
\theoremstyle{remark}

\DeclareMathOperator*{\argmin}{arg\,min}
\DeclareMathOperator{\E}{E}
\DeclareMathOperator{\Exp}{Exp}
\DeclareMathOperator{\Log}{Log}
\DeclareMathOperator{\Hess}{Hess}
\DeclareMathOperator{\tr}{tr}
\DeclareMathOperator{\diag}{diag}
\newcommand{\M}{\mathcal M}
\newcommand{\R}{\mathbb R}
\newcommand{\1}{\mathbf 1}
\newcommand{\ip}[2]{\langle #1,#2\rangle}
\newcommand{\norm}[1]{\lVert #1\rVert}

\newcommand{\calD}{\mathcal D}

\newcommand{\calB}{\mathcal B}
\title{Geodesic Causal Forests: Learning Heterogeneous Treatment Effects for Random Objects}
\author{Yidong Zhou\\{\normalsize School of Statistics, University of Minnesota}}
\date{}
\begin{document}
\maketitle
\begin{abstract}
Causal forests and orthogonal learners estimate heterogeneous treatment effects for scalar or Euclidean outcomes, yet outcomes are increasingly random objects, such as compositions, covariance matrices, distributions, or networks, in metric spaces without linear structure. We introduce geodesic causal forests for conditional geodesic treatment effects. Geodesic transport moves each geodesic between conditional Fr\'echet means of the potential outcomes to a common reference, making effects comparable across individuals. Because random objects cannot be added or subtracted, Euclidean residual corrections are unavailable; we instead correct the squared-distance loss through which the forest selects splits and estimates effects. The loss is real-valued for every candidate effect, so the correction requires only Fr\'echet means and leaves second-order nuisance error; for Euclidean outcomes, it recovers doubly robust learning. We establish consistency, convergence rates, and split-selection guarantees under conditions verified for curved and nonsmooth spaces. In simulations with covariance-matrix outcomes, the correction substantially reduces estimation error. For women's livelihood compositions in BRAC's randomized antipoverty program in Bangladesh, new livestock work mainly replaces casual wage work; after alignment, a single effect reproduces much of the variation in percentage-point changes across baseline activities, and the fitted effects are largest among former domestic workers and older women.
\end{abstract}
\noindent\textbf{Keywords:} conditional average treatment effect; Fr\'echet mean; geodesic transport; metric statistics; orthogonal learning.

\section{Introduction}\label{sec:intro}
Understanding how treatment effects vary with pretreatment characteristics is central to evaluating interventions and policies \citep{caron2022review}. For scalar or vector outcomes, the conditional average treatment effect (CATE) is the difference between mean outcomes under treatment and control given the pretreatment characteristics. It can be estimated by meta-learners \citep{kunzel2019metalearners}, causal trees and forests \citep{athey2016recursive,wager2018causal}, generalized random forests \citep{athey2019grf}, and Bayesian causal forests, which regularize baseline outcomes and treatment effects separately \citep{hahn2020bcf}. These methods rely on estimated outcome regressions and propensity scores, whose errors can obscure heterogeneity. Orthogonal scores and sample splitting reduce their influence \citep{chernozhukov2018dml}, as in the R-learner \citep{nie2021quasioracle}, the DR-learner \citep{kennedy2023dr}, and orthogonal random forests \citep{oprescu2019orthogonal}, with a general analysis by orthogonal statistical learning \citep{foster2023orthogonal}.

Almost all of this work concerns scalar or Euclidean outcomes, whereas outcomes increasingly have a complex structure that scalar or vector summaries may fail to preserve. For example, an asset-and-training program can change women's participation in work and the shares of time devoted to livestock rearing, wage labor, and other activities \citep{bandiera2017labor,balboni2022poor}. A livelihood composition records these features jointly, but its components are nonnegative and sum to one, so that unrestricted vector operations can produce invalid compositions. Similarly, an increase in the minimum wage may affect the entire income distribution within each region, revealing changes in inequality that mean income can conceal \citep{dube2019minimumwages}; here the outcome is a probability distribution, and the ordinary difference between two distributions is not a distribution. Such outcomes, together with networks, functional data, and symmetric positive-definite (SPD) matrices, can be viewed as random objects, that is, random elements of a metric space $(\M,d)$ \citep{muller2016randomobjects,marron2014overview,dubey2024metric,song2026metric}. Their analysis is based on distances between outcomes and does not require a vector-space structure, so averages and differences, the two operations underlying the CATE, need geometric counterparts.

Fr\'echet means and geodesics provide such counterparts: a Fr\'echet mean minimizes the expected squared distance to a random object, and a geodesic is a shortest path between two objects. \citet{lin2023distribution} and \citet{gunsilius2023distributional} study causal effects on distributions through quantile functions. For general outcomes, geodesics between potential-outcome Fr\'echet means describe directed treatment contrasts \citep{kurisu2024gate}, distances between Fr\'echet means or geometric medians define effect magnitudes \citep{shin2024absolute}, and related methods address difference-in-differences, regression discontinuity, synthetic controls, continuous treatments, outcome selection, and jumps in conditional Fr\'echet means \citep{zhou2025gdid,kurisu2025grdd,vandijcke2025rdd,kurisu2025gsc,bhattacharjee2025continuous,kurisu2026lee,vandijcke2025jumps}.

To learn how such contrasts vary with covariates, Fr\'echet regression \citep{petersen2019frechet}, Fr\'echet forests \citep{qiu2024forest,capitaine2024forest}, or Fr\'echet geodesic boosting \citep{zhou2025fgboost} can estimate the conditional means separately within treatment arms. Heterogeneous-effect methods exist for distributional \citep{katta2024addmalts} and functional outcomes \citep{salmaso2026focal}, which admit linear representations that general random objects lack, and causal forests handle manifold covariates \citep{huo2023manifold}.

We introduce the geodesic causal forest (GCF), which groups individuals with similar treatment responses when outcomes are random objects. The target at each covariate value is the geodesic connecting the conditional Fr\'echet means of the two potential outcomes. Two challenges arise in learning how this target varies with the covariates. First, the untreated means vary with the covariates, and geodesics with different starting points are not directly comparable. We use geodesic transport to express each geodesic at a common reference; within a group, one aligned effect serves as the prediction for all members, and a split is useful when separate predictions for the two subgroups reduce the squared-distance loss.

Second, the aligned effects are unobserved, and a loss computed from fitted conditional means inherits their first-order errors. Euclidean orthogonal methods remove these errors through outcome residuals, which the DR-learner adds to the fitted effect and the R-learner uses to build its loss; random objects admit neither the subtraction that forms a residual nor the addition that applies it. We instead observe that a forest uses the effects only through the losses of candidate effects, and each such loss is a real number. We therefore correct the loss rather than the outcome: an observed outcome receives a small additional weight in the fitted law of its treatment arm, and the resulting change in loss, weighted by the inverse propensity score, estimates the first-order nuisance error. \citet[Section~3.5]{foster2023orthogonal} construct orthogonal losses from pathwise derivatives of a non-orthogonal loss; our correction obtains the required derivative as a difference quotient of Fr\'echet means under contamination of the fitted laws, so it applies to random objects without tangent vectors or derivatives in the outcome space. Our contributions are as follows.

\begin{itemize}
\item We show that orthogonal estimation of conditional effects does not require adding or subtracting outcomes: correcting the effect loss itself leaves a conditional bias of second order in the nuisance errors. The correction requires only Fr\'echet means of reweighted fitted laws, with no tangent vectors or derivatives in the outcome space. For Euclidean outcomes, it reduces to the DR-learner, so correcting the outcome is the Euclidean special case of correcting the loss.
\item We develop the GCF, in which the same corrected loss selects the splits and estimates the aligned effects, directing the forest toward heterogeneity in treatment response. Applying the estimated effect at the estimated untreated mean, restricted to valid outcomes, gives the treatment-effect geodesic in the original outcome space. The construction applies to curved and nonsmooth spaces with stable transport, including spherical, SPD, and open-book outcomes.
\item We establish error bounds for the forest estimator, conditions on tree growth under which it is consistent, and guarantees for split selection. The bounds separate the variation of effects within forest neighborhoods, sampling error, and second-order nuisance error, and give the usual nonparametric rate when the neighborhoods localize in the covariates governing the effect.
\item In simulations with SPD outcomes, the correction reduces the error of the forest severalfold, and the GCF outperforms separate estimation of the two conditional means when the effects rotate. In BRAC's randomized asset-and-training program, new livestock work mainly replaces casual wage work, and for women mainly in wage work or poultry at baseline it replaces that activity most. A single aligned effect, applied at each woman's untreated mean, reproduces much of this dependence on baseline activities; the effects fitted by the forest are largest among former domestic workers and older women and have a smaller held-out loss than a common effect.
\end{itemize}

\section{Random objects and motivating examples}\label{sec:economic_examples}
Let $Y$ be a random object taking values in a metric space $(\M,d)$, where $d(u,v)$ measures the distance between outcomes $u,v\in\M$. A constant-speed geodesic from $u$ to $v$ is a path $\gamma_{u,v}:[0,1]\to\M$ with $\gamma_{u,v}(0)=u$ and $\gamma_{u,v}(1)=v$ that satisfies
\begin{equation}\label{eq:geodesic_speed}d\{\gamma_{u,v}(s),\gamma_{u,v}(t)\}=|s-t|d(u,v),\qquad s,t\in[0,1].\end{equation}
The space is uniquely geodesic if every pair of points is joined by exactly one geodesic \citep{burago2001metric}. In Euclidean space, $\gamma_{u,v}(t)=u+t(v-u)$ starts at $u$ and has displacement $v-u$. The directed geodesic $\gamma_{u,v}$ is a geometric analogue of this difference, keeping its starting point and magnitude $d(u,v)$ without subtraction \citep{kurisu2024gate}.

Fr\'echet means extend ordinary means through their least-squares characterization. For a random object $Y$ with $\E\{d^2(u_0,Y)\}<\infty$ for some $u_0\in\M$, the Fr\'echet mean is
\begin{equation}\label{eq:frechet_mean}\mu_\oplus=\argmin_{u\in\M}\E\{d^2(u,Y)\},\end{equation}
provided that the minimizer exists and is unique. The conditional Fr\'echet mean given pretreatment covariates $X=x$ is defined analogously by minimizing $\E\{d^2(u,Y)\mid X=x\}$ \citep{petersen2019frechet}. In Euclidean space, these are $\E(Y)$ and $\E(Y\mid X=x)$, respectively. Section~\ref{app:geometry} of the Supplement gives further geometric details.

\subsection{Motivating example: Participation and livelihood allocation}\label{subsec:livelihood_example}
BRAC's Targeting the Ultra-Poor program in rural Bangladesh seeks to expand poor women's livelihood opportunities by combining productive assets, principally livestock, with two years of training and support \citep{bandiera2017labor}. The question of interest extends beyond whether recipients spend more time rearing livestock. For a woman initially outside recorded work, the new activity may accompany entry into employment; for a casual wage worker, it may replace agricultural or domestic wage labor; for another woman, it may add to an existing mix of activities. These responses have different implications for occupational choice and labor supply, even when the increases in livestock hours are similar.

We study these responses using the 2011 follow-up of the branch-randomized evaluation. The outcome for each woman records both participation and the allocation of recorded work across activities, which distinguishes entry into work from a change in the mix of activities among workers. Section~\ref{sec:brac_application} examines how this joint treatment response varies with pretreatment work histories and resources.

\subsection{Examples of outcome spaces}\label{subsec:other_outcomes}
Examples~\ref{ex:spherical} and~\ref{ex:spd} are used in the data application and the simulations. Section~\ref{app:more_examples} of the Supplement gives further examples, including distributions, networks, functional data, and phylogenetic trees, and Section~\ref{app:openbook} treats open books, nonsmooth spaces of half-spaces joined along a common boundary.

\begin{example}[Compositional data]\label{ex:spherical}
A composition $s=(s_1,\ldots,s_m)$ of nonnegative proportions that sum to one belongs to the simplex $\Delta^{m-1}=\{s\in[0,1]^m:\sum_{k=1}^m s_k=1\}$. For the livelihood outcome in Section~\ref{subsec:livelihood_example}, let $h_{i1},\ldots,h_{i5}$ be woman $i$'s annual hours in livestock rearing, agricultural wage work, domestic wage work, poultry, and other recorded activities, with total $H_i=\sum_{k=1}^5h_{ik}$. The six-part composition is
\begin{equation}\label{eq:livelihood_composition}S_i=\begin{cases}(0,h_{i1}/H_i,\ldots,h_{i5}/H_i),&H_i>0,\\(1,0,\ldots,0),&H_i=0.\end{cases}\end{equation}
The first coordinate records nonparticipation; the others describe work allocation.

The component-wise square-root transformation $y=(\sqrt{s_1},\ldots,\sqrt{s_m})$ maps the simplex to the nonnegative orthant $\mathbb S^{m-1}_+$ of the unit sphere. The angular distance $d_S(s,r)=\arccos(\sum_{k=1}^m\sqrt{s_kr_k})$ becomes the geodesic distance $d(u,v)=\arccos(u^\top v)$ on the sphere. The inner product measures compositional overlap; the distance accommodates zero shares and equals $\pi/2$ for disjoint supports. Spherical representations also arise in economic compositional time series \citep{zhu2024spherical}. For $\alpha=d(u,v)>0$, the geodesic is
\begin{equation}\label{eq:composition_path}\gamma_{u,v}(t)=\frac{\sin\{(1-t)\alpha\}}{\sin\alpha}u+\frac{\sin(t\alpha)}{\sin\alpha}v,\qquad 0\le t\le1.\end{equation}
It stays in the orthant, so squaring its coordinates gives valid compositions.
\end{example}

\begin{example}[Symmetric positive-definite matrices]\label{subsec:risk_example}\label{ex:spd}
Covariance matrices of returns on a common set of assets record marginal variances and co-movement, whereas correlation matrices describe dependence after standardizing marginal scales \citep{chiriac2011volatility,engle2002dcc}. Both are relevant to portfolio allocation, since an intervention can alter diversification through dependence even when marginal variances remain unchanged.

Nonsingular covariance and correlation matrices belong to the space $\mathbb S_{++}^r$ of $r\times r$ SPD matrices, and we compute Fr\'echet means and geodesics in this full space. Under the affine-invariant metric \citep{pennec2006tensor}, the distance and geodesic are $d(U,V)=\norm{\log(U^{-1/2}VU^{-1/2})}_F$ and $\gamma_{U,V}(t)=U^{1/2}(U^{-1/2}VU^{-1/2})^tU^{1/2}$. For covariance matrices, the distance measures relative changes in joint risk and is invariant to a common nonsingular linear transformation of returns. For $r\ge2$, the geometry is curved, and Fr\'echet means generally differ from ordinary averages of matrices or their logarithms.
\end{example}

\section{Conditional geodesic treatment effects}\label{sec:geometry}
\subsection{Causal target and identification}
Let $O=(X,A,Y)$ denote a generic observation, where $X\in\mathcal X\subset\R^p$ contains pretreatment covariates, $A\in\{0,1\}$ is the treatment indicator, and the outcome $Y$ takes values in $(\M,d)$. For $a\in\{0,1\}$, let $Y(a)$ be the potential outcome under treatment $a$ and $m_a(x)$ its conditional Fr\'echet mean given $X=x$. We write $\ell_u(y)=d^2(u,y)/2$ for the squared-distance loss and $p_a(x)=P(A=a\mid X=x)$ for the propensity score of arm $a$.

\begin{assumption}\label{ass:causal}
Observations are independent copies of $O$, with consistency $Y=Y(A)$. For $a=0,1$, some $c_p>0$, and $u_0\in\M$: unconfoundedness, $Y(a)\perp A\mid X$; overlap, $p_a(X)\ge c_p$ almost surely; and a finite second moment, $\E\{d^2(Y(a),u_0)\}<\infty$.
\end{assumption}

These are the CATE identifying conditions \citep{kennedy2023dr}, and randomization gives unconfoundedness and overlap, so the conditional Fr\'echet risk of $Y(a)$ is identified as
\begin{equation}\label{eq:conditional_risk}R_a(u,x)=\E\{\ell_u(Y(a))\mid X=x\}=\E\{\ell_u(Y)\mid A=a,X=x\}.\end{equation}
The conditional Fr\'echet mean minimizes this risk, $m_a(x)=\argmin_{u\in\M}R_a(u,x)$. To state the continuity of geodesics in their endpoints, we measure the distance between continuous paths $\gamma,\widetilde\gamma:[0,1]\to\M$ by the uniform path metric
\begin{equation}\label{eq:path_metric}D_\infty(\gamma,\widetilde\gamma)=\sup_{0\le t\le1}d\{\gamma(t),\widetilde\gamma(t)\}.\end{equation}

\begin{assumption}\label{ass:geodesic}
The metric space $(\M,d)$ is complete, separable, and uniquely geodesic. Its constant-speed geodesics depend continuously on their endpoints: for any sequences $u_n\to u$ and $v_n\to v$ in $(\M,d)$, $D_\infty(\gamma_{u_n,v_n},\gamma_{u,v})\to0$. The conditional Fr\'echet means exist uniquely at all covariate values under consideration.
\end{assumption}

Assumption~\ref{ass:geodesic} holds for laws with finite second moments on Hadamard spaces, including nonsmooth ones \citep{sturm2003probability}. For the positively curved spherical representation of compositions, Section~\ref{app:spherical_profiles} of the Supplement verifies it for all laws on the simplex.

Extending the geodesic average treatment effect of \citet{kurisu2024gate}, we define the conditional geodesic treatment effect as the geodesic from $m_0(x)$ to $m_1(x)$,
\begin{equation}\label{eq:causal_path}\tau_G(x):=\gamma_{m_0(x),m_1(x)}.\end{equation}
In Euclidean space, its displacement $m_1(x)-m_0(x)$ is the usual CATE. Its endpoints depend only on the conditional laws of $Y(0)$ and $Y(1)$ given $X=x$, not on their joint law, so Assumption~\ref{ass:causal} identifies $\tau_G(x)$ through~\eqref{eq:conditional_risk}.

\subsection{Geodesic transport and reconstruction}\label{subsec:transport}
The geodesics $\tau_G(x)$ in~\eqref{eq:causal_path} start at the untreated means $m_0(x)$, which vary with the covariates. To compare them, we use geodesic transport \citep{zhu2026transport,zhou2025gdid} to move each geodesic to a common starting point, the reference point $o\in\M$. For untreated and treated endpoints $u$ and $v$, the alignment map $F_o$, determined by the geometry of $(\M,d)$, gives the aligned endpoint $e=F_o(u,v)$, so the change from $u$ to $v$ is represented by the change from $o$ to $e$; in Euclidean space, $F_o(u,v)=o+v-u$.

The reconstruction map $G_o$ applies an aligned effect at a baseline: $G_o(u,e)$ is the endpoint reached from $u$ by the change that $e$ represents at $o$, and $G_o(u,e)=u+e-o$ in Euclidean space. Aligned endpoints need not be possible outcomes; for compositions, they may lie outside the nonnegative orthant. We therefore allow them in a space $\overline{\M}\supset\M$ with the same distances, and Section~\ref{subsec:feasible_estimation} restricts estimates to effects reconstructing into $\M$.

\begin{assumption}\label{ass:metric_transport}
Let $(\overline{\M},d)$ be a metric space containing $(\M,d)$ isometrically, such as $\M$ itself. Fix $o\in\M$, bounded baseline and treated-mean sets $K_0,K_1\subset\M$, and a compact effect set $K_\theta\subset\overline{\M}$ containing $o$. The maps $F_o$ and $G_o$ take values in $\overline{\M}$ and satisfy
\begin{equation}\label{eq:metric_inverse}G_o\{u,F_o(u,v)\}=v,\qquad d\{o,F_o(u,v)\}=d(u,v),\qquad u\in K_0,\ v\in K_1.\end{equation}
For finite constants $L_F,L_G$,
\begin{equation}\label{eq:metric_stability}\begin{aligned}d\{F_o(u,v),F_o(u',v')\}&\le L_F\{d(u,u')+d(v,v')\},\\d\{G_o(u,e),G_o(u',e')\}&\le L_G\{d(u,u')+d(e,e')\},\end{aligned}\end{equation}
for $u,u'\in K_0$, $v,v'\in K_1$, and $e,e'\in K_\theta$. Moreover, $m_a(x)\in K_a$ for $a=0,1$, and, for a fixed $\eta_\theta>0$, $K_\theta$ contains $F_o\{u,m_1(x)\}$ whenever $u\in K_0$ and $d\{u,m_0(x)\}\le\eta_\theta$.
\end{assumption}

The identities in~\eqref{eq:metric_inverse} make reconstruction reverse the alignment and preserve the effect magnitude, and the Lipschitz bounds control how estimation errors propagate through both maps. The effect set must contain the aligned effects at the true baseline $m_0(x)$ and at nearby baselines, which enter when the baseline is estimated, but not those of other pairs in $K_0\times K_1$, so it can be much smaller than the image of $F_o$. Translation gives these maps for Hilbert-space representations, and parallel transport for Riemannian manifolds such as spheres and SPD matrices (Section~\ref{app:model_verification} of the Supplement).

Aligning the two conditional means defines the aligned effect,
\begin{equation}\label{eq:theta}\theta(x)=F_o\{m_0(x),m_1(x)\}.\end{equation}
The reference point $o$ represents no change: if $m_0(x)=m_1(x)$, the length identity gives $\theta(x)=o$. A constant $\theta$ describes a common aligned response even when the untreated means differ. By the reconstruction identity in~\eqref{eq:metric_inverse}, the geodesic~\eqref{eq:causal_path} can be written as
\begin{equation}\label{eq:metric_reconstruction}\tau_G(x)=\gamma_{m_0(x),G_o\{m_0(x),\theta(x)\}}.\end{equation}
In Euclidean space, the choice of $o$ is immaterial: changing $o$ shifts all aligned effects and the effect set by the same vector. On curved spaces, we use a canonical reference point: the identity matrix for SPD outcomes, which gives $\theta(x)=m_0(x)^{-1/2}m_1(x)m_0(x)^{-1/2}$, and the uniform composition, which treats all parts symmetrically and is farthest from zero shares.

\section{Geodesic causal forests}\label{sec:forest}
\Needspace{5\baselineskip}
\subsection{Effect loss and its orthogonal correction}\label{sec:split}
Each tree of the GCF recursively splits the covariate space at thresholds of the pretreatment variables. For a node $B\subset\mathcal X$ with $P(X\in B)>0$, a common prediction $z\in K_\theta$ incurs the loss $\ell_z\{\theta(x)\}=d^2\{z,\theta(x)\}/2$ at covariate value $x$. The node mean $e_B$, the Fr\'echet mean of the aligned effects in the node, is the best common prediction, and the minimized risk $V(B)$ measures the variation of effects within the node:
\begin{equation}\label{eq:node_risk}e_B=\argmin_{z\in K_\theta}\E\{\ell_z(\theta(X))\mid X\in B\},\qquad V(B)=\E\{\ell_{e_B}(\theta(X))\mid X\in B\}.\end{equation}
Splitting $B$ into children $L$ and $R$ of positive probability allows each child to use its own prediction. With $\pi_C=P(X\in C\mid X\in B)$, the split criterion is the reduction in risk,
\begin{equation}\label{eq:gain}\Delta(B;L,R)=V(B)-\pi_LV(L)-\pi_RV(R).\end{equation}
With unique node means, the gain is positive exactly when the children's mean effects differ.

The effects $\theta(x)$ are unobserved. Fitted effects give a loss that can be computed, but errors in the preliminary outcome regressions enter it at first order. Residual corrections such as the DR-learner \citep{kennedy2023dr} are undefined in a metric space, but the forest uses the effects only through the losses of candidate effects in~\eqref{eq:node_risk} and~\eqref{eq:gain}. Each loss is a real-valued functional of the conditional laws of the outcomes in the two arms, so we apply the one-step correction for such functionals \citep{vandervaart1998asymptotic} to the loss itself, estimating its first-order change from the fitted to the true laws.

We represent each conditional law by its Fr\'echet risk function. The method uses risk functions only through their minimizers, so we fix their additive constant by setting the loss to zero at a fixed point $u_*\in\M$:
\begin{equation}\label{eq:metric_k}k_y(u)=\frac12\{d^2(u,y)-d^2(u_*,y)\},\qquad u\in\M.\end{equation}
The conditional risk functions $Q_a(\cdot,x)=\E\{k_Y(\cdot)\mid A=a,X=x\}$ differ from~\eqref{eq:conditional_risk} by a term that does not depend on $u$, so they have the same minimizers, and the corrected loss below does not depend on $u_*$. Fixing the constant lets us measure their errors by a norm.

For a pair $Q=(Q_0,Q_1)$ with unique minimizers, write $m_a^Q(x)=\argmin_{u\in\M}Q_a(u,x)$ for $a=0,1$, so that the true risk functions give $m_a^Q(x)=m_a(x)$. On a preliminary nuisance sample $\calD_0$, we fit risk functions $\widetilde Q_a$ with minimizers $\widetilde m_a(x)=m_a^{\widetilde Q}(x)$. For example, nonnegative regression weights within arm $a$ give
\begin{equation}\label{eq:profilefit}\widetilde Q_a(u,x)=\sum_{j\in\calD_0:A_j=a}w_{aj}(x)k_{Y_j}(u),\qquad w_{aj}(x)\ge0,\quad\sum_{j:A_j=a}w_{aj}(x)=1.\end{equation}
Any such weights may be used; arm-specific Fr\'echet forests \citep{qiu2024forest,capitaine2024forest} and local Fr\'echet regressions on the adjustment covariates \citep{petersen2019frechet} give weights that require only the metric. Alignment gives the aligned effect and its prediction loss:
\begin{equation}\label{eq:metric_T}\begin{aligned}
\theta^Q(x)&=F_o\{m_0^Q(x),m_1^Q(x)\},\\
\mathcal T_z^Q(x)&=\frac12\{d^2(z,\theta^Q(x))-d^2(o,\theta^Q(x))\},\qquad z\in K_\theta.
\end{aligned}\end{equation}
The true risk functions give $\theta^Q(x)=\theta(x)$, and the fitted ones give $\widetilde\theta(x)=\theta^{\widetilde Q}(x)$ and the fitted effect loss $\mathcal T_z^{\widetilde Q}(x)$. Subtracting the loss at the no-change reference $o$ fixes the additive constant of the effect loss; the subtracted term does not depend on $z$, so it leaves minimizers and split gains unchanged. The effect loss depends on $Q$ only through the minimizers $m_a^Q$. We write $\mathcal T^Q(x)$ for the function $z\mapsto\mathcal T_z^Q(x)$.

Suppressing $x$, write $\dot{\mathcal T}_z^{\widetilde Q}[H]$ for the derivative of the fitted loss at $z$ when the pair of risk functions moves from $\widetilde Q$ in the direction of a pair $H$. The directions we need move the fitted law toward an observed outcome. For an outcome $y$ and arm $a$, replacing a fraction $t$ of the fitted law in arm $a$ by a point mass at $y$ gives the pair $\widetilde Q^{a,t,y}$ with arm-$a$ component
\begin{equation}\label{eq:contaminated_risk}\widetilde Q^{a,t,y}_a=(1-t)\widetilde Q_a+t k_y\end{equation}
and the other component unchanged. The direction $\widetilde Q^{a,1,y}-\widetilde Q$ of this perturbation has arm-$a$ component $k_y-\widetilde Q_a$, whose conditional mean given $A=a$ and $X=x$ is $Q_a-\widetilde Q_a$. Linearity of the derivative therefore gives, conditionally on the nuisance sample,
\begin{equation}\label{eq:loss_one_step}
\begin{aligned}
\mathcal T_z^{Q}&\approx\mathcal T_z^{\widetilde Q}+\dot{\mathcal T}_z^{\widetilde Q}[Q-\widetilde Q],\\
\dot{\mathcal T}_z^{\widetilde Q}[Q-\widetilde Q]&=\E\biggl\{\sum_{a=0}^1\frac{\1(A=a)}{p_a(X)}\,\dot{\mathcal T}_z^{\widetilde Q}\bigl[\widetilde Q^{a,1,Y}-\widetilde Q\bigr]\,\bigg|\,X=x\biggr\},
\end{aligned}
\end{equation}
where the first relation holds up to an error quadratic in $Q-\widetilde Q$ under Assumption~\ref{ass:metric_profiles}. The unknown first-order change in the loss is thus the conditional mean of the change in loss caused by the observed outcome, weighted by the inverse propensity score of its arm.

This derivative can be computed without differentiation in the outcome space. It is the limit as $t\downarrow0$ of the difference quotient $(\mathcal T_z^{\widetilde Q^{a,t,y}}-\mathcal T_z^{\widetilde Q})/t$, which requires only the Fr\'echet mean of the contaminated law, its alignment with the fitted mean of the other arm, and the resulting loss \citep{carone2019computerized,jordan2022gateaux}. Replacing the derivative by this quotient with a small contamination size $0<\varepsilon\le1$, and the propensity scores by fitted values $\widetilde p_a(x)>0$, gives the corrected loss
\begin{equation}\label{eq:metric_psi}\psi_z^{\varepsilon}(O)=\mathcal T_z^{\widetilde Q}(X)+\sum_{a=0}^1\frac{\1(A=a)}{\widetilde p_a(X)}\frac{\mathcal T_z^{\widetilde Q^{a,\varepsilon,Y}}(X)-\mathcal T_z^{\widetilde Q}(X)}{\varepsilon}.\end{equation}
The first term is the fitted effect loss, and the second is the change in loss in the observed arm, scaled by the inverse of its fitted propensity score. We write $\psi^\varepsilon(O)$ for the function $z\mapsto\psi_z^\varepsilon(O)$. Section~\ref{subsec:loss_regularity} gives conditions under which its remaining bias is of second order in the nuisance errors, up to a finite-difference error of order $\varepsilon$.

With the weights in~\eqref{eq:profilefit}, the perturbation multiplies the fitted weights by $1-\varepsilon$ and gives weight $\varepsilon$ to the observed outcome; each original and perturbed mean is computed once per observation and reused for all candidate effects and splits. The contamination size enters the nuisance remainder $\rho_\varepsilon$ in~\eqref{eq:metric_remainders} additively, whereas errors in the perturbed means are amplified by $1/\varepsilon$, so we take $\varepsilon$ of the order of the inverse sample size and compute these means to correspondingly high precision.

For Euclidean outcomes, the correction reduces exactly to DR learning \citep{kennedy2023dr}. Consider fitted laws with finite second moments and means $\widetilde m_a(x)$, the translation $F_o(u_0,u_1)=o+u_1-u_0$, and the pseudo-outcome $\phi(O)=\widetilde m_1(X)-\widetilde m_0(X)+A\{Y-\widetilde m_1(X)\}/\widetilde p_1(X)-(1-A)\{Y-\widetilde m_0(X)\}/\widetilde p_0(X)$. The effect loss is linear in the fitted means, so the difference quotient in~\eqref{eq:metric_psi} is exact and, for every admissible contamination size,
\begin{equation}\label{eq:metric_euclidean}\psi_z^{\varepsilon}(O)=\frac12\norm{z-o}^2-\ip{z-o}{\phi(O)}=\frac12\norm{z-o-\phi(O)}^2-\frac12\norm{\phi(O)}^2.\end{equation}
Weighted minimization and splitting thus use ordinary squared-error loss on DR pseudo-outcomes, so correcting the loss is equivalent to correcting the outcome. The corrected loss is conditionally unbiased for $\mathcal T_z^Q(X)$ when either both conditional means or both propensity scores are correct. Section~\ref{app:euclidean_reduction_proof} of the Supplement derives this reduction.

\subsection{Forest construction and estimation}\label{subsec:forest_estimation}
We use three disjoint samples: the nuisance sample $\calD_0$ to fit the risk functions and propensity scores; the structure sample $\calD_S$, of size $N$, to select splits; and the estimation sample $\calD_E$, of size $n$, to estimate effects within the resulting neighborhoods. The nuisance sample also determines the tuning of the nuisance fits. Separating structure and estimation makes the forest honest \citep{wager2018causal}: conditional on covariates, its estimation weights are independent of the estimation treatments and outcomes.

For each node $B$, we fit one aligned effect by minimizing the total corrected loss of its structure observations. With the common normalization $N$, the minimized node loss is
\begin{equation}\label{eq:empnode}\widehat W(B)=\inf_{z\in K_\theta}\frac1N\sum_{i\in\calD_S}\1(X_i\in B)\psi_z^{\varepsilon}(O_i).\end{equation}
For a candidate split $s=(B,L,R)$, the estimated reduction in risk is
\begin{equation}\label{eq:empgain}\widehat G(s)=\widehat W(B)-\widehat W(L)-\widehat W(R).\end{equation}
We select the candidate split with the largest $\widehat G(s)$. The common normalization in~\eqref{eq:empnode} makes $\widehat G(s)$ estimate $P(X\in B)\Delta(B;L,R)$, with $\Delta$ defined in~\eqref{eq:gain}.

Candidate thresholds are taken from the structure covariates, and each child must contain minimum numbers of observations in total and from each treatment arm, so the correction in every node uses both arms. As in the random-split trees of \citet{wager2018causal}, a split is drawn uniformly from the admissible candidates with a small fixed probability, so every covariate is split with positive probability. Each tree is grown on a random subsample of $\calD_S$, to which the sum and the normalization $N$ in~\eqref{eq:empnode} then refer.

Fix a covariate value $x$, and let $B_t(x)$ be the leaf of tree $t$ containing $x$ and $N_t(x)=\sum_{i\in\calD_E}\1\{X_i\in B_t(x)\}$ its estimation-sample count. Each tree assigns equal weight to the estimation observations in this leaf, and averaging over $T$ trees gives
\begin{equation}\label{eq:weights}\alpha_i(x)=\frac1T\sum_{t=1}^T\frac{\1\{X_i\in B_t(x)\}}{N_t(x)},\qquad \sum_{i\in\calD_E}\alpha_i(x)=1.\end{equation}
If a leaf containing $x$ has no estimation observations, we use its nearest ancestor that does. The aligned-effect estimate minimizes the resulting weighted corrected loss:
\begin{equation}\label{eq:estimator}\widehat\theta(x)\in\argmin_{z\in K_\theta}\widehat L_x(z),\qquad \widehat L_x(z)=\sum_{i\in\calD_E}\alpha_i(x)\psi_z^{\varepsilon}(O_i).\end{equation}

\subsection{Feasible geodesic reconstruction}\label{subsec:feasible_estimation}
To express the estimate in the outcome space, we apply it, as in~\eqref{eq:metric_reconstruction}, at an untreated mean $\widehat b(x)\in K_0$ estimated from the nuisance sample with all adjustment covariates. An aligned effect is possible at baseline $u$ only if it reconstructs into $\M$, so the feasible effects form
\begin{equation}\label{eq:feasible_domain}\Gamma(u)=\{e\in K_\theta:G_o(u,e)\in\M\}.\end{equation}
We therefore minimize the weighted corrected loss over the effects feasible at $\widehat b(x)$:
\begin{equation}\label{eq:feasible_estimator}\widehat\theta_{\mathrm{F}}(x)\in\argmin_{z\in\Gamma\{\widehat b(x)\}}\widehat L_x(z).\end{equation}
By~\eqref{eq:metric_inverse}, $\theta(x)\in\Gamma\{m_0(x)\}$, and $\Gamma\{\widehat b(x)\}$ contains the no-change effect $o$, since each construction in Section~\ref{app:model_verification} of the Supplement satisfies $G_o(u,o)=u$. The reconstructed treated endpoint and the estimated geodesic are
\begin{equation}\label{eq:reconstruct}\widehat m_1^{\mathrm{rec}}(x)=G_o\{\widehat b(x),\widehat\theta_{\mathrm F}(x)\},\qquad \widehat\tau_G(x)=\gamma_{\widehat b(x),\widehat m_1^{\mathrm{rec}}(x)}.\end{equation}

\begin{figure}[!t]
\centering
\includegraphics[width=\linewidth]{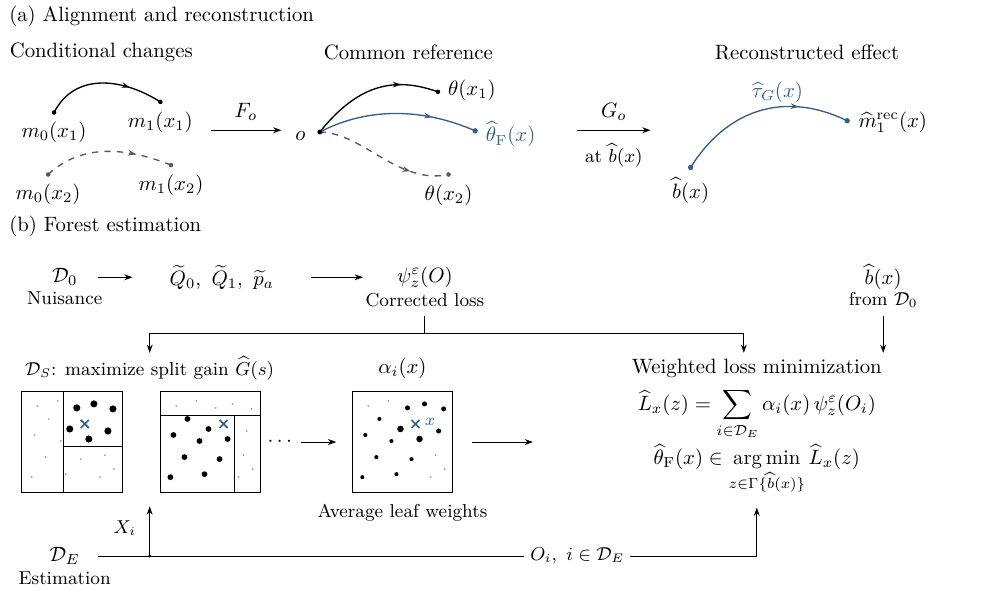}
\caption{Geometry and estimation for the geodesic causal forest. (a) Conditional treatment effects are aligned at the reference point $o$ through $F_o$. The paths ending at $\widehat\theta_{\mathrm F}(x)$ and $\widehat m_1^{\mathrm{rec}}(x)$ show the estimated aligned effect and its reconstruction at $\widehat b(x)$ through $G_o$. (b) Nuisance estimates from $\calD_0$ define the corrected loss used for splitting on $\calD_S$ and estimation on $\calD_E$. Crosses mark the covariate value $x$, and filled points identify the estimation observations sharing its leaf. Averaging leaf weights over trees gives $\alpha_i(x)$, represented by point area in the pooled panel. The baseline $\widehat b(x)$ is the untreated mean estimated from $\calD_0$, and $\Gamma\{\widehat b(x)\}$ contains the effects that reconstruct to valid outcomes there.}\label{fig:forest_schematic}
\end{figure}

For compositions, Proposition~\ref{prop:spherical_feasibility} in the Supplement writes nonnegativity of the reconstructed endpoint as convex constraints and shows that the connecting geodesic stays in the orthant. Figure~\ref{fig:forest_schematic} illustrates the procedure, which is summarized in Algorithm~\ref{alg:forest}.

\begin{algorithm}[t]
\DontPrintSemicolon
\caption{Geodesic causal forest (GCF)}\label{alg:forest}
\KwIn{Disjoint nuisance ($\calD_0$), structure ($\calD_S$), and estimation ($\calD_E$) samples; covariate value $x$; reference point $o$; number of trees $T$; admissible splits and tree-growth rules.}
\KwOut{Feasible aligned-effect estimate $\widehat\theta_{\mathrm F}(x)$ and estimated conditional geodesic treatment effect $\widehat\tau_G(x)$.}
\textbf{1. Fit nuisance functions.} Using $\calD_0$, estimate the conditional Fr\'echet risk functions $\widetilde Q_a$, propensity scores $\widetilde p_a$, and untreated reconstruction baseline $\widehat b(x)\in K_0$. Use all covariates needed for causal adjustment, and set $\varepsilon$ as in Section~\ref{sec:split}.\;
\BlankLine
\textbf{2. Align and correct the loss.} For each $i\in\calD_S\cup\calD_E$, compute the fitted mean pair at $X_i$ and align it at $o$. Replace a fraction $\varepsilon$ of the fitted risk function in arm $A_i$ with $k_{Y_i}$, recompute that arm's mean, and align the perturbed pair. Use the two aligned effects to evaluate the corrected loss $\psi_z^{\varepsilon}(O_i)$ in~\eqref{eq:metric_psi}.\;
\BlankLine
\textbf{3. Grow the forest on $\calD_S$.}\;
\For{$t=1,\ldots,T$}{
Draw a subsample of $\calD_S$. At each node, choose the admissible split that gives the largest reduction in minimized corrected loss~\eqref{eq:empgain}, or, with a small fixed probability, an admissible split drawn uniformly at random.\;
Continue until no admissible split remains or the maximum depth is reached.\;
}
\BlankLine
\textbf{4. Estimate the aligned effect using $\calD_E$.} Find the leaf of each tree containing $x$. Average the within-leaf weights as in~\eqref{eq:weights}. Obtain $\widehat\theta_{\mathrm F}(x)$ by minimizing $\sum_{i\in\calD_E}\alpha_i(x)\psi_z^\varepsilon(O_i)$ over $\Gamma\{\widehat b(x)\}$, up to a small objective tolerance, as in~\eqref{eq:feasible_estimator}.\;
\BlankLine
\textbf{5. Reconstruct the treatment-effect geodesic.} Set $\widehat m_1^{\mathrm{rec}}(x)=G_o\{\widehat b(x),\widehat\theta_{\mathrm F}(x)\}$ and let $\widehat\tau_G(x)$ be the geodesic from $\widehat b(x)$ to $\widehat m_1^{\mathrm{rec}}(x)$.\;
\end{algorithm}

\Needspace{6\baselineskip}
\section{Statistical theory}\label{sec:theory}
\subsection{Regularity of the corrected loss}\label{subsec:loss_regularity}
The accuracy of the corrected loss depends on how errors in the fitted risk functions propagate to the effect loss. A Fr\'echet mean depends on a risk function only via differences $Q_a(u,x)-Q_a(u',x)$, and the forest uses the effect loss only through differences $\mathcal T_z^Q(x)-\mathcal T_{z'}^Q(x)$, so we use Lipschitz norms. For a metric space $(S,d)$ and $s_0\in S$, let $\operatorname{Lip}_0(S;s_0)$ be the Banach space of Lipschitz functions $f:S\to\R$ with $f(s_0)=0$, normed by
\begin{equation}\label{eq:metric_B}\norm f_{\mathrm{Lip}}=\sup_{s,s'\in S,\,s\ne s'}\frac{|f(s)-f(s')|}{d(s,s')}.\end{equation}

We measure errors in effect losses in $\mathbb Z=\operatorname{Lip}_0(K_\theta;o)$. The effect loss in~\eqref{eq:metric_T} and the corrected loss in~\eqref{eq:metric_psi} vanish at the no-change reference $z=o$, and $\|\mathcal T^Q\|_{\mathbb Z}\le\sup_{z\in K_\theta}d\{z,\theta^Q(x)\}\le D_\theta+D_K$ whenever $m_a^Q(x)\in K_a$ for $a=0,1$, where $D_\theta$ is the diameter of $K_\theta$ and $D_K=\sup_{u\in K_0,v\in K_1}d(u,v)$. Errors in risk functions are measured in a Banach space $\mathbb B$ of functions on $\M$ that vanish at $u_*$. Risk functions differ only through the part of the anchored loss~\eqref{eq:metric_k} that depends on the outcome. Writing
\begin{equation}\label{eq:centered_risk_kernel}\bar k_y(u)=k_y(u)-k_{u_*}(u)=\frac12\{d^2(u,y)-d^2(u,u_*)-d^2(y,u_*)\},\qquad k_{u_*}(u)=\frac12d^2(u,u_*),\end{equation}
we have $Q_a=k_{u_*}+\E\{\bar k_Y\mid A=a,X=x\}$, so $k_{u_*}$ cancels in differences of risk functions. Section~\ref{app:model_verification} of the Supplement specifies $\mathbb B$ for each outcome space; for smooth models, its norm also bounds derivatives near the minimizers. The contribution of a single outcome to a risk function has size $A_Q(y)=\|\bar k_y\|_{\mathbb B}$.

\begin{assumption}\label{ass:metric_profiles}
The nuisance functions are fitted on an independent sample. Fitted propensity scores satisfy $\widetilde p_a(x)\ge\epsilon_0>0$, and the fitted risk functions have the form $\widetilde Q_a=k_{u_*}+\int\bar k_y\,d\widetilde P_a(y\mid x)$ for laws $\widetilde P_a(\cdot\mid x)$ on $\M$. For fixed constants $M_Q$, $L_0$, $L$, $C$, and $0<\epsilon_*\le1$ and contamination size $0<\varepsilon\le\epsilon_*$, the following hold uniformly in $x$ and fits.
\begin{enumerate}
\item The functions $\bar k_y$ belong to $\mathbb B$, and, for $a=0,1$,
\begin{equation}\label{eq:risk_moment}\E\{A_Q(Y)^2\mid A=a,X=x\}\le M_Q^2,\qquad \int A_Q(y)^2\,d\widetilde P_a(y\mid x)\le M_Q^2.\end{equation}
\item The true and fitted risk pairs, and the contaminated pairs~\eqref{eq:contaminated_risk} with $0<t\le\epsilon_*$ and $y$ in the support of the conditional law of $Y$ given $A=a$ and $X=x$, have unique minimizers, at which $F_o$ is defined. The fitted minimizers satisfy $\widetilde m_a(x)\in K_a$.
\item For each fitted $\widetilde Q$, there is a linear map $\dot{\mathcal T}^{\widetilde Q}:\mathbb B^2\to\mathbb Z$ with operator norm at most $L$ for the sum norm on $\mathbb B^2$ such that, for $Q'$ the true pair or a contaminated pair,
\begin{align}
\|\mathcal T^{Q'}-\mathcal T^{\widetilde Q}\|_{\mathbb Z}&\le L_0\|Q'-\widetilde Q\|_{\mathbb B^2},\label{eq:metric_first_order}\\
\norm{\mathcal T^{Q'}-\mathcal T^{\widetilde Q}-\dot{\mathcal T}^{\widetilde Q}[Q'-\widetilde Q]}_{\mathbb Z}&\le C\norm{Q'-\widetilde Q}_{\mathbb B^2}^{2}.\label{eq:metric_vonmises}
\end{align}
\end{enumerate}
\end{assumption}

Independent nuisance samples allow the fits to be held fixed. The moment condition~\eqref{eq:risk_moment} allows unbounded outcomes, the bound~\eqref{eq:metric_first_order} controls the change in loss caused by an observation, and the von Mises-type expansion~\eqref{eq:metric_vonmises} makes the first relation in~\eqref{eq:loss_one_step} precise, as in orthogonal learning \citep{oprescu2019orthogonal}. Section~\ref{app:model_verification} of the Supplement verifies Assumption~\ref{ass:metric_profiles} for Hilbert space-valued and open-book outcomes with unbounded support, compact smooth SPD models, and compositions with unrestricted support.

We measure nuisance errors at $x$ by the $\mathbb B$-norm error $r(x)$ of the fitted risk functions and the error $q(x)$ of the fitted propensity scores, combined in the remainder $\rho_\varepsilon(x)$:
\begin{equation}\label{eq:metric_remainders}\begin{gathered}r(x)=\sum_{a=0}^1\norm{\widetilde Q_a(\cdot,x)-Q_a(\cdot,x)}_{\mathbb B},\qquad q(x)=\sum_{a=0}^1|\widetilde p_a(x)-p_a(x)|,\\\rho_\varepsilon(x)=r(x)^2+r(x)q(x)+\varepsilon.\end{gathered}\end{equation}

\begin{proposition}\label{prop:metric_bias}
Under Assumption~\ref{ass:metric_profiles}, there are constants $C_0,C_1<\infty$, not depending on $\varepsilon$ or the fitted functions, such that, given the nuisance sample and for every $0<\varepsilon\le\epsilon_*$,
\begin{equation}\label{eq:metric_scorelip}\E\{\|\psi^{\varepsilon}(O)\|_{\mathbb Z}^2\mid X=x\}\le C_0^2,\end{equation}
\begin{equation}\label{eq:metric_bias}\norm{\E\{\psi^{\varepsilon}(O)\mid X=x\}-\mathcal T^Q(x)}_{\mathbb Z}\le C_1\rho_\varepsilon(x).\end{equation}
Hence a loss comparison between $z,z'\in K_\theta$ has conditional bias at most $C_1\rho_\varepsilon(x)d(z,z')$.
\end{proposition}

The first bound controls the random variation of loss comparisons. The second shows that the corrected loss has second-order bias: the terms $r^2$ and $rq$ are products of nuisance errors, as for doubly robust estimators, and $\varepsilon$ reflects the finite-difference approximation. Section~\ref{app:orthogonality_proof} of the Supplement proves Proposition~\ref{prop:metric_bias} and gives the constants in~\eqref{eq:loss_constants}.

\subsection{Effect-directed split selection}\label{subsec:localized_splits}
Strong convexity links the split gain~\eqref{eq:gain} to the separation of the children's mean effects. Suppose that $K_\theta$ is geodesically convex and the half squared distance is $\kappa_Z$-strongly convex along its geodesics for a fixed $0<\kappa_Z\le1$ (Assumption~\ref{ass:effect_curvature} in Section~\ref{app:forest} of the Supplement). This holds with $\kappa_Z=1$ on compact convex subsets of Hadamard spaces \citep{sturm2003probability} and, for compositions with unrestricted support, with $\kappa_Z=2R\cot(2R)$ on spherical balls about $o$ of radius $R<\pi/4$ (Proposition~\ref{prop:effect_curvature_domains}). Each node then has a unique mean effect, and $\Delta(B;L,R)\ge\kappa_Z\pi_L\pi_Rd^2(e_L,e_R)/2$; for Euclidean outcomes, equality holds with $\kappa_Z=1$, and $e_L-e_R$ is the difference between the child-average CATEs.

The estimated gain $\widehat G(s)$ in~\eqref{eq:empgain} estimates the gain computed from the true effects $\theta(X_i)$ at the structure covariates; write $\Delta_X(s)$ for the gain~\eqref{eq:gain} with the law of $X$ in $B$ replaced by the empirical law of the $m_B$ structure covariates in $B$, and $\widehat\Delta_X(s)=(N/m_B)\widehat G(s)$ for its estimate. Theorem~\ref{thm:localized_split} in Section~\ref{app:forest} of the Supplement shows that, conditionally on the nuisance sample, the structure covariates, and the tree randomization, and for bounded contributions $A_Q(Y)$, with probability at least $1-\delta$ simultaneously over all splits among $M$ candidate nodes, including nodes selected during tree growth,
\[|\widehat\Delta_X(s)-\Delta_X(s)|\le a_s\{2\Delta_X(s)/\kappa_Z\}^{1/2}+a_s^2/(2\kappa_Z)+b_s.\]
Here $a_s$ is the largest, over the parent and children $D$, of $C_0(2t_S/m_D)^{1/2}+C_1\bar\rho_D$, where $m_D$ is the number of observations in $D$, $\bar\rho_D$ their average nuisance remainder, and $t_S=\log\{2M(H+M)/\delta\}$ for a finite set of $H$ candidate effects used in numerical minimization; $b_s$ accounts for that approximation. The error thus adapts to the gain: a split that does not separate effects has only quadratic sampling and nuisance error, apart from $b_s$.

\subsection{Pointwise estimation and convergence rates}\label{subsec:metric_rates}
The accuracy of the forest estimator~\eqref{eq:estimator} at $x$ depends on the number of weighted observations, the variation of their true effects around $\theta(x)$, and the nuisance bias. Let $(X_i,A_i,Y_i)_{i=1}^n$ be the estimation sample, and let $\mathcal F_n$ be the $\sigma$-field generated by the nuisance and structure samples, the tree randomization, and $(X_i)_{i=1}^n$. Because the forest is honest \citep{wager2018causal,athey2019grf}, the weights $\alpha_i(x)$ in~\eqref{eq:weights}, the nuisance fits, and $\varepsilon$ are fixed given $\mathcal F_n$, whereas the pairs $(A_i,Y_i)$ are conditionally independent, each with the conditional law of $(A,Y)$ given $X=X_i$. The three sources of error are measured by
\begin{equation}\label{eq:metric_local}s_n^2=\sum_i\alpha_i(x)^2,\qquad b_n=\sum_i\alpha_i(x)d\{\theta(X_i),\theta(x)\},\qquad B_n=\sum_i\alpha_i(x)\rho_\varepsilon(X_i).\end{equation}
The sampling scale $s_n$ is the inverse square root of the effective sample size; equal weights on $k$ observations give $s_n=k^{-1/2}$. The neighborhood bias $b_n$ measures how far the weighted effects $\theta(X_i)$ are from $\theta(x)$. The term $B_n$ averages the nuisance remainder $\rho_\varepsilon$ from~\eqref{eq:metric_remainders}.

We allow approximate minimization in~\eqref{eq:estimator}: the estimate $\widehat\theta=\widehat\theta(x)\in K_\theta$ satisfies
\begin{equation}\label{eq:metric_approx}\widehat L_x(\widehat\theta)\le\inf_{z\in K_\theta}\widehat L_x(z)+\zeta_n\end{equation}
for a tolerance $\zeta_n\ge0$ chosen without the estimation treatments and outcomes. Exact minimization gives $\zeta_n=0$. Otherwise, Theorem~\ref{thm:metric_rate} below includes an error of order $\sqrt{\zeta_n}$; minimization over a finite set of candidates within distance $h$ of every effect gives $\zeta_n\le C_0h$ when $A_Q(Y)$ is bounded, since $\widehat L_x$ is then $C_0$-Lipschitz.

\begin{assumption}\label{ass:metric_entropy}
There is $J\ge1$ such that, for $\theta=\theta(x)$ and every $0<\delta\le2D_{\theta}$,
\begin{equation}\label{eq:metric_entropy}\int_0^1\sqrt{1+\log N\{t\delta,K_\theta\cap B(\theta,\delta),d\}}\,dt\le J.\end{equation}
Here $N(\eta,S,d)$ is the covering number of $S$ by metric balls of radius $\eta$.
\end{assumption}

This entropy condition bounds the local fluctuations of the criterion, as in Fr\'echet regression and barycenter estimation \citep{petersen2019frechet,ahidarcoutrix2020barycenters}. A polynomial covering bound $(c_0/t)^{d_0}$ with fixed $c_0$ and $d_0$ suffices; compact smooth subsets of finite-dimensional manifolds and bounded open books satisfy it.

\begin{theorem}\label{thm:metric_rate}
Under Assumptions~\ref{ass:causal}--\ref{ass:metric_transport}, \ref{ass:metric_profiles}, and~\ref{ass:metric_entropy}, there are fixed constants $C_2,C_3<\infty$, depending only on $D_\theta$, $D_K$, and the constants in Assumption~\ref{ass:metric_profiles}, such that, for every fixed $x$, every estimate satisfying~\eqref{eq:metric_approx}, and every $v\ge1$,
\begin{equation}\label{eq:metric_rate}P\left\{d(\widehat\theta,\theta)>C_2\left[b_n+B_n+v s_nJ+\sqrt{\zeta_n}\right]\,\middle|\,\mathcal F_n\right\}\le C_3v^{-2}.\end{equation}

Consequently, $d(\widehat\theta,\theta)=O_p\{b_n+B_n+s_nJ+\sqrt{\zeta_n}\}$.
\end{theorem}

The four terms in~\eqref{eq:metric_rate} are neighborhood bias, residual nuisance bias, sampling error, and optimization error. With bounded local entropy, the sampling error is of the order of the inverse square root of the effective sample size, and the estimator is consistent at $x$ when all four terms vanish. Nuisance estimation contributes a lower-order term when $B_n=o_p(b_n+s_n)$. In terms of the nuisance errors, let $\bar r_\alpha^2=\sum_i\alpha_i r(X_i)^2$ and $\bar q_\alpha^2=\sum_i\alpha_i q(X_i)^2$ be the weighted mean squared errors of the fitted risk functions and propensity scores. The Cauchy--Schwarz inequality gives $B_n\le \bar r_\alpha^2+\bar r_\alpha \bar q_\alpha+\varepsilon$, so the nuisance contribution is determined by the squared and product errors and the contamination size. Because $b_n$ and $B_n$ are unknown, the bound does not directly give confidence regions; comparisons of the corrected loss give regions with finite-sample coverage when bounds on these biases are available (Section~\ref{sec:inference} of the Supplement).

Theorem~\ref{thm:metric_rate} holds conditionally on the forest weights. Corollary~\ref{cor:forest_consistency} in Section~\ref{app:pointwise_proofs} of the Supplement shows that the estimator is consistent at $x$ if $\theta$ is continuous at $x$, the nuisance remainder vanishes uniformly, the optimization tolerance vanishes, and every leaf containing $x$ retains a growing number of estimation observations while its diameter tends to zero in probability. By Lemma~2 of \citet{wager2018causal}, the leaf conditions hold for random-split trees with uniform covariates whose splits keep a fixed fraction of the estimation observations in each child; Section~\ref{subsec:forest_estimation} imposes leaf sizes on the structure sample instead. For H\"older continuous effects, the neighborhood bias then decreases at a polynomial rate in the ratio of sample size to leaf size, with an exponent that deteriorates with the number of covariates; the overall rate also depends on the sampling, nuisance, and optimization terms in~\eqref{eq:metric_rate}. If the neighborhoods instead shrink only in the $p_0$ covariates governing a $\beta$-H\"older effect, $0<\beta\le1$, the error attains the faster rate $n^{-\beta/(2\beta+p_0)}$ when $B_n+\sqrt{\zeta_n}$ is no larger (Section~\ref{app:pointwise_proofs} of the Supplement). This rate assumes a localization that neither Corollary~\ref{cor:forest_consistency} nor Theorem~\ref{thm:localized_split} establishes.

Algorithm~\ref{alg:forest} minimizes the criterion over the feasible set in~\eqref{eq:feasible_estimator}, which depends on the estimated baseline. Theorem~\ref{thm:feasible_rate} in the Supplement shows that this adds a term $L_F\delta_n$ to the bound, where $\delta_n=d\{\widehat b(x),m_0(x)\}$ is the baseline error and $L_F$ is the Lipschitz constant of alignment in~\eqref{eq:metric_stability}. Through the corresponding bound for reconstruction, consistent baseline and effect estimates yield a reconstructed geodesic~\eqref{eq:reconstruct} that is consistent for $\tau_G(x)$ in the path metric~\eqref{eq:path_metric} under Assumption~\ref{ass:geodesic}.

\section{Simulation studies}\label{sec:simulations}
\subsection{Simulation scenarios}\label{subsec:simulation_models}
In a noncommuting setting and a matched commuting control with diagonal matrices, all outcomes belong to $\mathbb S_{++}^2$ with the affine-invariant metric of Example~\ref{ex:spd}, and all effects are aligned at $o=I$. The six pretreatment covariates are independent uniform variables on $[-1,1]$, and treatment is randomized independently with probability $1/2$. Given a symmetric matrix $\Omega(x)$ and an aligned effect $\theta(x)$, the potential outcomes are
\begin{equation}\label{eq:simulation_outcomes}\begin{aligned}
m_0(x)&=\exp\{\Omega(x)\},\qquad m_1(x)=m_0(x)^{1/2}\theta(x)m_0(x)^{1/2},\\
Y(a)&=m_a(X)^{1/2}\exp(\sigma Z_a)m_a(X)^{1/2},\qquad a\in\{0,1\},
\end{aligned}\end{equation}
where $\sigma=0.25$ and $Z_0,Z_1$ are independent random symmetric matrices, specified below, with $\E(Z_a)=0$ and $\norm{Z_a}_F=1$. The observed outcome is $Y=Y(A)$. The treated mean is the reconstruction $m_1(x)=G_I\{m_0(x),\theta(x)\}$ of $\theta(x)$ at the untreated mean. Each potential outcome lies at distance $\sigma$ from $m_a(X)$, and, because $Z_a$ has mean zero, $m_a(x)$ is its conditional Fr\'echet mean. The target~\eqref{eq:theta} is thus exactly $\theta(x)=m_0(x)^{-1/2}m_1(x)m_0(x)^{-1/2}$.

In the noncommuting setting, let $R(\phi)$ be the $2\times2$ rotation by angle $\phi$ and set
\begin{equation}\label{eq:simulation_directional}\begin{gathered}
\theta(x)=R(\alpha x_2)\diag\{e^{\lambda(x)},e^{-\lambda(x)}\}R(\alpha x_2)^\top,\qquad \lambda(x)=1.3+\delta x_1,\\
\Omega(x)=\frac{0.4}{\sqrt6}\begin{pmatrix}u_1(x)+u_2(x)&u_3(x)\\u_3(x)&u_1(x)-u_2(x)\end{pmatrix},\qquad Z_a=\frac1{\sqrt3}\begin{pmatrix}\xi_{a1}&\xi_{a3}/\sqrt2\\\xi_{a3}/\sqrt2&\xi_{a2}\end{pmatrix},
\end{gathered}\end{equation}
where $u_1(x)=(x_3+0.5x_5)/1.5$, $u_2(x)=(x_4+0.5x_6)/1.5$, $u_3(x)=(x_5-x_6)/2$, and the signs $\xi_{a1},\xi_{a2},\xi_{a3}$ are independent Rademacher variables, drawn separately for each arm and observation. The coordinate $x_1$ changes the effect magnitude $d\{I,\theta(x)\}=\sqrt2\,\lambda(x)$, and $x_2$ rotates the eigenvectors of the effect. The covariates $x_3,\ldots,x_6$ change only the untreated mean, whose distance $\norm{\Omega(x)}_F$ from $I$ is at most $0.4$. We take $\delta=0.6$ and $\alpha=\delta/[2\{\E\sinh^2\lambda(X)\}^{1/2}]\approx0.1548$, so that equal small changes in $x_1$ and $x_2$ contribute equally, in expectation, to the squared distance.

The commuting control uses diagonal matrices: $\theta(x)=\diag\{e^{\lambda(x)+v(x)},e^{-\lambda(x)+v(x)}\}$, $\Omega(x)=(0.4/\sqrt2)\diag\{u_1(x),u_2(x)\}$, and $Z_a=\diag(\xi_{a1},\xi_{a2})/\sqrt2$, with $v(x)=\beta x_2/\sqrt2$, so that $x_2$ multiplies both eigenvalues of the effect by $e^{v(x)}$ instead of rotating its eigenvectors. We take $\beta\approx0.7382$, which makes the Fr\'echet variance of the aligned effects, their expected squared distance from their Fr\'echet mean, equal to that in the noncommuting setting, approximately $0.4216$. For diagonal matrices, the affine-invariant distance is the Euclidean distance between log-diagonal coordinates, so the GCF solves the squared-error DR learning problem in~\eqref{eq:metric_euclidean} in these coordinates. Section~\ref{app:simulation_laws} of the Supplement derives $\alpha$ and $\beta$ and verifies the stated properties of the outcome laws.

\subsection{Implementation and evaluation}
The study uses total sample sizes $n_{\mathrm{tot}}\in\{600,1200,2400\}$ and 200 independent Monte Carlo replications per configuration. For the GCF, each sample is divided equally into nuisance, structure, and estimation observations. The risk functions are fitted as in~\eqref{eq:profilefit}, with weights from arm-specific regression forests for the entries of $\log Y$; these coordinates only make the weights fast to compute (Section~\ref{app:simulations} of the Supplement). The untreated baseline is $\widehat b=\widetilde m_0$, and the known propensity score is used. The GCF uses 200 trees, grown until the leaf-size and split-admissibility rules prevent further splitting.

We compare the GCF with three estimators. The uncorrected forest replaces the corrected loss $\psi_z^\varepsilon(O_i)$ by the fitted effect loss $\mathcal T_z^{\widetilde Q}(X_i)$ in both split selection and final estimation, with the same samples, nuisance fits, tree randomization, and tuning, so that it differs from the GCF only through the correction. The plug-in estimator applies Fr\'echet regression separately within the treatment arms: it refits the arm-specific nuisance forests to all $n_{\mathrm{tot}}$ observations and aligns their estimated conditional means at $I$. The common effect minimizes the corrected loss with equal weights on the estimation observations. Section~\ref{app:simulation_protocol} of the Supplement gives the remaining tuning parameters.

All estimators use the same sample in each replication and are evaluated at the same 400 independent covariate values $X_j^*$ by the integrated squared geodesic error (ISE),
\begin{equation}\label{eq:simulation_ISE}\operatorname{ISE}(\widehat\theta)=\frac1{400}\sum_{j=1}^{400}d^2\{\widehat\theta(X_j^*),\theta(X_j^*)\}.\end{equation}

\subsection{Results}\label{subsec:simulation_results}
Table~\ref{tab:simulation_primary} reports the integrated squared geodesic error of the four estimators. The error of the GCF decreases with the sample size in both settings, from $0.068$ to $0.030$ in the noncommuting setting and from $0.055$ to $0.024$ in the commuting control. The error of the common effect stays near $0.42$, the population Fr\'echet variance of the aligned effects and hence the smallest expected error of any single effect, so the forest recovers most of the variation in the effects even at the smallest sample size.

\begin{table}[!htbp]
\centering
\begingroup
\small
\setlength{\tabcolsep}{4pt}
\renewcommand{\arraystretch}{1.15}
\begin{tabular}{@{}lr*{4}{r@{\hspace{0.3em}}l}@{}}
\toprule
Setting & $n_{\mathrm{tot}}$ & \multicolumn{2}{c}{GCF} & \multicolumn{2}{c}{Uncorrected} & \multicolumn{2}{c}{Plug-in} & \multicolumn{2}{c}{Common effect} \\
\midrule
Noncommuting & 600 & \textbf{6.799} & (1.247) & 21.553 & (1.632) & 12.962 & (1.026) & 42.593 & (1.575) \\
 & 1{,}200 & \textbf{4.274} & (0.791) & 16.543 & (1.248) & 10.582 & (0.808) & 42.550 & (1.515) \\
 & 2{,}400 & \textbf{3.011} & (0.410) & 13.182 & (1.004) & 8.723 & (0.656) & 42.208 & (1.639) \\
\addlinespace[5pt]
Commuting & 600 & 5.493 & (0.927) & 14.912 & (1.376) & \textbf{5.376} & (0.621) & 42.550 & (1.401) \\
 & 1{,}200 & \textbf{3.312} & (0.470) & 9.136 & (0.823) & 3.343 & (0.345) & 42.503 & (1.342) \\
 & 2{,}400 & 2.386 & (0.284) & 6.025 & (0.523) & \textbf{2.101} & (0.207) & 42.207 & (1.506) \\
\bottomrule
\end{tabular}

\endgroup
\caption{Integrated squared geodesic error in the noncommuting SPD setting and its commuting control, matched in effect Fr\'echet variance. The uncorrected forest omits the correction, and the plug-in aligns separately estimated conditional means. Each cell reports the mean over 200 replications and the standard deviation across replications in parentheses, both in units of $10^{-2}$. Boldface indicates the smallest mean error in each row.}\label{tab:simulation_primary}
\end{table}

The correction accounts for most of this accuracy. The uncorrected forest, whose loss inherits the first-order errors of the fitted means, has 3.2 to 4.4 times the error of the GCF in the noncommuting setting and 2.5 to 2.8 times in the commuting control, and is less accurate in every replication. The plug-in, although it fits the conditional means on three times as many observations, has 1.9 to 2.9 times the error of the GCF in the noncommuting setting, again in every replication. There the log coordinates of the treated mean combine the baseline and the rotating effect through matrix products, and the arm-specific fits target these coordinates rather than the effect. In the commuting control, these coordinates are sums of those of the untreated mean and the effect, which coordinatewise forests estimate well, and the plug-in is as accurate as the GCF or slightly more so, with mean paired differences below $0.003$. Section~\ref{app:simulation_protocol} of the Supplement separates the roles of the correction in splitting and final estimation; with coarser nuisance fits (noncommuting, $n_{\mathrm{tot}}=1{,}200$), the error of the GCF increases by 12\% and that of the uncorrected forest by 56\%.

A least-squares line fitted to the logarithm of the mean GCF error against $\log n_{\mathrm{tot}}$ has slope $-0.59$ in the noncommuting setting and $-0.60$ in the commuting control. Over these sample sizes the error thus decreases at least as fast as $n_{\mathrm{tot}}^{-1/2}$, the square of the localized rate in Section~\ref{subsec:metric_rates} for a Lipschitz effect of $p_0=2$ covariates; three sample sizes cannot establish an asymptotic rate, and that rate assumes a localization not established for the forest. The noncommuting setting, in which the effects also rotate, has a 24\% to 29\% larger error at every sample size but a similar rate of decrease.

\begin{figure}[!htb]
\centering
\includegraphics[width=\textwidth]{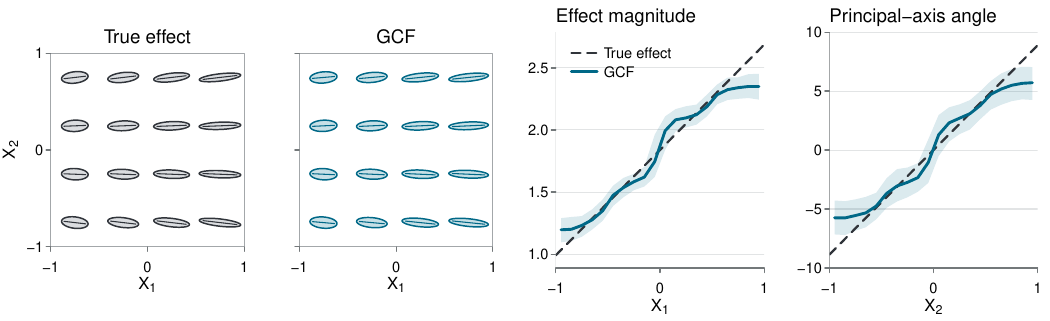}
\caption{Recovery of aligned SPD effects in the noncommuting setting at $n_{\mathrm{tot}}=1200$. Each ellipse in the first two panels represents the Fr\'echet mean of the true effects (first panel) or GCF estimates (second panel) in one of 16 equal covariate cells, with equal weight per replication. Its semiaxes follow the eigenvectors, with lengths proportional to the square roots of the eigenvalues on the same scale in both panels. The last two panels compare the true magnitude $d(I,\theta)$ and principal-axis angle, in degrees, with GCF estimates averaged in 20 bins of the indicated covariate. Shading spans the 5th--95th percentiles of the 200 replication-specific bin means. Angles are averaged modulo $180^\circ$.}\label{fig:simulation_recovery}
\end{figure}

Figure~\ref{fig:simulation_recovery} represents the true and estimated aligned effects in the noncommuting setting by ellipses whose axes follow the eigenvectors of the effect matrices. As $X_1$ increases from left to right, the true ellipses become longer and narrower; as $X_2$ increases from bottom to top, their long axes rotate. The fitted ellipses reproduce both patterns. The last two panels show that the estimated magnitude and principal-axis angle track their true values but flatten near the extremes of the corresponding covariates.

The mean squared errors in the effect magnitude $d(I,\widehat\theta)$ and in the reconstructed treated mean $G_I(\widehat b,\widehat\theta)$, computed with the same nuisance-sample baseline $\widehat b$ for all estimators, follow the same pattern (Table~\ref{tab:simulation_secondary} of the Supplement).

\Needspace{9\baselineskip}
\section{Asset transfers and livelihood change in Bangladesh}\label{sec:brac_application}
\subsection{Program, sample, and outcome}
The BRAC evaluation introduced in Section~\ref{subsec:livelihood_example} randomized forty branches in twenty pairs, assigning one branch in each pair to receive the program in 2007 and the other after 2011. We study outcomes in 2011, four years after the intervention and before control branches entered the program. The public replication data contain 6,732 baseline-eligible women in a balanced 2007, 2009, and 2011 panel: 4,045 in treated branches and 2,687 in control branches. We weight branches equally and women equally within branches. Interpretation as an assignment effect in this released cohort assumes that follow-up selection preserves the randomized comparison; the release does not recover the complete original eligible roster.

The six-part livelihood composition $S_i$ in~\eqref{eq:livelihood_composition} includes the 382 women reporting no work in 2011 through its nonparticipation coordinate, avoiding selection on posttreatment participation. Estimation uses its square-root representation $Y_i\in\mathbb S^5_+$ with angular distance, while total annual hours describe the scale of labor supply. Section~\ref{app:spherical_profiles} of the Supplement verifies the conditions of Section~\ref{sec:geometry}, Section~\ref{subsec:loss_regularity}, and Assumption~\ref{ass:metric_entropy} for this outcome, including women who report no work or a single activity. The statistical guarantees of Section~\ref{sec:theory}, however, assume independent observations, whereas the program was assigned to branches. We therefore keep randomized pairs intact, treat the fitted effects as descriptive summaries, and assess the held-out comparison across pairs at the end of Section~\ref{subsec:brac_responses}.

The aligned effect may vary with pretreatment assets, total hours, livelihood shares, age, adult-equivalent household size, and female headship. Section~\ref{app:brac_analysis} of the Supplement gives the data construction and estimation details.

\subsection{Heterogeneous livelihood responses}\label{subsec:brac_responses}
Paired branch comparisons of scalar outcomes (Section~\ref{app:brac_paired} of the Supplement) show the scale of the response. From 2007 to 2011, annual livestock hours increase by 420.7 hours more in treated than in control branches, with a 95\% confidence interval of $[345.9,495.5]$ based on the twenty randomized pairs and a $t_{19}$ reference distribution. Total recorded work rises by only 233.7 hours, with interval $[70.3,397.1]$, so livestock work partly replaces other activities, and work participation increases by 8.28 percentage points, with interval $[3.46,13.09]$. These contrasts do not show which activities livestock work replaces, or for whom.

The GCF addresses both questions through the joint composition. For each woman, applying the estimated aligned effect at her fitted untreated mean through~\eqref{eq:reconstruct} gives estimates of her untreated and treated conditional Fr\'echet means, whose squared coordinates are livelihood compositions. Unlike arithmetic mean shares, these compositions describe a central allocation of time; their nonparticipation coordinate, for example, is smaller than the proportion of women not working. We average these compositions over women, weighting branches equally, and over three sample partitions in which each prediction is computed without data from its own randomized pair (Section~\ref{app:brac_validation} of the Supplement). Across nonparticipation, livestock, agricultural wage work, domestic wage work, poultry, and other activities, the untreated composition $(3.3,16.5,18.3,25.3,22.4,14.3)\%$ moves to $(1.1,59.1,8.3,6.9,17.8,6.9)\%$ under treatment. The livestock share rises by 42.7 percentage points, and declines in casual wage work offset two thirds of this gain: domestic and agricultural wage work fall by 18.3 and 10.0 points, other activities by 7.5, poultry by 4.6, and nonparticipation by 2.2. This shift from casual wage labor to livestock rearing agrees with the findings of \citet{bandiera2017labor}.

\begin{figure}[!htbp]
\centering
\includegraphics[width=\textwidth]{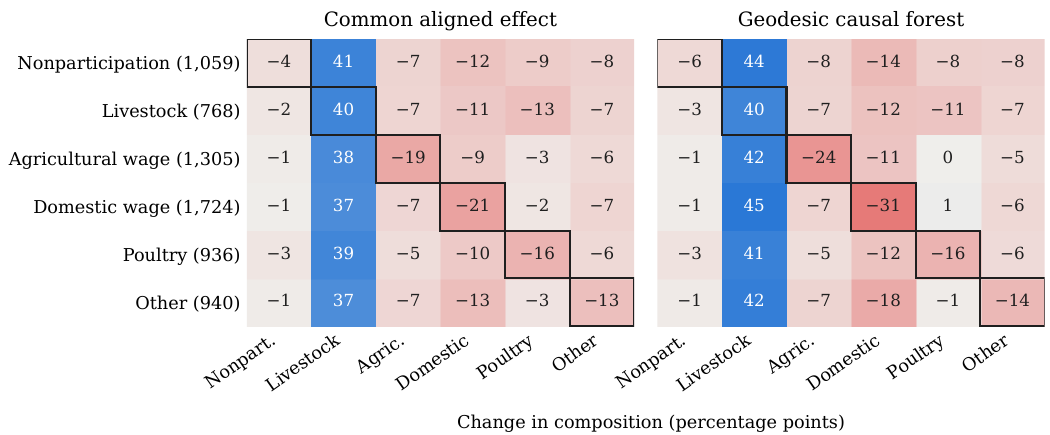}
\caption{Changes in conditional Fr\'echet means of the livelihood composition, in percentage points, by main 2007 activity, the category with the largest baseline share; numbers of women are in parentheses. Each cell averages, over women in the row group, the difference between the reconstructed treated and untreated compositions in the column activity, weighting branches equally and averaging over three sample partitions. Outlined cells give the change in each group's main 2007 activity. The left panel applies a single common aligned effect at each woman's untreated mean, and the right panel applies the effect estimated by the GCF.}\label{fig:brac_livelihood}
\end{figure}

Figure~\ref{fig:brac_livelihood} shows that the activities replaced depend on a woman's main 2007 activity, the category with the largest baseline share. The livestock share rises by 40 to 45 points in every group. Among women mainly in agricultural wage work, domestic wage work, or poultry, the largest reduction falls on that activity: 31.4 points of domestic wage work among women mainly in domestic service, 24.2 points of agricultural wage work among agricultural laborers, and 15.9 points of poultry among poultry keepers. Women mainly in other activities give up 14.5 points of these activities and 18.4 points of domestic wage work. Among women not working in 2007, livestock rearing replaces domestic wage work most, by 14.4 points, while nonparticipation falls by 5.8 points.

Part of this pattern follows from where women start. A woman can give up only activities she performs: since shares are squared coordinates on the sphere, a change $\Delta$ in the square root of a share $s$ changes the share by approximately $2s^{1/2}\Delta$. The same aligned effect therefore removes more of an activity that occupies a larger part of a woman's time. The left panel of Figure~\ref{fig:brac_livelihood} applies a single aligned effect fitted for all women, hereafter the common effect, at each woman's untreated mean. It already reduces the main activity of wage, poultry, and other workers by 13.0 to 20.7 points, in the same order as the forest. Paired componentwise contrasts within these groups also show large reductions in the main activity (Table~\ref{tab:brac_main_activity} of the Supplement). These percentage-point differences are effects on the share scale and are meaningful in their own right. Under the spherical geometry and transport to the uniform composition, however, much of their variation is consistent with a common aligned effect, so the aligned effects describe heterogeneity beyond what differences in baseline activities produce on this scale. Relative to the common effect, the forest estimates stronger exits from casual wage work: 31.4 rather than 20.7 points among former domestic workers and 24.2 rather than 19.2 points among agricultural laborers, whereas the estimated response of women mainly engaged in poultry is close to the common effect. The magnitude $d\{o,\widehat\theta_{\mathrm F}(x)\}$ of the estimated aligned effect, 0.46 for the common effect, averages 0.56 for former domestic workers, the largest among the six groups, and 0.49 to 0.54 in the others. It also rises with age, from 0.49 for women younger than 32 to 0.58 for women aged 47 or older, and age is the covariate most often chosen for splitting. These subgroup comparisons describe the fitted effects; we do not attach measures of uncertainty to them.

To check that this heterogeneity improves prediction, we compare the forest with the common effect and with separate aligned effects for four prespecified groups, defined by any baseline productive assets and any baseline casual wage work, on randomized pairs held out from fitting (Section~\ref{app:brac_validation} of the Supplement). On held-out pairs, the corrected loss~\eqref{eq:metric_psi} estimates the effect loss $\mathcal T_z^Q$ in~\eqref{eq:metric_T}, so a difference in held-out loss between two predictors estimates half the difference in their mean squared distances from the true aligned effects, up to the second-order nuisance remainder of Proposition~\ref{prop:metric_bias}, which known treatment probabilities do not remove. Averaged over three partitions of the pairs, the forest reduces the mean held-out loss by $0.0098$ relative to the common effect, and the reduction is positive in 16 of the 20 pairs. The prespecified groups reduce the loss by $0.0051$; the additional reduction by the forest, $0.0048$, is positive in 13 pairs and small relative to its variation across pairs. Each pair also helps fit the predictors that score other pairs, so these pair-level differences are not independent; we report them as descriptive summaries, and they do not measure uncertainty in the subgroup patterns above.

\section{Discussion}\label{sec:discussion}
Orthogonal learners for Euclidean outcomes remove first-order nuisance errors through outcome residuals, which are undefined for random objects because they cannot be added or subtracted. A forest, however, uses the effects only through the losses of candidate effects, and each loss is a real number, so the GCF corrects the loss instead of the outcome. The correction requires only Fr\'echet means, leaves a second-order nuisance bias in both split gains and effect estimates, and reduces to DR learning for Euclidean outcomes. In the BRAC analysis, a single aligned effect produces different percentage-point changes at different baseline activities, because it removes more of an activity that occupies more of a woman's time. Alignment separates this variation from heterogeneity in the aligned effects; percentage-point changes and aligned effects are both valid summaries, on different scales.

Several directions remain open. The faster rate in Section~\ref{subsec:metric_rates} assumes that the forest neighborhoods shrink only in the covariates that govern the effect. Whether splits chosen by the corrected loss produce such neighborhoods is not yet known. Resampling the loss comparisons could bring the coverage of the confidence regions in Section~\ref{sec:inference} of the Supplement closer to the nominal level, and extending them to clustered assignment, such as the randomization of BRAC branches, would attach measures of uncertainty to fitted patterns like those in Section~\ref{sec:brac_application}. The forest could also be extended beyond binary treatments with unconfounded assignment, to multi-valued or continuous treatments \citep{bhattacharjee2025continuous} and to difference-in-differences and regression discontinuity designs for random objects \citep{zhou2025gdid,kurisu2025grdd}.

\bibliographystyle{plainnat-nodoi}
\bibliography{references}

\clearpage
\begin{center}
\bfseries SUPPLEMENTARY MATERIAL
\end{center}
\setcounter{section}{0}
\renewcommand{\thesection}{S.\arabic{section}}
\renewcommand{\theHsection}{supp.\arabic{section}}
\counterwithout{theorem}{section}
\setcounter{theorem}{0}
\renewcommand{\thetheorem}{S\arabic{theorem}}
\renewcommand{\theHtheorem}{supp.\arabic{theorem}}
\counterwithout{equation}{section}
\setcounter{equation}{0}
\renewcommand{\theequation}{S\arabic{equation}}
\renewcommand{\theHequation}{supp.\arabic{equation}}
\setcounter{figure}{0}
\renewcommand{\thefigure}{S\arabic{figure}}
\renewcommand{\theHfigure}{supp.\arabic{figure}}
\setcounter{table}{0}
\renewcommand{\thetable}{S\arabic{table}}
\renewcommand{\theHtable}{supp.\arabic{table}}

\section{Geometric preliminaries and further examples}\label{app:geometry}

\subsection{Geodesics and Fr\'echet means}

The length of a continuous curve $c:[0,1]\to\M$ is
\[L(c)=\sup_{0=t_0<\cdots<t_k=1}\sum_{j=1}^k d\{c(t_{j-1}),c(t_j)\},\]
where the supremum is over all finite partitions of $[0,1]$. The constant-speed identity~\eqref{eq:geodesic_speed} gives $L(\gamma_{u,v})=d(u,v)$, the smallest possible length of a curve joining $u$ and $v$. When geodesics between points of $K$ are unique in the ambient space, $K$ is geodesically convex if $\gamma_{u,v}(t)\in K$ for all $u,v\in K$ and $t\in[0,1]$. This extends convexity of Euclidean sets by replacing line segments with geodesics \citep{sturm2003probability}.

Finite second moments do not depend on the reference point, because $d^2(u,Y)\le2d^2(u,u_0)+2d^2(u_0,Y)$. In Euclidean space, the identity $\E\|Y-u\|^2=\E\|Y-\E(Y)\|^2+\|u-\E(Y)\|^2$ shows that the Fr\'echet mean in~\eqref{eq:frechet_mean} is the ordinary mean; conditioning gives the same identity for conditional means. In a general metric space, existence and uniqueness depend on both the space and the outcome law. Assumption~\ref{ass:geodesic} imposes these properties on the causal targets, and Proposition~\ref{prop:spherical_profiles} verifies them for all laws of compositions.

A Hadamard space is a complete geodesic metric space satisfying the CAT(0) nonpositive-curvature comparison inequality. In such a space, squared distance satisfies
\[d^2\{\gamma_{u,v}(t),w\}\le(1-t)d^2(u,w)+td^2(v,w)-t(1-t)d^2(u,v),\qquad t\in[0,1].\]
This inequality makes squared-distance risks strongly convex along geodesics. Every probability measure with finite second moment has a unique Fr\'echet mean, and for the half squared distance $\ell_z(v)=d^2(z,v)/2$, the variance inequality is
\[\E\{\ell_z(Z)-\ell_{\bar z}(Z)\}\ge\frac12d^2(z,\bar z),\qquad \bar z=\argmin_u\E\ell_u(Z).\]
These properties also hold for minimization over a closed convex set \citep[Propositions~2.3, 4.3, and~4.4]{sturm2003probability}.

\subsection{Further examples of outcome spaces}\label{app:more_examples}
\begin{example}[Univariate probability distributions]
A region's income distribution distinguishes broad gains from changes concentrated in either tail, as in the analysis of wage densities by \citet{dinardo1996labor}. In a regional policy evaluation, each region contributes a distribution-valued outcome, and the question is how its response varies with pretreatment regional characteristics.

For probability measures $\mu,\nu$ on $\R$ with finite second moments and quantile functions $Q_\mu,Q_\nu$, the $2$-Wasserstein metric and geodesics are \citep{panaretos2020wasserstein}:
\[d_{\mathcal W}^2(\mu,\nu)=\int_0^1\{Q_\mu(q)-Q_\nu(q)\}^2\,dq,\qquad Q_{\gamma_{\mu,\nu}(t)}=(1-t)Q_\mu+tQ_\nu.\]
This representation places the outcome space in a convex subset of $L^2(0,1)$ and preserves the interpretation in income units at each quantile.
\end{example}

\begin{example}[Networks]
A country's production network records interdependencies among sectors, which contribute to aggregate fluctuations \citep{acemoglu2012network}. Trade policy can reorganize production through input substitution, with changes that aggregate output or trade may not reveal.

For an undirected weighted network on $r$ common labeled nodes, let $a_{jk}=a_{kj}\ge0$ denote the edge weights. Its Laplacian has entries $L_{jk}=-a_{jk}$ for $j\ne k$ and $L_{jj}=\sum_{k\ne j}a_{jk}$ \citep{zhou2022network,severn2022networks}. Symmetrizing production links gives this representation. Laplacians form a convex subset of a matrix space: the Frobenius metric $d_F(L_1,L_2)=\|L_1-L_2\|_F$ gives geodesics $\gamma_{L_1,L_2}(t)=(1-t)L_1+tL_2$ that interpolate link weights while preserving the Laplacian constraints.
\end{example}

\begin{example}[Functional data]
A government-bond yield curve records borrowing costs across maturities \citep{diebold2006yields}, and a policy may affect short- and long-term rates differently. Age-specific fertility and mortality curves are further examples of functional outcomes \citep{hyndman2007functional}.

Functional data can be studied as random elements of $L^2(\mathcal T)$, where $\mathcal T$ is a compact interval \citep{wang2016functional,hsing2015functional}. The metric is $d(f,g)=\{\int_{\mathcal T}[f(v)-g(v)]^2\,dv\}^{1/2}$ and the geodesic is $\gamma_{f,g}(t)=(1-t)f+tg$. Conditional Fr\'echet means are ordinary conditional mean functions, and geodesic displacement is the functional mean difference.
\end{example}

\begin{example}[Phylogenetic trees]\label{ex:phylo_trees}
Evolutionary trees record relatedness through their branching patterns and divergence through branch lengths. \citet{nye2011trees} analyzes gene trees from animal species to identify variation in both features. The Billera--Holmes--Vogtmann space consists of trees on a common set of labeled leaves \citep{billera2001trees}. Each topology corresponds to a Euclidean orthant of nonnegative edge lengths; orthants meet where edges contract to zero. The metric is shortest-path length through these orthants. Geodesics are unique and interpolate edge lengths, possibly contracting and expanding edges as topology changes. This is a nonsmooth Hadamard space.
\end{example}

\section{Transport and regularity for specific outcome spaces}\label{app:model_verification}
This section constructs the alignment and reconstruction maps $F_o$ and $G_o$ in Assumption~\ref{ass:metric_transport} for the outcome spaces of Section~\ref{subsec:other_outcomes} and verifies the regularity conditions of Section~\ref{sec:theory} for them. The alignment map expresses a change at the reference point $o$, the reconstruction map applies an aligned effect at a baseline $u$, and the feasible set $\Gamma(u)$ in~\eqref{eq:feasible_domain} contains the effects that reconstruct to valid outcomes. The analysis of the corrected loss rests on the moment bound~\eqref{eq:risk_moment} and the first-order expansion~\eqref{eq:metric_first_order}--\eqref{eq:metric_vonmises} in Assumption~\ref{ass:metric_profiles}: the former bounds the random size of the centered loss $\bar k_Y$, and the latter controls how errors in the risk functions change the effect loss.

For Hilbert space-valued outcomes, $\bar k_y(u)=-\langle u-u_*,y-u_*\rangle$ and $\mathbb B$ is the subspace of affine functions in $\operatorname{Lip}_0(\M;u_*)$; for the smooth SPD and spherical models, the norm of $\mathbb B$ also bounds first and second derivatives near the minimizers, which govern how the minimizers of smooth risk functions move; and for open-book outcomes, $\mathbb B=\operatorname{Lip}_0(\M;u_*)$. On a Hadamard space, the quadruple inequality gives $\|\bar k_y\|_{\mathrm{Lip}}\le d(y,u_*)$, so that uniformly bounded second moments of $d(Y,u_*)$ under the true and fitted laws imply~\eqref{eq:risk_moment} for $\mathbb B=\operatorname{Lip}_0(\M;u_*)$ \citep{sturm2003probability,schotz2019quadruple}.

Expectations of Banach space-valued maps, such as $y\mapsto\bar k_y$ and $O\mapsto\psi^\varepsilon(O)$, are Bochner integrals; they require these maps to be strongly measurable, which we assume and verify for the models below. We fix $x$ and suppress it in the notation.

\subsection{Hilbert-space outcomes}\label{app:hilbert_profiles}
For an outcome space represented isometrically as a convex subset of a Hilbert space $\mathcal H$, we take $\overline{\M}=\mathcal H$ and use translation, $F_o(u,v)=o+v-u$ and $G_o(u,e)=u+e-o$. The inverse and length identities are immediate, and both stability bounds hold with constant one. This applies to quantile functions in $L^2(0,1)$, graph Laplacians with the Frobenius norm, and functional outcomes in $L^2(\mathcal T)$. The aligned endpoint $o+v-u$ may lie outside the set of quantile functions or graph Laplacians. Reconstruction at $u$ is feasible when $u+e-o$ belongs to the corresponding outcome set. For unrestricted $L^2$ functional outcomes, every reconstructed function is an admitted outcome.

The conditional means depend linearly on their risks, and the effect loss is affine in the mean difference, so the finite-difference correction is exact.

\begin{proposition}\label{prop:hilbert_profiles}
Let $\M=\overline{\M}=\mathcal H$ be a separable Hilbert space, with translation transport $F_o(u,v)=o+v-u$. Take $\mathbb B=\{f_w:u\mapsto\langle u-u_*,w\rangle,\ w\in\mathcal H\}$, with $\|f_w\|_{\mathbb B}=\|w\|$. Suppose the true and fitted laws are measurable probability kernels, the fits use an independent nuisance sample, and these laws have uniformly bounded second moments about $u_*$, and their means lie in sets $K_a$ whose aligned images are contained in a compact candidate set $K_\theta$ containing $o$. Then Assumption~\ref{ass:metric_profiles} holds, with $L_0=L=1$, $C=0$, and $\epsilon_*=1$, provided the fitted propensity scores satisfy their stated lower bound. The laws may have unbounded support.
\end{proposition}

\begin{proof}
Expanding squared norms gives $\bar k_y=f_{-(y-u_*)}$, hence $A_Q(y)=\|y-u_*\|$ and strong measurability. A risk is $Q_a=k_{u_*}+f_{w_a}$ and has unique minimizer $m_a=u_*-w_a$. Its aligned effect loss is
\[\mathcal T_z^Q=\frac12\|z-o\|^2-\langle z-o,m_1-m_0\rangle.\]
For a perturbation $h_a=f_{v_a}$, the exact loss increment is $\langle z-o,v_1-v_0\rangle$. Its Lipschitz norm over $K_\theta$ is at most $\|v_0\|+\|v_1\|$, proving both regularity bounds with zero remainder. All contaminated laws have finite second moments and ordinary means, and translation is defined at every resulting pair. Continuity of these formulas proves the remaining measurability assertions.
\end{proof}

\subsection{Riemannian manifolds and SPD matrices}\label{subsec:smooth_profile_verification}
When the ambient space $\overline{\M}$ is a Riemannian manifold, its metric tensor $g$ equips each tangent space $T_p\overline{\M}$ with an inner product $g_p$ and norm $\norm{w}_p=\{g_p(w,w)\}^{1/2}$. A tangent vector represents the initial velocity of a smooth curve through $p$, and the distance $d$ is the infimum of the lengths of curves joining two points. The exponential map $\Exp_p(w)$ is the endpoint at time one of the geodesic with initial point $p$ and velocity $w$, whenever that geodesic is defined up to time one \citep[Chapters~5--6]{lee2018riemannian}.

Let $\mathcal D_p\subset T_p\overline{\M}$ be a star-shaped open neighborhood of zero on which $\Exp_p$ is a diffeomorphism and its radial geodesics minimize length. The inverse on $\Exp_p(\mathcal D_p)$ is the logarithm map $\Log_p$. For $q$ in this normal neighborhood,
\[\gamma_{p,q}(t)=\Exp_p\{t\Log_p(q)\},\qquad \norm{\Log_p(q)}_p=d(p,q).\]
The vector $\Log_p(q)\in T_p\overline{\M}$ represents the direction and magnitude of the change from $p$ to $q$.

Parallel transport moves a tangent vector along a path while preserving its length and its inner products with other transported vectors. Let $c$ be the unique minimizing geodesic from $p$ to $q$. For $w\in T_p\overline{\M}$, the parallel vector field $V$ along $c$ satisfies $\nabla_{\dot c}V=0$ and $V(0)=w$, where $\nabla$ denotes the Levi--Civita covariant derivative associated with $g$. Write $\mathsf P_{p\to q}w=V(1)$, and take $\mathsf P_{q\to p}$ along the reversed path. These maps preserve inner products and are inverses:
\[g_q(\mathsf P_{p\to q}w_1,\mathsf P_{p\to q}w_2)=g_p(w_1,w_2),\qquad \mathsf P_{q\to p}\mathsf P_{p\to q}w=w.\]
In particular, parallel transport preserves velocity length. In Euclidean space, $\Exp_p(w)=p+w$, $\Log_p(q)=q-p$, and parallel transport is the identity under the common identification of tangent spaces with $\R^d$.

On a Riemannian manifold, alignment and reconstruction can be built from these operations. For each admitted baseline $u$, let $\mathsf P_{u\to o}$ denote parallel transport along the unique minimizing geodesic from $u$ to $o$ and $\mathsf P_{o\to u}$ parallel transport along its reverse. The initial velocity of a geodesic is moved to a new starting point by parallel transport, and the geodesic with the transported velocity is then followed from that point \citep{zhou2025gdid}. The alignment and reconstruction maps are thus given by
\begin{equation}\label{eq:spherical_transport}F_o(u,v)=\Exp_o\{\mathsf P_{u\to o}\Log_u(v)\},\qquad G_o(u,e)=\Exp_u\{\mathsf P_{o\to u}\Log_o(e)\}.\end{equation}
The working sets are chosen such that the logarithms are defined and
\[\mathsf P_{u\to o}\Log_u(v)\in\mathcal D_o,\qquad \mathsf P_{o\to u}\Log_o(e)\in\mathcal D_u,\qquad u\in K_0,\ v\in K_1,\ e\in K_\theta,\]
with $K_\theta$ containing $o$ and the image of $F_o$. These conditions on the logarithms and transported vectors hold globally on smooth Hadamard manifolds, where $\mathcal D_p=T_p\overline{\M}$.

The first map expresses the geodesic change at $o$ while preserving its magnitude:
\[d\{o,F_o(u,v)\}=\norm{\mathsf P_{u\to o}\Log_u(v)}_o=\norm{\Log_u(v)}_u=d(u,v).\]
The condition $\mathsf P_{u\to o}\Log_u(v)\in\mathcal D_o$ gives $\Log_o\{F_o(u,v)\}=\mathsf P_{u\to o}\Log_u(v)$, so reversing parallel transport recovers the treated endpoint:
\[G_o\{u,F_o(u,v)\}=\Exp_u\{\mathsf P_{o\to u}\mathsf P_{u\to o}\Log_u(v)\}=\Exp_u\{\Log_u(v)\}=v.\]
These are the length and inverse identities in~\eqref{eq:metric_inverse}. If the maps are smooth on neighborhoods of compact geodesically convex working sets, bounded derivatives give the Lipschitz bounds in~\eqref{eq:metric_stability} by integration along geodesics in their product sets.

With the affine-invariant geometry of Example~\ref{ex:spd} and reference $o=I$, transport standardizes the treated covariance by the untreated covariance, and reconstruction reverses this congruence transformation:
\begin{equation}\label{eq:spdrelative}F_I(U,V)=U^{-1/2}VU^{-1/2},\qquad G_I(U,E)=U^{1/2}EU^{1/2}.\end{equation}
For $U=m_0(x)$ and $V=m_1(x)$, the aligned contrast is $\theta(x)=F_I(U,V)$. Its smallest and largest eigenvalues bound the relative portfolio variance $w^\top Vw/(w^\top Uw)$ for every $w\ne0$, and its determinant gives the generalized-variance ratio.

The affine-invariant geometry in Example~\ref{ex:spd} has metric tensor $g_U(V,W)=\tr(U^{-1}VU^{-1}W)$ and logarithm and exponential maps \citep[Sections~3.3--3.4]{pennec2006tensor}
\[\Log_U(V)=U^{1/2}\log(U^{-1/2}VU^{-1/2})U^{1/2},\qquad \Exp_U(W)=U^{1/2}\exp(U^{-1/2}WU^{-1/2})U^{1/2}.\]
Parallel transport along the unique affine-invariant geodesic from $U$ to $I$ sends $W$ to $U^{-1/2}WU^{-1/2}$, and transport along the reverse path sends $W$ to $U^{1/2}WU^{1/2}$. Substitution into~\eqref{eq:spherical_transport} gives the covariance standardization and reconstruction maps in~\eqref{eq:spdrelative}.

Congruence preserves positive definiteness, so every admitted effect reconstructs to an SPD matrix and $\Gamma(U)=K_\theta$. The SPD maps in~\eqref{eq:spdrelative} satisfy Assumption~\ref{ass:metric_transport} on compact endpoint sets and a compact geodesically convex effect set containing $I$ and the transport image. The inverse identity is immediate. The length identity holds because the congruence $V\mapsto U^{-1/2}VU^{-1/2}$ is an isometry of the affine-invariant metric that maps $U$ to $I$. The maps are smooth, so the Lipschitz bounds follow by integrating bounded derivatives, as for the general construction above.

For smooth outcomes, the mean equation $\nabla Q_a(m_a)=0$ shows how a risk perturbation moves its minimizing mean. Applying the transport map to the two perturbed means then changes the effect loss. The calculation below bounds the linear response and its remainder in the Lipschitz norm required by Assumption~\ref{ass:metric_profiles}.

Let $\overline{\M}=\M$ be a finite-dimensional smooth Hadamard manifold. For the bounded-support models in this subsection, we choose a compact geodesically convex neighborhood $C_U$ containing all minimizing means and a compact effect set $K_\theta$, both with nonempty interiors. We evaluate risk derivatives on $C_U$ while defining the means by minimization over $\M$. The functions below are smooth on neighborhoods of these sets. Fix $u_*\in C_U$ and $o\in K_\theta$, and define
\[\norm{h}_{2}=\sup_{u\in C_U}|h(u)|+\sup_{u\in C_U}\norm{\nabla h(u)}+\sup_{u\in C_U}\norm{\nabla^2h(u)}_{\mathrm{op}},\qquad \mathbb B_2=\{h\in C^2(C_U):h(u_*)=0\}.\]
One may take $C_U$ to be a closed geodesic ball, so that $C^2(C_U)$ has its usual Banach-space interpretation on a compact smooth set with boundary. The embedding $\mathbb B_2\hookrightarrow\operatorname{Lip}_0(C_U;u_*)$ is continuous, since integration of the gradient along the geodesic between two points gives $\norm h_{\mathrm{Lip}}\le\sup_{C_U}\norm{\nabla h}\le\norm h_2$. Equip pairs of functions with the sum norm.

For an endpoint pair $u=(u_0,u_1)$ at which transport is defined, set
\begin{equation}\label{eq:endpoint_loss}J(z,u)=\frac12\left[d^2\{z,F_o(u_0,u_1)\}-d^2\{o,F_o(u_0,u_1)\}\right],\qquad z\in K_\theta.\end{equation}
Here $u=(u_0,u_1)$ collects the two conditional means, and $J(z,u)$ is the anchored loss for predicting their aligned effect by $z$. Derivatives in $u$ describe how errors in these means change that loss. We use operator norms and parallel transport to compare the derivatives at different endpoint pairs.

\begin{assumption}\label{ass:smooth_profile_verification}
All relevant risk functions $Q_a$ induced by probability laws, their fitted versions, and the contaminations used in the correction have unconstrained Fr\'echet minimizers in a fixed compact subset of $\operatorname{int}(C_U)$, and the alignment $F_o$ is defined at all relevant endpoint pairs. Their restrictions to $C_U$ satisfy, for fixed $c,M_3>0$,
\begin{equation}\label{eq:smooth_risk_bounds}\nabla^2Q_a(u)\succeq cI,\qquad \sup_{u\in C_U}\norm{\nabla^3Q_a(u)}\le M_3.\end{equation}
The centered losses $\bar k_y=k_y-k_{u_*}$ satisfy $\sup_y\|\bar k_y\|_2\le M_Q<\infty$ on the specified outcome support. The function $J$ is twice continuously differentiable in the endpoints on the geodesic hull of all pairs under consideration. There is a fixed $C_J<\infty$ such that, uniformly over these pairs and $z,z'\in K_\theta$,
\begin{equation}\label{eq:smooth_J_bounds}\norm{D_uJ(z,u)-D_uJ(z',u)}_{\mathrm{op}}+\norm{D_u^2J(z,u)-D_u^2J(z',u)}_{\mathrm{op}}\le C_Jd(z,z').\end{equation}
Here $D_u^2$ denotes the covariant Hessian on the product of the two endpoint sets. These bounds hold, for example, when $F_o$ is smooth on a neighborhood of the compact endpoint sets and its required mixed derivatives are bounded there.
\end{assumption}

The Hessian lower bound prevents a small risk perturbation from moving the mean too far. The third-derivative bound controls the remainder in the expansion of $\nabla Q_a(m_a)=0$, following intrinsic-mean asymptotics \citep[Section~2]{bhattacharya2005large}. The bounds for $J$ ensure that the resulting approximation controls loss differences between any two candidate effects.

\begin{proposition}\label{prop:smooth_profile_verification}
Suppose Assumption~\ref{ass:smooth_profile_verification} holds. For a relevant pair of risk functions $Q=(Q_0,Q_1)$, write $u_a=m_a^Q$, $u=(u_0,u_1)$, $H_a=\nabla^2Q_a(u_a)$, and $\mathcal T_z^Q=J(z,u)$. Define the bounded linear operator from $\mathbb B_2^2$ to $\operatorname{Lip}_0(K_\theta;o)$ by
\begin{equation}\label{eq:smooth_profile_derivative}\dot{\mathcal T}_{z}^{Q}[h]=-\sum_{a=0}^1D_{u_a}J(z,u)\left[H_a^{-1}\nabla h_a(u_a)\right].\end{equation}
There are fixed constants $L,C<\infty$ such that, for every relevant pair $Q'=Q+h$,
\begin{equation}\label{eq:smooth_profile_remainder}\norm{\dot{\mathcal T}^{Q}[h]}_{\mathrm{Lip}}\le L\norm h_{\mathbb B_2^2},\qquad \norm{\mathcal T^{Q'}-\mathcal T^Q-\dot{\mathcal T}^{Q}[h]}_{\mathrm{Lip}}\le C\norm h_{\mathbb B_2^2}^{2}.\end{equation}
The mean-displacement bound in the proof and the first endpoint-derivative bound for $J$ also give $\|\mathcal T^{Q'}-\mathcal T^Q\|_{\mathrm{Lip}}\le L_0\|h\|_{\mathbb B_2^2}$. Thus these models satisfy both bounds in Assumption~\ref{ass:metric_profiles}.
\end{proposition}

\begin{proof}
We first bound the displacement of one arm's mean. Let $u$ and $v$ minimize $Q$ and $Q+h$, respectively, and put $\xi=\operatorname{Log}_u(v)$. Both means are interior, so $\nabla Q(u)=0$ and $\nabla(Q+h)(v)=0$. Strong convexity of $Q+h$ along the geodesic $\gamma(t)=\operatorname{Exp}_u(t\xi)$ gives
\[c\norm\xi^2\le-\langle\nabla h(u),\xi\rangle\le\norm{\nabla h(u)}\norm\xi,\qquad \norm\xi\le c^{-1}\norm h_2.\]
The first inequality follows by integrating the second derivative of $(Q+h)\{\gamma(t)\}$: its derivative at $t=1$ is zero and its derivative at $t=0$ is $\langle\nabla h(u),\xi\rangle$. It also holds trivially when $\xi=0$.

To expand the perturbed mean equation in one tangent space, transport its gradient at $v$ back to $T_u\M$. Integrating the covariant derivative of $\nabla Q$ and using the third-derivative bound yields
\[\mathsf P_{v\to u}\nabla Q(v)=H\xi+R_Q,\qquad \norm{R_Q}\le\frac12M_3\norm\xi^2,\qquad H=\nabla^2Q(u).\]
Similarly, the Hessian bound for $h$ gives
\[\mathsf P_{v\to u}\nabla h(v)=\nabla h(u)+R_h,\qquad \norm{R_h}\le\norm h_2\norm\xi.\]
Since the two transported gradients sum to zero, $\xi=-H^{-1}\{\nabla h(u)+R_Q+R_h\}$. The bounds $\norm{H^{-1}}_{\mathrm{op}}\le c^{-1}$ and $\norm\xi\le c^{-1}\norm h_2$ then imply
\begin{equation}\label{eq:smooth_mean_expansion}\xi=-H^{-1}\nabla h(u)+r_h,\qquad \norm{r_h}\le C\norm h_2^2.\end{equation}
All constants are uniform over the admitted risk functions.

We now propagate the two mean displacements through the effect loss. Apply~\eqref{eq:smooth_mean_expansion} to both arms, and write $\xi=(\xi_0,\xi_1)$ for the resulting product-manifold displacement. Along the product geodesic from $u=(u_0,u_1)$ to $v=(v_0,v_1)$, Taylor's formula gives
\[J(z,v)-J(z,u)=D_uJ(z,u)[\xi]+\int_0^1(1-t)D_u^2J\{z,\gamma(t)\}[\dot\gamma(t),\dot\gamma(t)]\,dt.\]
Subtract this formula at $z'$ from the one at $z$. Equation~\eqref{eq:smooth_J_bounds} bounds the resulting integral by $C d(z,z')(\norm{\xi_0}+\norm{\xi_1})^2$. Substituting~\eqref{eq:smooth_mean_expansion} into the linear term adds at most $C d(z,z')\sum_a\norm{h_a}_2^2$. Dividing by $d(z,z')$ proves the remainder bound in~\eqref{eq:smooth_profile_remainder}. All terms vanish at $z=o$, so the remainder belongs to the anchored Lipschitz space. The same increment bound for $D_uJ$, the inverse-Hessian bound, and $\norm{\nabla h_a(u_a)}\le\norm{h_a}_2$ prove boundedness of the linear operator in~\eqref{eq:smooth_profile_derivative}.
\end{proof}

On compact subsets of smooth Hadamard manifolds, squared distance is smooth and has a uniformly positive Hessian in its first argument. These properties, together with smooth transport, verify the preceding assumptions for the SPD construction above.

\begin{corollary}\label{cor:smooth_SPD_profiles}
Let the outcome support and relevant endpoint and effect sets be compact subsets of a finite-dimensional smooth Hadamard manifold, with all relevant means in the interior of the optimization set. Suppose the specified transport $F_o$ is smooth on a neighborhood of the endpoint sets. Then Assumption~\ref{ass:smooth_profile_verification} holds after enlarging the fixed compact working sets to contain the endpoint geodesics, provided the transport remains defined on that enlargement. In particular, this applies to SPD matrices with the affine-invariant metric and the smooth fixed-reference transport used in the manifold construction. The scalar functional expansion is valid in the $C^2$ norm.
\end{corollary}

\begin{proof}
On a smooth Hadamard manifold, the Cartan--Hadamard theorem and Hessian comparison imply that the squared-distance function is smooth on $\M\times\M$ and that the Hessian of $u\mapsto d^2(u,y)/2$ is bounded below by the identity \citep[Chapters~11--12]{lee2018riemannian}. Its covariant derivatives through order three are bounded on a fixed compact product of the optimization and outcome sets. Averaging with respect to any probability law supported on the outcome set preserves these derivative bounds and gives~\eqref{eq:smooth_risk_bounds} with $c=1$, together with the bound on $\|\bar k_y\|_2$. Smoothness also makes $y\mapsto\bar k_y$ continuous in the $C^2$ norm on a compact outcome set. Its range is therefore separable, and the centered observation losses are strongly measurable and Bochner integrable in $\mathbb B_2$.

Smoothness of the transport and squared distance bounds the endpoint derivatives of $J$ and their covariant derivatives in $z$ on the fixed compact sets. Integrating those $z$ derivatives along geodesics in $K_\theta$ proves~\eqref{eq:smooth_J_bounds}. Since $J(o,u)=0$, its endpoint derivatives are anchored as required. SPD matrices with the affine-invariant metric form a smooth Hadamard manifold; the usual exponential, logarithm, and parallel transport along the unique connecting geodesic are smooth. Consequently, the fixed-reference map built from these operations satisfies the asserted smoothness on compact sets, and the proposition applies.
\end{proof}

\subsection{Spherical compositions}\label{app:spherical_profiles}
For square-root compositions, we take $\M=\mathbb S^{m-1}_+$ and $\overline{\M}=\mathbb S^{m-1}$. For $\alpha=d(u,v)<\pi$ and a tangent vector $w\perp u$, the spherical logarithm and exponential maps are \citep[Section~2]{dai2022sphere}
\[\Log_u(v)=\frac{\alpha}{\sin\alpha}(v-\cos\alpha\,u),\qquad \Exp_u(w)=\cos(\|w\|)u+\frac{\sin(\|w\|)}{\|w\|}w,\]
using the continuous values at zero. Take a reference $o$ with strictly positive coordinates, such as the uniform composition $o=(1,\ldots,1)/\sqrt m$. Then $\langle u,o\rangle>0$ for every composition $u$, so the parallel transport below is defined on the whole outcome space. Parallel transport along the shorter geodesic from $u$ to $o$ is
\[\mathsf P_{u\to o}w=w-\frac{\langle w,o\rangle}{1+\langle u,o\rangle}(u+o).\]
Substituting these operations into~\eqref{eq:spherical_transport} gives the spherical transport and reconstruction. Here $\mathcal D_p$ is the tangent ball of radius $\pi$, so the transported velocities are restricted to lengths less than $\pi$. Aligned endpoints may lie outside the nonnegative orthant; Proposition~\ref{prop:spherical_feasibility} below gives the constraint for valid reconstructed compositions, including at zero shares.

Two properties of the nonnegative orthant allow outcome laws with unrestricted support. Any two compositions have a nonnegative inner product and are therefore at most $\pi/2$ apart, which keeps the logarithm, parallel transport, and squared distance smooth on the orthant. In addition, the Euclidean mean of a law on the orthant bounds the curvature of its Fr\'echet risk, so every such law has a nondegenerate Fr\'echet mean. The assumption below therefore restricts only the mean sets and the effect set.

\begin{assumption}\label{ass:spherical_domains}
Let $m\ge2$, $\M=\mathbb S^{m-1}_+$, $\overline{\M}=\mathbb S^{m-1}$, and fix a reference $o$ with strictly positive coordinates. Let $K_0,K_1\subset\M$ be closed geodesically convex sets containing the conditional and fitted means of the respective arms, and put $D_K=\max_{u\in K_0,v\in K_1}d(u,v)$. The effect set is the closed ball $K_\theta=\overline B(o,R)$ in the ambient sphere, with $0<R\le\pi/2$ and $R+D_K<\pi$. For a fixed $0<\eta_\theta<R$, the conditional effects satisfy $d\{m_0(x),m_1(x)\}\le R-\eta_\theta$ at the covariate values under consideration. The conditional potential-outcome laws and fitted probability laws are measurable probability kernels on $\M$.
\end{assumption}

The outcome laws may place mass on every face of the simplex, including the vertices that represent a single part. Since $D_K\le\pi/2$, the choice $K_0=K_1=\M$ is admissible for every $R<\pi/2$. The closed hemisphere $R=\pi/2$ contains every aligned effect and requires $D_K<\pi/2$, that is, no untreated mean in $K_0$ and treated mean in $K_1$ with disjoint supports. The condition $R+D_K<\pi$ keeps candidate effects away from the antipodes of aligned effects of fitted means, where squared distance is not smooth. Unique intrinsic means on positively curved spaces are usually obtained by confining the outcome support to a convexity radius \citep{afsari2011center}; here, the sign structure of the orthant replaces that restriction.

\begin{proposition}\label{prop:spherical_profiles}
Under Assumption~\ref{ass:spherical_domains}, the outcome space and the transport~\eqref{eq:spherical_transport} satisfy Assumptions~\ref{ass:geodesic} and~\ref{ass:metric_transport}, with $L_F=L_G=5\pi$. For a probability law $P$ on $\M$, write $Q_P(u)=\E_P\{d^2(u,Y)\}/2$ for its Fr\'echet risk and $\bar y_P=\E_P(Y)\in\R^m$ for the Euclidean mean of its square-root representation, and put $c_m=1-(2/\pi)\arccos(m^{-1/2})>0$. Then $P$ has a unique Fr\'echet mean $\mu_P\in\M$, and
\begin{equation}\label{eq:orthant_curvature}\begin{gathered}\nabla^2Q_P(u)\succeq\ip u{\bar y_P}I,\qquad\ip{\mu_P}{\bar y_P}\ge c_m,\\Q_P(v)-Q_P(\mu_P)\ge\frac{c_m}3d^2(v,\mu_P),\qquad u,v\in\M.\end{gathered}\end{equation}
Take $u_*=o$ and the anchored $C^2$ function space $\mathbb B_2$ of Section~\ref{subsec:smooth_profile_verification} with $C_U=\M$ and derivatives on the ambient sphere. Then the moment condition~\eqref{eq:risk_moment} holds, and, for a sufficiently small fixed $\epsilon_*>0$, the map $Q\mapsto\mathcal T^Q$ has the bounded linear expansion and quadratic Lipschitz remainder in~\eqref{eq:smooth_profile_derivative}--\eqref{eq:smooth_profile_remainder}, uniformly over pairs of risk functions induced by the true and fitted laws and their contaminations by outcomes in $\M$. With the stated independence and overlap conditions on the nuisance fits, these risk functions satisfy Assumption~\ref{ass:metric_profiles}. The effect set also satisfies Assumption~\ref{ass:metric_entropy} with a fixed entropy bound depending only on $m$.
\end{proposition}

For six-part compositions, as in Section~\ref{sec:brac_application}, $c_6\approx0.27$. The bounds in~\eqref{eq:orthant_curvature} hold for every law on the orthant, which gives the uniformity over fitted and contaminated laws required by Assumption~\ref{ass:metric_profiles}.

\begin{proof}
The orthant is a closed subset of the compact unit sphere and is therefore complete and separable. Any two of its points have a nonnegative inner product and hence distance at most $\pi/2$. The shorter spherical geodesic is unique and stays in the orthant, since its interior points are positive linear combinations of its endpoints followed by normalization. Its constant-speed parametrization depends continuously on the endpoints, including coincident endpoints.

We next prove~\eqref{eq:orthant_curvature}. For the unit sphere, with $a=d(u,y)<\pi$ and $e$ the unit tangent from $u$ toward $y$, the constant-curvature Hessian formula \citep[Chapter~11]{lee2018riemannian} gives $\Hess_u\{d^2(u,y)/2\}=e\otimes e+a\cot(a)(I-e\otimes e)$, with value $I$ at $u=y$. If $u,y\in\M$, then $a\le\pi/2$, and both eigenvalues are at least $\cos a=\ip uy$, because $a\cot a=(a/\sin a)\cos a\ge\cos a$. Averaging over $P$ gives the first inequality. The risk is continuous on the compact sphere and attains its minimum there. Replacing a minimizer by its coordinatewise absolute value gives a unit vector whose inner product with every $y\in\M$ is at least as large, so the risk does not increase. Hence some minimizer $\mu$ over the sphere lies in $\M$, and the gradient of $Q_P$ on the sphere vanishes at $\mu$. Let $\bar o=(1,\ldots,1)/\sqrt m$. For $y\in\M$, $\sum_jy_j\ge(\sum_jy_j^2)^{1/2}=1$, so $\ip{\bar o}y\ge m^{-1/2}$ and $d(\bar o,y)\le\arccos(m^{-1/2})$. Concavity of the cosine on $[0,\pi/2]$ gives $\cos a\ge1-2a/\pi$ there. Jensen's inequality and minimality of $\mu$ therefore yield
\[\ip\mu{\bar y_P}=\E_P\cos d(\mu,Y)\ge1-\frac2\pi\left\{\E_Pd^2(\mu,Y)\right\}^{1/2}\ge1-\frac2\pi\left\{\E_Pd^2(\bar o,Y)\right\}^{1/2}\ge c_m.\]
For $v\in\M$ with $\delta=d(\mu,v)>0$, the geodesic $\gamma(t)=[\sin\{(1-t)\delta\}\mu+\sin(t\delta)v]/\sin\delta$ lies in $\M$. Since $\ip v{\bar y_P}\ge0$ and $\sin\{(1-t)\delta\}\ge(1-t)\sin\delta$ by concavity of the sine on $[0,\pi]$, the previous display gives $\ip{\gamma(t)}{\bar y_P}\ge c_m(1-t)$. Taylor's formula along $\gamma$, whose initial derivative vanishes, and the first inequality in~\eqref{eq:orthant_curvature} give
\[Q_P(v)-Q_P(\mu)\ge c_m\delta^2\int_0^1(1-t)^2\,dt=\frac{c_m}3\delta^2.\]
Thus $\mu_P=\mu$ is the unique Fr\'echet mean in $\M$, and~\eqref{eq:orthant_curvature} holds for every probability law on $\M$. This verifies the mean condition in Assumption~\ref{ass:geodesic}.

For the transport, write $(a\otimes b)h=\ip bh\,a$ and, for $u\in\M$,
\[\mathsf R_u=\left\{I-\frac{(u+o)\otimes(u+o)}{1+\ip uo}\right\}(I-2u\otimes u)=I+2o\otimes u-\frac{(u+o)\otimes(u+o)}{1+\ip uo}.\]
Since $\norm{u+o}^2=2(1+\ip uo)$ and $\ip uo\ge0$, $\mathsf R_u$ is the product of the reflections in the hyperplanes perpendicular to $u$ and $u+o$. It satisfies $\mathsf R_uu=o$ and $\mathsf R_uw=\mathsf P_{u\to o}w$ for $w\perp u$. Orthogonal maps are isometries of the sphere and commute with its exponential maps, so~\eqref{eq:spherical_transport} gives $F_o(u,v)=\mathsf R_uv$ for $v\in\M$ and $G_o(u,e)=\mathsf R_u^*e$ for $d(o,e)<\pi$. Consequently, $G_o\{u,F_o(u,v)\}=v$, $d\{o,F_o(u,v)\}=d(u,v)$, and $G_o(u,o)=u$. For $u,u'\in\M$, put $c=\ip uo$ and $c'=\ip{u'}o$. Subtracting the rank-one expressions, whose denominators are at least one, gives
\[\norm{\mathsf R_u-\mathsf R_{u'}}_{\mathrm{op}}\le2\norm{u-u'}+4\norm{u-u'}+4|c-c'|\le10\norm{u-u'}.\]
Orthogonal maps preserve angular distance, and $\norm{a-b}\le d(a,b)\le(\pi/2)\norm{a-b}$ for unit vectors $a,b$. Hence
\[\begin{aligned}d\{F_o(u,v),F_o(u',v')\}&\le d(v,v')+5\pi d(u,u'),\\d\{G_o(u,e),G_o(u',e')\}&\le d(e,e')+5\pi d(u,u'),\end{aligned}\]
which gives~\eqref{eq:metric_stability} with $L_F=L_G=5\pi$. Finally, if $u\in K_0$ and $d\{u,m_0(x)\}\le\eta_\theta$, the length identity gives $d[o,F_o\{u,m_1(x)\}]=d\{u,m_1(x)\}\le d\{m_0(x),m_1(x)\}+\eta_\theta\le R$. This verifies Assumption~\ref{ass:metric_transport}.

For the regularity of the risks, let $U$ be the set of unit vectors within distance $\pi/8$ of $\M$. For $u\in U$ and $y\in\M$, $d(u,y)\le5\pi/8$ and $1+\ip uo>0$. Squared distance to outcomes, the spherical logarithm, parallel transport, and $F_o$ are therefore smooth on $U$ and its products, with bounded derivatives. Consequently, $\sup_{y\in\M}\|\bar k_y\|_2<\infty$, the map $y\mapsto\bar k_y$ is continuous in $\mathbb B_2$, and geodesic integration of the gradient gives $\norm h_{\mathrm{Lip}}\le\norm h_2$. These observations verify the moment condition~\eqref{eq:risk_moment} and strong measurability. Averaging the derivatives of $k_y$ bounds the third derivatives of every risk function on $\M$ by a fixed constant $M_3$. Anchoring does not affect the means, the Hessians, or differences of risk values.

The proof of Proposition~\ref{prop:smooth_profile_verification} applies with $C_U=\M$ and $c=c_m$, with one change. The means $u$ and $v$ of two relevant risk functions $Q$ and $Q+h$ minimize over the sphere, so their gradients vanish even on a face of the orthant, and the variance inequality in~\eqref{eq:orthant_curvature} replaces strong convexity along the connecting geodesic in the displacement bound:
\[\frac{c_m}3d^2(u,v)\le Q(v)-Q(u)\le h(u)-h(v)\le\norm h_2d(u,v).\]
Thus $d(u,v)\le3c_m^{-1}\norm h_2$. In particular, a contaminated mean lies within $6c_m^{-1}M_Q\varepsilon$ of the fitted mean, where $M_Q=\sup_{y\in\M}\|\bar k_y\|_2$. For the effect loss $J$ in~\eqref{eq:endpoint_loss}, every $z\in K_\theta$ and $(u_0,u_1)\in K_0\times K_1$ satisfy $d\{z,F_o(u_0,u_1)\}\le R+D_K<\pi$, and this separation from $\pi$ persists for the contaminated pairs when $\epsilon_*$ is small. Since squared distance is smooth away from antipodal pairs, $J$ is twice continuously differentiable on the geodesic hull of the relevant pairs, and integrating its bounded $z$-derivatives along geodesics in the convex ball $K_\theta$ gives~\eqref{eq:smooth_J_bounds}. The Hessian bound $\nabla^2Q(u)\succeq c_mI$ at the mean, the third-derivative bound, and the displacement bound then give the expansion~\eqref{eq:smooth_mean_expansion} and, as in that proof, the bounds~\eqref{eq:smooth_profile_derivative}--\eqref{eq:smooth_profile_remainder} and~\eqref{eq:metric_first_order}, with constants that do not depend on the laws. Continuity of the means and losses gives strong measurability of the fitted and contaminated losses; the probability-kernel assumption gives strong measurability of the risks.

Finally, normal coordinates at $o$ are uniformly bi-Lipschitz on $K_\theta$, since $R\le\pi/2$. The image of $K_\theta\cap B(\theta,\delta)$ lies in a Euclidean ball of radius $C\delta$ in $\R^{m-1}$. A Euclidean packing argument followed by the inverse coordinate map gives $N\{t\delta,K_\theta\cap B(\theta,\delta),d\}\le(C'/t)^{m-1}$ for $0<t\le1$, with constants uniform in $\theta$ and $\delta$. Substitution into~\eqref{eq:metric_entropy} proves the entropy assertion.
\end{proof}

Smaller effect sets also satisfy the curvature conditions on the effect loss.

\begin{proposition}\label{prop:effect_curvature_domains}
Under Assumption~\ref{ass:spherical_domains} with $R<\pi/2$, the effect set $K_\theta$ satisfies the mixed-increment bound~\eqref{eq:effect_quadruple} in Section~\ref{sec:inference} with $\Lambda_Z=2R/\sin(2R)$. If $R<\pi/4$, it also satisfies Assumption~\ref{ass:effect_curvature} in Section~\ref{app:forest} with $\kappa_Z=2R\cot(2R)$.
\end{proposition}

\begin{proof}[Proof of Proposition~\ref{prop:effect_curvature_domains}]
The ball $K_\theta$ is geodesically convex because $R<\pi/2$, and every distance between two of its points is at most $2R$. If $R<\pi/4$, the Hessian formula used in the proof of Proposition~\ref{prop:spherical_profiles} gives radial eigenvalue $1$ and transverse eigenvalue $a\cot a$ for $\ell_v$, where $a=d(z,v)\le2R<\pi/2$, with the continuous value $1$ when $a=0$. Thus its Hessian is bounded below by $2R\cot(2R)>0$ on $K_\theta$. Integrating twice along $\gamma_{z,u}$ gives~\eqref{eq:effect_strong_convexity}.

For the mixed-increment bound, differentiation of the spherical exponential gives radial singular value $1$ and transverse singular value $\sin(a)/a$ at a tangent vector of length $a$. Its inverse therefore gives
\[\norm{D_v\Log_z(v)}_{\mathrm{op}}\le\frac{a}{\sin a}\le\frac{2R}{\sin(2R)},\]
with the continuous interpretation at $a=0$. Set $z_t=\gamma_{z,u}(t)$ and $v_s=\gamma_{v,w}(s)$. Since $\nabla_z\ell_z(v)=-\Log_z(v)$, differentiating $f(t,s)=\ell_{z_t}(v_s)$ yields
\[|\partial_s\partial_t f(t,s)|=\left|\left\langle D_{v_s}\Log_{z_t}(v_s)[\dot v_s],\dot z_t\right\rangle\right|\le\Lambda_Zd(z,u)d(v,w).\]
Integration over $[0,1]^2$ proves~\eqref{eq:effect_quadruple}. All derivatives are smooth at coincident points and throughout these sets.

\end{proof}

The feasible set in~\eqref{eq:feasible_domain} requires the reconstructed composition to have nonnegative coordinates. Writing an aligned effect by its initial velocity at $o$ converts this requirement into convex inequalities for that velocity.

\begin{proposition}\label{prop:spherical_feasibility}
Let $u\in\mathbb S^{m-1}_+$, let $B_u:\R^{m-1}\to T_u\mathbb S^{m-1}$ be an orthonormal frame obtained by parallel transport of a fixed frame at $o$, and write an effect as $e=\Exp_o(B_o v)$ with $r=\norm v\le\pi/2$. Define $f(0)=1$ and $f(r)=r\cot r$ for $0<r\le\pi/2$. Then
\begin{equation}\label{eq:spherical_feasibility}G_o(u,e)=\cos(r)u+\frac{\sin r}{r}B_uv\in\mathbb S^{m-1}_+\quad\Longleftrightarrow\quad B_uv+f(r)u\ge0,\end{equation}
with coordinatewise inequality and the continuous convention at zero. The set $\{v:\norm v\le\pi/2,\ B_uv+f(\norm v)u\ge0\}$ is closed and convex. Every reconstructed geodesic from $u$ to a feasible endpoint remains in $\mathbb S^{m-1}_+$.
\end{proposition}

\begin{proof}[Proof of Proposition~\ref{prop:spherical_feasibility}]
The spherical exponential gives the equality in~\eqref{eq:spherical_feasibility}. For $0<r\le\pi/2$, division by the positive factor $\sin r/r$ gives the equivalent coordinate inequalities; continuity gives the result at zero. For $0<r\le\pi/2$, we have $f'(r)=\cot r-r\csc^2r<0$ and $f''(r)=2\csc^2r\{r\cot r-1\}<0$. The inequalities follow from $\sin r\cos r<r$ and $r\cot r<1$. Thus the radial function $v\mapsto f(\norm v)$ is concave: away from zero its radial and tangential Hessian eigenvalues are $f''(r)$ and $f'(r)/r$, and its Hessian at zero is $-2I/3$. Since each $u_j\ge0$, the coordinate constraint $(B_uv)_j+u_jf(\norm v)\ge0$ is a superlevel set of a concave function. Their intersection with the closed radius-$\pi/2$ ball is closed and convex and contains zero. Intersections for several baselines have these properties. Finally, the shorter geodesic between two nonnegative spherical points is a normalized positive linear combination of the endpoints, or is constant when they coincide. It remains in the orthant.
\end{proof}

\subsection{Open books and tree-valued outcomes}\label{app:openbook}
For tree-valued outcomes within a fixed topology, nonnegative edge-length vectors form a convex Euclidean orthant, so the translations of Section~\ref{app:hilbert_profiles} apply in edge-length coordinates, and feasibility requires nonnegative reconstructed edge lengths. Changes of topology require a branching space. An open book consists of Euclidean half-spaces joined along a common boundary, called the spine. It is the simplest branching space and supplies a nonsmooth example of the method: untreated means lie on the spine, while treated means can lie on a page. The branch conditions below keep each mean on the same page or on the spine when its risk is perturbed.

Let $T_k$, $k\ge3$, consist of $k$ copies of $[0,\infty)$ joined at their origins. Denote the common vertex by $v_*$ and a point on branch $j$ by $(r,j)$, where all points $(0,j)$ represent $v_*$. Its metric and the product metric on $\M=\R^d\times T_k$ are
\[d_T\{(r,j),(s,l)\}=\begin{cases}|r-s|,&j=l,\\r+s,&j\ne l,\end{cases}\qquad d^2\{(b,t),(c,s)\}=\norm{b-c}^2+d_T^2(t,s).\]
Each branch becomes a Euclidean half-space, and the half-spaces meet along the spine $\R^d\times\{v_*\}$. The resulting space is complete and CAT(0), with a singularity at every spine point. Its Fr\'echet means can remain on the spine under perturbations of the outcome law.

Unrooted trees with four labeled leaves give a concrete instance. Their single interior edge separates the leaves in one of three ways, so its length and split form $T_3$, with $v_*$ the star tree, whose interior edge has length zero. The four pendant edge lengths lie in $[0,\infty)^4$. The tree space of Example~\ref{ex:phylo_trees} with four leaves is therefore the convex subset $[0,\infty)^4\times T_3$ of $\R^4\times T_3$ \citep{billera2001trees}. Take $\M=[0,\infty)^4\times T_3$ and $\overline{\M}=\R^4\times T_3$, and intersect the sets $K_0$ and $K_1$ in~\eqref{eq:openbook_domains} with $\M$. Fr\'echet means of laws on $\M$ remain in $\M$, so the results below apply with $d=4$ and $k=3$; aligned effects lie in $\overline{\M}$, and the feasible reconstruction of Section~\ref{subsec:feasible_estimation} keeps reconstructed edge lengths nonnegative. Under Assumption~\ref{ass:openbook} below, the untreated mean is the star tree, and the treated mean either remains the star tree or resolves the interior edge into one of the three splits.

\begin{assumption}\label{ass:openbook}
Outcomes have the form $Y=(Z,(R,J))\in\R^d\times T_k$, with $R\le R_{\max}$ almost surely. True conditional laws and fitted laws satisfy $\E\|Z\|^2\le L^2$ uniformly, for fixed $L,R_{\max}>0$. The horizontal coordinate $Z$ may be unbounded. Set $u_*=o=(0,v_*)$ and $\mathbb B=\operatorname{Lip}_0(\M;o)$. Let $B_L=\{b\in\R^d:\norm b\le L\}$ and $T_{k,R_{\max}}=\{t\in T_k:d_T(t,v_*)\le R_{\max}\}$, and use
\begin{equation}\label{eq:openbook_domains}K_0=B_L\times\{v_*\},\qquad K_1=B_L\times T_{k,R_{\max}},\qquad K_\theta=B_{2L}\times T_{k,R_{\max}}.\end{equation}
For $j=1,\ldots,k$, define $f_j(r,l)=r\{2\mathbf1(l=j)-1\}$, with $f_j(v_*)=0$, and $q_{aj}(x)=\E\{f_j(R,J)\mid A=a,X=x\}$. There is a constant $\kappa>0$ such that $q_{0j}(x)\le-\kappa$ for every $j$. For the treated arm, either $q_{1j_*}(x)\ge\kappa$ for one branch $j_*$ or $q_{1j}(x)\le-\kappa$ for every $j$. These conditions hold at all covariate values to which the result is applied. Fitted risk functions are generated by probability laws satisfying the same radial and moment bounds. For the finite-contamination statements, $\norm{\widetilde Q_a-Q_a}_{\mathrm{Lip}}\le\kappa/4$, $0<\varepsilon\le\min\{1,\kappa/(8R_{\max})\}$, and the fitted propensity scores satisfy $\widetilde p_a\ge\epsilon_0>0$. Fits are independent of the observations used to evaluate the loss.
\end{assumption}

The folded moment $q_{aj}(x)$ is the expected radial distance contributed by branch $j$ minus that contributed by all other branches. The untreated inequalities place $m_0(x)$ on the spine. A positive treated folded moment places the treated mean on the corresponding page; negative moments on every page place it on the spine. The strict margins preserve this page or spine membership under fitted-risk and contamination perturbations, as in the sticky and nonsticky regimes of \citet{hotz2013sticky}. For example, on a three-branch tree with $R=1$, untreated probabilities $(1/3,1/3,1/3)$ give folded moments $(-1/3,-1/3,-1/3)$, while treated probabilities $(0.70,0.15,0.15)$ give $(0.40,-0.70,-0.70)$.

The alignment below translates the horizontal coordinates, moving the untreated spine mean to $o$. The formulas for the means follow the folded-moment characterization in \citet[Sections~2--4]{hotz2013sticky}; the accompanying norm bound relates risk estimation error to errors in these moments.

\begin{proposition}\label{prop:openbook_geometry}
Under Assumption~\ref{ass:openbook}, for $u=(b,v_*)\in\R^d\times\{v_*\}$, $v=(c,t)\in\M$, and $e=(a,s)\in\overline{\M}$, define
\begin{equation}\label{eq:openbook_transport}F_o\{(b,v_*),(c,t)\}=(c-b,t),\qquad G_o\{(b,v_*),(a,s)\}=(b+a,s).\end{equation}
These maps satisfy Assumption~\ref{ass:metric_transport}, with $L_F=L_G=1$ on the sets in~\eqref{eq:openbook_domains}. For any finite-second-moment law $P$ on $\M$, write $\mu(P)=\E_P Z$ and $q_j(P)=\E_P f_j(R,J)$. Its risk function anchored at $o$ is
\begin{equation}\label{eq:openbook_profile}Q_P(b,r,j)=\frac12\E_P\{d^2((b,(r,j)),Y)-d^2(o,Y)\}=\frac12\norm b^2+\frac12r^2-b^\top\mu(P)-r q_j(P).\end{equation}
At most one $q_j(P)$ is positive, and the unique Fr\'echet mean is
\begin{equation}\label{eq:openbook_mean}m(P)=\begin{cases}(\mu(P),(q_j(P),j)),&q_j(P)>0\text{ for some }j,\\(\mu(P),v_*),&q_j(P)\le0\text{ for every }j.\end{cases}\end{equation}
The Lipschitz norm~\eqref{eq:metric_B} satisfies
\begin{equation}\label{eq:openbook_profile_norm}\begin{aligned}\max\left\{\norm{\mu(P)-\mu(P')},\max_j|q_j(P)-q_j(P')|\right\}&\le\norm{Q_P-Q_{P'}}_{\mathrm{Lip}},\\\norm{Q_P-Q_{P'}}_{\mathrm{Lip}}&\le\norm{\mu(P)-\mu(P')}+\max_j|q_j(P)-q_j(P')|.\end{aligned}\end{equation}
\end{proposition}

\begin{proof}[Proof of Proposition~\ref{prop:openbook_geometry}]
For each baseline on the spine, both maps in~\eqref{eq:openbook_transport} are horizontal translations, hence isometries on $\M$. They are inverses, send the baseline to $o$, and preserve its distance to the treated endpoint. Comparing $(c-b,t)$ with $(c'-b',t')$ through the intermediate point $(c-b',t)$, the triangle inequality gives $d\{F_o(u,v),F_o(u',v')\}\le d(u,u')+d(v,v')$, and the same argument applies to $G_o$. This is~\eqref{eq:metric_stability} with constants one, and the horizontal displacement has norm at most $2L$, giving the required image inclusion. Formula~\eqref{eq:openbook_profile} follows by expanding the horizontal squared distance and using $d_T^2\{(r,j),(R,J)\}=r^2+R^2-2r f_j(R,J)$. For distinct $j,l$, $q_j(P)+q_l(P)=-2\E_P[R\mathbf1\{J\notin\{j,l\}\}]\le0$, so two folded moments cannot both be positive. Minimization within branch $j$ gives horizontal coordinate $\mu(P)$ and radial coordinate $(q_j(P))_+$. This proves~\eqref{eq:openbook_mean}, including uniqueness at the vertex. Since $\norm{\mu(P)}\le(\E_P\norm Z^2)^{1/2}\le L$ and $q_j(P)\le R_{\max}$, Assumption~\ref{ass:openbook} places the untreated mean in $K_0$ and the treated mean in $K_1$.

Let $h=Q_P-Q_{P'}$, $\Delta\mu=\mu(P)-\mu(P')$, and $\Delta q_j=q_j(P)-q_j(P')$. Then $h(b,r,j)=-b^\top\Delta\mu-r\Delta q_j$. Comparing the origin with a short horizontal displacement in the direction of $\Delta\mu$, and then with a short radial displacement on each branch, proves the lower bound in~\eqref{eq:openbook_profile_norm}. The function $(r,j)\mapsto r\Delta q_j$ is Lipschitz with constant $\max_j|\Delta q_j|$: on different branches its increment is bounded by that constant times $r+s$. The upper bound follows by combining this inequality with Cauchy--Schwarz for the horizontal coordinate. Since every difference between risk functions vanishes at $o$, the Lipschitz seminorm is a norm on these differences.
\end{proof}

The mean and effect sets in~\eqref{eq:openbook_domains} are compact and convex, while the outcome law can have unbounded horizontal support. By~\eqref{eq:openbook_profile}, $\bar k_y(b,r,j)=-b^\top Z_y-r f_j(t_y)$. Reshetnyak's quadruple inequality \citep[Proposition~2.4]{sturm2003probability} gives $\|\bar k_y-\bar k_{y'}\|_{\mathrm{Lip}}\le d(y,y')$ and hence $A_Q(y)\le d(y,o)$. The kernel is continuous in this norm, so separability gives strong measurability, and the moment condition holds with $M_Q^2=L^2+R_{\max}^2$. Measurability of ordinary and folded moments gives measurable risks and means. For the entropy bound, view each page in its Euclidean coordinates. A metric ball intersects the page inside a Euclidean ball of the same radius; when the original center lies on another page, reflect its radial coordinate across the spine to obtain the Euclidean center. Covering each page separately therefore gives
\[N\{t\delta,K_\theta\cap B(\theta,\delta),d\}\le k(C/t)^{d+1},\qquad 0<t\le1,\]
for a constant depending only on the dimension. Integration verifies Assumption~\ref{ass:metric_entropy}.

Within a fixed page or spine, the mean formula~\eqref{eq:openbook_mean} is linear in the ordinary and folded moments. The anchored effect loss is affine in those moments, so its finite-difference correction is exact. The next result verifies the expansion required by Assumption~\ref{ass:metric_profiles}.
\begin{proposition}\label{prop:openbook_affinity}
Suppose Assumption~\ref{ass:openbook} holds, and fix a covariate value $x$. Suppress $x$ in the notation, let $Q=(Q_{P_0},Q_{P_1})$ and $\delta=\mu(P_1)-\mu(P_0)$, and write a candidate effect as $z=(c,t)\in K_\theta$. In the treated-page regime with positive folded moment $q_{1j_*}$,
\begin{equation}\label{eq:openbook_affine}\mathcal T_z^Q=\frac12d^2(z,o)-c^\top\delta-q_{1j_*}f_{j_*}(t).\end{equation}
In the strictly sticky treated regime, the final term is absent. Let $h_a$ be a difference between risk functions and denote its moment differences by $\dot\mu_a$ and $\dot q_{aj}$. Within either fixed regime, define the linear functionals
\begin{equation}\label{eq:openbook_D}D_{0,z}[h_0]=c^\top\dot\mu_0,\qquad D_{1,z}[h_1]=-c^\top\dot\mu_1-f_{j_*}(t)\dot q_{1j_*},\end{equation}
omitting the final term in the sticky treated regime. For any two pairs of risk functions $Q,\widetilde Q$ with untreated means on the spine and the same treated regime and page,
\[\mathcal T_z^Q-\mathcal T_z^{\widetilde Q}=\sum_{a=0}^1D_{a,z}[Q_a-\widetilde Q_a],\qquad |(D_{a,z}-D_{a,z'})[h_a]|\le\sqrt2\,d(z,z')\norm{h_a}_{\mathrm{Lip}}.\]
Thus the scalar von Mises remainder and its increment in $z$ are identically zero.

Under the fitted-risk and contamination bounds in Assumption~\ref{ass:openbook}, for $k_y(u)=\{d^2(u,y)-d^2(o,y)\}/2$, define $\widetilde Q_a^{a,\varepsilon,y}=(1-\varepsilon)\widetilde Q_a+\varepsilon k_y$, and let $\widetilde Q^{a,\varepsilon,y}$ denote the pair $\widetilde Q$ with arm $a$ replaced by this contaminated risk function. Every such contamination remains in the same regime, with all relevant margins at least $\kappa/2$, and
\begin{equation}\label{eq:openbook_finite_difference}\frac{\mathcal T_z^{\widetilde Q^{a,\varepsilon,y}}-\mathcal T_z^{\widetilde Q}}{\varepsilon}=D_{a,z}[k_y-\widetilde Q_a].\end{equation}
The same regime is preserved along convex mixtures of the true and fitted risk functions. The finite-contamination correction therefore has no approximation error in this example.
\end{proposition}

\begin{proof}[Proof of Proposition~\ref{prop:openbook_affinity}]
In the treated-page regime,~\eqref{eq:openbook_mean} and~\eqref{eq:openbook_transport} give the aligned endpoint $\theta^Q=(\delta,(q_{1j_*},j_*))$. Expanding~\eqref{eq:metric_T} cancels the terms $\norm\delta^2/2$ and $q_{1j_*}^2/2$, leaving~\eqref{eq:openbook_affine}. If the treated mean is sticky, its radial coordinate is zero and the same argument removes the final term. Moment differences are linear in differences between risk functions by~\eqref{eq:openbook_profile}, so~\eqref{eq:openbook_D} gives the exact expansion. Each folding function $f_j$ is 1-Lipschitz: increments on a common branch are bounded by the radial difference, while increments between branches are bounded by the sum of the radii. Consequently, for arm one,
\[|(D_{1,z}-D_{1,z'})[h_1]|\le\norm{c-c'}\norm{\dot\mu_1}+d_T(t,t')|\dot q_{1j_*}|\le\sqrt2\,d(z,z')\norm{h_1}_{\mathrm{Lip}}.\]
The arm-zero inequality follows with constant one.

By~\eqref{eq:openbook_profile_norm}, every fitted folded moment differs from its true value by at most $\kappa/4$. The relevant fitted margins are therefore at least $3\kappa/4$. In the active-page case, all other true folded moments are at most $-\kappa$, because their sum with the active moment is nonpositive. Convex mixtures of true and fitted risk functions preserve all these linear inequalities. A contamination replaces a fraction $\varepsilon$ of the fitted law by an outcome with $R\le R_{\max}$. Since $|f_j|\le R_{\max}$, each folded moment changes by at most $2R_{\max}\varepsilon\le\kappa/4$. All relevant margins thus remain at least $\kappa/2$. In particular the contaminated untreated mean stays on the spine, where the transport is defined. Applying the affine expansion within this unchanged regime proves~\eqref{eq:openbook_finite_difference}.
\end{proof}

To obtain the operator on $\mathbb B$ required by Assumption~\ref{ass:metric_profiles}, we extend the moment differences in~\eqref{eq:openbook_D} from risk differences to arbitrary anchored Lipschitz functions. Evaluating a function along short coordinate displacements gives such an extension. We choose $a_0\in(0,\min\{L,R_{\max}\}]$, let $e_l$ be the $l$th Euclidean coordinate vector, and, for $h\in\operatorname{Lip}_0(\M;o)$, define
\[\dot\mu_l(h)=-\frac{h(a_0e_l,v_*)}{a_0},\quad l=1,\ldots,d,\qquad \dot q_j(h)=-\frac{h(0,(a_0,j))}{a_0}.\]
These evaluation functionals recover the moment differences when $h$ is a difference of risk functions induced by probability laws. For general $h$, they satisfy $\norm{\dot\mu(h)}\le\sqrt d\,\norm h_{\mathrm{Lip}}$ and $|\dot q_j(h)|\le\norm h_{\mathrm{Lip}}$. Substitution into~\eqref{eq:openbook_D} therefore defines bounded linear maps into $\operatorname{Lip}_0(K_\theta;o)$, with operator norm at most $\sqrt{d+1}$ for either arm. The exact expansion proves~\eqref{eq:metric_first_order} as well as~\eqref{eq:metric_vonmises}, with $L_0=\sqrt{d+1}$ and $C=0$, and with $\sqrt{d+1}$ also bounding the operator norm in Assumption~\ref{ass:metric_profiles}. Contaminated horizontal means may leave $B_L$, but the translations in~\eqref{eq:openbook_transport} remain defined on the full spine. Thus Assumption~\ref{ass:metric_profiles} holds for these unbounded outcomes. Within each admitted regime, the means depend continuously on the ordinary and folded moments. The CAT(0) quadruple inequality makes the anchored effect loss continuous in its endpoint in the effect Lipschitz norm, so measurable fitted and contaminated moments also yield strongly measurable losses.

\section{Supporting results and proofs for Sections~\ref{sec:forest} and~\ref{sec:theory}}\label{app:proofs}

\subsection{Exact Euclidean reduction}\label{app:euclidean_reduction_proof}

For Euclidean outcomes, the identity~\eqref{eq:metric_euclidean} follows directly from translation transport. Suppose the fitted laws have finite second moments and means $\widetilde m_a(x)$, with minimization over $\R^q$. Fix $x$ and suppress it in the notation. The fitted effect loss is
\[\mathcal T_z^{\widetilde Q}=\frac12\norm{z-o}^2-\ip{z-o}{\widetilde m_1-\widetilde m_0}.\]
Contaminating arm $a$ changes its mean to $(1-\varepsilon)\widetilde m_a+\varepsilon Y$, so the finite difference in~\eqref{eq:metric_psi} is exactly $-\ip{z-o}{(2a-1)(Y-\widetilde m_a)}$. Since $\sum_a\1(A=a)(2a-1)(Y-\widetilde m_a)/\widetilde p_a=\phi(O)-\widetilde m_1+\widetilde m_0$, the weighted finite differences sum to $-\ip{z-o}{\phi(O)-\widetilde m_1+\widetilde m_0}$. Adding $\mathcal T_z^{\widetilde Q}$ gives~\eqref{eq:metric_euclidean} for every admissible $\varepsilon>0$. Its final term is independent of $z$ and cancels between a parent and its children when empirical losses use their sample proportions. Thus weighted minimization and splitting agree with squared-error DR learning, with the same constraint on the candidate effect when one is imposed. Conditional expectation gives the true anchored effect loss when either both conditional means or both propensity scores are correct, since then $\E(\phi\mid X)=m_1-m_0$.

\Needspace{5\baselineskip}
\subsection{Conditional orthogonality}\label{app:orthogonality_proof}

The correction in~\eqref{eq:metric_psi} estimates the first-order change from the fitted to the true effect loss. Taking its conditional expectation leaves a second-order error in the nuisance estimates and an error from the finite perturbation size. The proof follows the one-step correction in numerical influence-function methods \citep{carone2019computerized,jordan2022gateaux}, with the remainder controlled in the Lipschitz norm so that the same bound applies to loss comparisons.

The proof below also shows that, if $A_Q(Y)\le M_Q$ almost surely, then $\|\psi^\varepsilon(O)\|_{\mathbb Z}\le C_0$ almost surely.

\begin{proof}[Proof of Proposition~\ref{prop:metric_bias}]
Fix the nuisance fits and $x$, and suppress $x$ in the risk notation. Let $r=r(x)$ and $q=q(x)$ be the errors of the fitted risk functions and propensity scores in~\eqref{eq:metric_remainders}. The perturbation $H_a(y)\in\mathbb B^2$ has $a$th component $k_y-\widetilde Q_a$ and other component zero. The common quadratic in~\eqref{eq:centered_risk_kernel} cancels, giving
\[H_a(y)_a=\bar k_y-\int\bar k_v\,d\widetilde P_a(v\mid x),\qquad \|H_a(y)\|_{\mathbb B^2}\le A_Q(y)+M_Q.\]
Minkowski's inequality and~\eqref{eq:risk_moment} imply $\E\{\|H_a(Y)\|_{\mathbb B^2}^2\mid A=a,X=x\}\le4M_Q^2$. The bound $\|\mathcal T^{\widetilde Q}\|_{\mathbb Z}\le D_\theta+D_K$ of Section~\ref{subsec:loss_regularity} and the first-order bound~\eqref{eq:metric_first_order} give
\begin{equation}\label{eq:score_envelope}\|\psi^\varepsilon(O)\|_{\mathbb Z}\le D_\theta+D_K+\frac{L_0}{\epsilon_0}\{A_Q(Y)+M_Q\}.\end{equation}
Taking the conditional $L^2$ norm gives the bound $D_\theta+D_K+2L_0M_Q/\epsilon_0$ in~\eqref{eq:metric_scorelip}. When $A_Q(Y)\le M_Q$ almost surely,~\eqref{eq:score_envelope} gives the same bound almost surely.

The quadratic expansion gives
\[\frac{\mathcal T^{\widetilde Q+\varepsilon H_a(y)}-\mathcal T^{\widetilde Q}}{\varepsilon}=\dot{\mathcal T}^{\widetilde Q}[H_a(y)]+R_{a,\varepsilon}(y),\qquad \|R_{a,\varepsilon}(y)\|_{\mathbb Z}\le C\varepsilon\|H_a(y)\|_{\mathbb B^2}^2.\]
Let $\Delta_a$ have $a$th component $Q_a-\widetilde Q_a$ and other component zero. Strong measurability, integrability, and bounded linearity justify taking expectations through $\dot{\mathcal T}^{\widetilde Q}$, yielding $\E\{\dot{\mathcal T}^{\widetilde Q}[H_a(Y)]\mid A=a,X=x\}=\dot{\mathcal T}^{\widetilde Q}[\Delta_a]$. Hence
\[\E(\psi^\varepsilon\mid X=x)=\mathcal T^{\widetilde Q}+\sum_a\frac{p_a}{\widetilde p_a}\dot{\mathcal T}^{\widetilde Q}[\Delta_a]+R_\varepsilon,\qquad \|R_\varepsilon\|_{\mathbb Z}\le\frac{4CM_Q^2}{\epsilon_0}\varepsilon,\]
where $\sum_a p_a/\widetilde p_a\le1/\epsilon_0$. Applying~\eqref{eq:metric_vonmises} to the true pair gives $\mathcal T^Q=\mathcal T^{\widetilde Q}+\sum_a\dot{\mathcal T}^{\widetilde Q}[\Delta_a]+R_Q$, with $\|R_Q\|_{\mathbb Z}\le Cr^2$. Subtraction gives
\[\E(\psi^\varepsilon\mid X=x)-\mathcal T^Q=\sum_a\frac{p_a-\widetilde p_a}{\widetilde p_a}\dot{\mathcal T}^{\widetilde Q}[\Delta_a]+R_\varepsilon-R_Q.\]
The first term has norm at most $(L/\epsilon_0)rq$. Combining the three terms proves~\eqref{eq:metric_bias}; the increment assertion follows from the definition of $\mathbb Z$. In particular, valid constants are
\begin{equation}\label{eq:loss_constants}C_0=D_\theta+D_K+\frac{2L_0M_Q}{\epsilon_0},\qquad C_1=\max\{C,L/\epsilon_0,4CM_Q^2/\epsilon_0\}.\end{equation}
\end{proof}

\subsection{Split selection}\label{app:forest}
This section proves the split-selection results summarized in Section~\ref{subsec:localized_splits}. They use the following curvature condition.

\begin{assumption}\label{ass:effect_curvature}
$K_\theta$ is geodesically convex and, for a fixed $0<\kappa_Z\le1$, $\ell$ satisfies
\begin{equation}\label{eq:effect_strong_convexity}\ell_{\gamma_{z,u}(t)}(v)\le(1-t)\ell_z(v)+t\ell_u(v)-\frac{\kappa_Z}{2}t(1-t)d^2(z,u),\qquad z,u,v\in K_\theta,\quad 0\le t\le1.\end{equation}
\end{assumption}

Strong convexity ensures a unique mean effect in each node. A compact convex subset of a Hadamard space satisfies it with $\kappa_Z=1$ \citep{sturm2003probability}. For compositions, Proposition~\ref{prop:effect_curvature_domains} in Section~\ref{app:spherical_profiles} gives the constant when $K_\theta$ is a spherical ball about $o$ of radius less than $\pi/4$; the outcomes themselves are unrestricted.

We now derive the properties of population split gains stated in Sections~\ref{sec:split} and~\ref{subsec:localized_splits}. For a split with positive child probabilities and unique node means, write $V_C(z)=\E\{\ell_z(\theta(X))\mid X\in C\}$, using the effect loss and node means in~\eqref{eq:node_risk}. Then
\[\Delta(B;L,R)=\sum_{C=L,R}\pi_C\{V_C(e_B)-V_C(e_C)\}\ge0.\]
Equality forces $e_B=e_L=e_R$ by uniqueness of the child minimizers. Conversely, a common child minimizer minimizes the parent mixture and gives zero gain.

Under Assumption~\ref{ass:effect_curvature}, let $z\in K_\theta$ and $z_t=\gamma_{e_C,z}(t)$. Assumption~\ref{ass:metric_transport} places $\theta(X)$ in $K_\theta$, so averaging~\eqref{eq:effect_strong_convexity} with $v=\theta(X)$ over the conditional law of $X$ given $X\in C$ gives $V_C(z_t)\le(1-t)V_C(e_C)+tV_C(z)-\kappa_Zt(1-t)d^2(z,e_C)/2$. Since $V_C(e_C)\le V_C(z_t)$, dividing by $t$ gives $V_C(z)-V_C(e_C)\ge\kappa_Z(1-t)d^2(z,e_C)/2$ for every $0<t\le1$. Letting $t\downarrow0$ gives the variance inequality $V_C(z)-V_C(e_C)\ge\kappa_Zd^2(z,e_C)/2$. Set $a=d(e_L,e_B)$ and $b=d(e_R,e_B)$. Applying the variance inequality at $z=e_B$, followed by $\pi_La^2+\pi_Rb^2\ge\pi_L\pi_R(a+b)^2$ and the triangle inequality, yields
\[\Delta(B;L,R)\ge\frac{\kappa_Z}{2}\sum_{C=L,R}\pi_Cd^2(e_C,e_B)\ge\frac{\kappa_Z\pi_L\pi_R}{2}d^2(e_L,e_R).\]
In Euclidean space with translation transport and convex effect set, put $\tau(x)=m_1(x)-m_0(x)$ and $\bar\tau_C=\E\{\tau(X)\mid X\in C\}$. Since $e_C=o+\bar\tau_C$ and $\bar\tau_B=\pi_L\bar\tau_L+\pi_R\bar\tau_R$, expansion of the squared norms gives
\[\Delta(B;L,R)=\frac12\sum_{C=L,R}\pi_C\|\bar\tau_C-\bar\tau_B\|^2=\frac{\pi_L\pi_R}{2}\|\bar\tau_L-\bar\tau_R\|^2.\]

Candidate thresholds can depend on the structure covariates. We therefore analyze the gains conditionally on $\mathcal F_S=\sigma\{\calD_0,(X_i:i\in\calD_S),U_S\}$, where the tree randomization $U_S$ is independent of the data. Given $\mathcal F_S$, the structure pairs $(A_i,Y_i)$ are independent, each with the conditional law of $(A,Y)$ given $X=X_i$. Let $\calB_S$ be a family of $M$ candidate nodes; it can include every node obtainable from thresholds at the structure covariates, so the result below covers nodes selected during tree growth. For $D\in\calB_S$, let $I_D=\{i\in\calD_S:X_i\in D\}$ and $m_D=|I_D|>0$. A candidate split $s=(B,L,R)$ consists of nodes in $\calB_S$ such that $I_B$ is the disjoint union of $I_L$ and $I_R$, and $\mathcal S_S$ is the set of candidate splits. For a tree grown on a subsample, $\calD_S$ and $N$ refer to that subsample, which is fixed given $\mathcal F_S$.

Given $\mathcal F_S$, the target gain averages the true effects $\theta_i=\theta(X_i)$ at the observed covariates rather than over the population as in~\eqref{eq:gain}. For a node $D$, let $R_D^X$ be the average loss with respect to its true effects and $e_D^X$ its node mean. The target gain is
\begin{equation}\label{eq:design_split_target}\begin{aligned}
R_D^X(z)&=\frac1{m_D}\sum_{i\in I_D}\frac12d^2(z,\theta_i),&e_D^X&=\argmin_{z\in K_\theta}R_D^X(z),\\
\Delta_X(s)&=R_B^X(e_B^X)-\sum_{C=L,R}\widehat\pi_C R_C^X(e_C^X),&\widehat\pi_C&=\frac{m_C}{m_B}.
\end{aligned}\end{equation}
The estimated gain uses the corrected losses $\widehat R_D(z)=m_D^{-1}\sum_{i\in I_D}\psi_z^\varepsilon(O_i)$ in place of $R_D^X$. For numerical minimization, each node prediction $\widehat e_D\in K_\theta$ need only satisfy
\begin{equation}\label{eq:localized_node_optimization}\widehat R_D(\widehat e_D)\le\min_{z\in\mathcal G_S}\widehat R_D(z)+\xi_D,\end{equation}
where $\xi_D\ge0$ is a tolerance and $\mathcal G_S\subset K_\theta$ is a finite $h_S$-net with $H$ points: every point of $K_\theta$ lies within distance $h_S$ of $\mathcal G_S$. The candidate nodes, the net, and the tolerances are chosen without the structure treatments and outcomes. The estimated gain per parent observation is $\widehat\Delta_X(s)=\widehat R_B(\widehat e_B)-\sum_{C=L,R}\widehat\pi_C\widehat R_C(\widehat e_C)$; under exact minimization over $K_\theta$, $(m_B/N)\widehat\Delta_X(s)$ equals $\widehat G(s)$ in~\eqref{eq:empgain}.

Uniform control over candidate splits uses an almost-sure bound: if $A_Q(Y)\le M_Q$, then $\|\psi^\varepsilon(O)\|_{\mathbb Z}\le C_0$, both almost surely, with $C_0$ as in~\eqref{eq:loss_constants}. For $0<\delta<1$, set
\begin{equation}\label{eq:localized_split_errors}\begin{aligned}
t_S&=\log\frac{2M(H+M)}\delta,\qquad \bar\rho_D=\frac1{m_D}\sum_{i\in I_D}\rho_\varepsilon(X_i),\\
a_D&=C_0\sqrt{\frac{2t_S}{m_D}}+C_1\bar\rho_D,\\
a_s&=\max_{D\in\{B,L,R\}}a_D,\qquad b_s=2C_0h_S+\max_{D\in\{B,L,R\}}\xi_D.
\end{aligned}\end{equation}
Here $a_D$ combines sampling fluctuation within node $D$ with its remaining nuisance bias. The quantity $a_s$ uses the largest of these errors across the parent and children, while $b_s$ accounts for the net resolution and numerical minimization.

\begin{theorem}\label{thm:localized_split}
Suppose that Assumptions~\ref{ass:causal}--\ref{ass:metric_transport}, \ref{ass:metric_profiles}, and~\ref{ass:effect_curvature} hold and that $A_Q(Y)\le M_Q$ almost surely. Then, with conditional probability at least $1-\delta$ given $\mathcal F_S$,
\begin{equation}\label{eq:localized_split_bound}|\widehat\Delta_X(s)-\Delta_X(s)|\le a_s\sqrt{\frac{2\Delta_X(s)}{\kappa_Z}}+\frac{a_s^2}{2\kappa_Z}+b_s\qquad\text{for every }s\in\mathcal S_S.\end{equation}
\end{theorem}

Theorem~\ref{thm:localized_split} shows that the error of the estimated gain adapts to the gain itself. When $\Delta_X(s)=0$, the estimated gain is at most $a_s^2/(2\kappa_Z)+b_s$ in absolute value on the stated event. A split without effect separation thus has only quadratic sampling and nuisance error, apart from numerical approximation.

The parent and child criteria use the same observations, so their errors partly cancel in the split gain. The proof preserves this cancellation by comparing each fitted node prediction with its true node mean before subtracting the child risks. This gives the gain-dependent bound in Theorem~\ref{thm:localized_split}.
\begin{proof}[Proof of Theorem~\ref{thm:localized_split}]
Condition on $\mathcal F_S$. The node risks $R_D^X$ and their node means $e_D^X$ in~\eqref{eq:design_split_target} are then fixed. Use the concentration level $t_S$ and node error bound $a_D$ in~\eqref{eq:localized_split_errors}. Let $\mathcal E_S=\mathcal G_S\cup\{e_D^X:D\in\calB_S\}$, which has at most $H+M$ members. For $D\in\calB_S$ and $z\in\mathcal E_S$, the summands $\psi_z^\varepsilon(O_i)-\psi_{e_D^X}^\varepsilon(O_i)$ are conditionally independent and, by the almost-sure bound $\|\psi^\varepsilon(O_i)\|_{\mathbb Z}\le C_0$ following~\eqref{eq:score_envelope}, belong to the interval $[-C_0d(z,e_D^X),C_0d(z,e_D^X)]$. Hoeffding's inequality \citep[Chapter~2]{boucheron2013concentration} therefore gives, except on an event of conditional probability at most $2e^{-t_S}$,
\[\left|\{\widehat R_D(z)-\widehat R_D(e_D^X)\}-\E\{\widehat R_D(z)-\widehat R_D(e_D^X)\mid\mathcal F_S\}\right|\le C_0d(z,e_D^X)\sqrt{\frac{2t_S}{m_D}}.\]
Proposition~\ref{prop:metric_bias} bounds the difference between the conditional expectation in this display and $R_D^X(z)-R_D^X(e_D^X)$ by $C_1\bar\rho_Dd(z,e_D^X)$; the anchoring terms cancel. A union bound over at most $M(H+M)$ pairs yields an event of conditional probability at least $1-\delta$ on which
\begin{equation}\label{eq:localized_increment_event}\left|\{\widehat R_D(z)-R_D^X(z)\}-\{\widehat R_D(e_D^X)-R_D^X(e_D^X)\}\right|\le a_Dd(z,e_D^X)\qquad(D\in\calB_S,\ z\in\mathcal E_S).\end{equation}
Including the true node means $e_D^X$ in the comparison set makes this event control both the fitted node values and the loss differences between parent and child predictions.

The variance inequality under Assumption~\ref{ass:effect_curvature} gives $R_D^X(z)-R_D^X(e_D^X)\ge\kappa_Zd^2(z,e_D^X)/2$. Combining it with~\eqref{eq:localized_increment_event}, for every $z\in\mathcal G_S$,
\[\widehat R_D(z)-\widehat R_D(e_D^X)\ge\frac{\kappa_Z}{2}d^2(z,e_D^X)-a_Dd(z,e_D^X)\ge-\frac{a_D^2}{2\kappa_Z}.\]
The empirical risk is $C_0$-Lipschitz. A grid point within $h_S$ of $e_D^X$ and~\eqref{eq:localized_node_optimization} give the upper bound below. For the lower bound, use a grid point within $h_S$ of $\widehat e_D$ and the preceding inequality. Writing $\delta_D$ for the criterion change from the true node mean to the fitted prediction, we obtain
\begin{equation}\label{eq:localized_profile_interval}-\frac{a_D^2}{2\kappa_Z}-C_0h_S\le\delta_D:=\widehat R_D(\widehat e_D)-\widehat R_D(e_D^X)\le C_0h_S+\xi_D.\end{equation}

Fix a split $s=(B,L,R)$ and, within this proof, abbreviate $e_D=e_D^X$. Because both the empirical parent risk and the parent risk conditional on the observed covariates are the child-weighted averages of their respective child risks,
\[\begin{aligned}\widehat\Delta_X(s)-\Delta_X(s)={}&\sum_{C=L,R}\widehat\pi_C\left[\{\widehat R_C(e_B)-R_C^X(e_B)\}-\{\widehat R_C(e_C)-R_C^X(e_C)\}\right]\\&+\delta_B-\sum_{C=L,R}\widehat\pi_C\delta_C.\end{aligned}\]
All the parent node means belong to $\mathcal E_S$. Hence~\eqref{eq:localized_increment_event} and Cauchy--Schwarz bound the absolute value of the first term by
\[\sum_{C=L,R}\widehat\pi_Ca_Cd(e_B,e_C)\le a_s\left\{\sum_{C=L,R}\widehat\pi_Cd^2(e_B,e_C)\right\}^{1/2}\le a_s\sqrt{2\Delta_X(s)/\kappa_Z}.\]
The last inequality follows by applying the variance inequality to each child risk at $e_B$. For the remaining term, the lower bounds in~\eqref{eq:localized_profile_interval} for the parent and the upper bounds for its children give
\[\delta_B-\sum_{C=L,R}\widehat\pi_C\delta_C\ge-\frac{a_B^2}{2\kappa_Z}-2C_0h_S-\sum_{C=L,R}\widehat\pi_C\xi_C\ge-\frac{a_s^2}{2\kappa_Z}-b_s.\]
Using the upper bound for the parent and the lower bounds for its children gives the matching upper bound $a_s^2/(2\kappa_Z)+b_s$. This proves~\eqref{eq:localized_split_bound} simultaneously over $\mathcal S_S$.

\end{proof}

\subsection{Pointwise estimation and convergence rates}\label{app:pointwise_proofs}

The pointwise rate compares the increase in ideal loss away from $\theta(x)$ with random fluctuations of the fitted loss. Conditioning on $\mathcal F_n$ fixes the forest weights, so the fluctuations can be bounded on successive balls around $\theta(x)$. The proof uses the localization and peeling approach to Fr\'echet estimation in \citet{petersen2019frechet,schotz2019quadruple}.

\begin{proof}[Proof of Theorem~\ref{thm:metric_rate}]
Condition on $\mathcal F_n$ and suppress $x$. Recall the weights $\alpha_i$, sampling scale $s_n$, and biases $b_n,B_n$ from~\eqref{eq:metric_local}. Set $V_i=\|\psi^\varepsilon(O_i)\|_{\mathbb Z}$, so that $\E(V_i^2\mid\mathcal F_n)\le C_0^2$ by Proposition~\ref{prop:metric_bias}. Subtracting the conditional mean from each loss contrast defines
\begin{equation}\label{eq:local_centered_process}Z_n(z)=\sum_i\alpha_i\big[\psi_z^\varepsilon(O_i)-\psi_\theta^\varepsilon(O_i)-\E\{\psi_z^\varepsilon(O_i)-\psi_\theta^\varepsilon(O_i)\mid\mathcal F_n\}\big].\end{equation}
For independent Rademacher signs $\epsilon_i$, conditional on the observations the process $\sum_i\epsilon_i\alpha_i(\psi_z^\varepsilon-\psi_\theta^\varepsilon)(O_i)$ has sub-Gaussian increments with scale at most $\Lambda_n d(z,z')$, where $\Lambda_n^2=\sum_i\alpha_i^2V_i^2$. On the ball $K_\theta\cap B(\theta,\delta)$, chaining over nets of radii $2^{-j}\delta$ bounds the $L^2$ norm of its supremum by $C \Lambda_n\delta J$. Indeed, the $L^2$ norm of the maximum of the level-$j$ increments is bounded by their sub-Gaussian scale times the square root of one plus the logarithm of the number of successive-net pairs. Minkowski's inequality and summation over levels give the entropy integral in~\eqref{eq:metric_entropy}. The processes equal zero at $\theta$, and continuity extends the bounds from the nets to the full ball.

Conditional $L^2$ symmetrization and $\E(\Lambda_n^2\mid\mathcal F_n)\le C_0^2s_n^2$ therefore yield
\begin{equation}\label{eq:metric_local_expect}\E\left\{\sup_{d(z,\theta)\le\delta}|Z_n(z)|^2\,\middle|\,\mathcal F_n\right\}\le C s_n^2\delta^2J^2.\end{equation}
These are the symmetrization and sub-Gaussian chaining arguments of \citet[Chapter~2]{vandervaart2023weak}, applied conditionally to the weighted observations. Compactness and pathwise Lipschitz continuity allow all suprema to be taken over countable dense subsets.

Write $\theta_i=\theta(X_i)$. The reverse triangle inequality bounds the ideal loss contrast by
\[\frac12\{d^2(z,\theta_i)-d^2(\theta,\theta_i)\}\ge\frac12d^2(z,\theta)-d(z,\theta)d(\theta_i,\theta).\]
The anchor cancels in this contrast. For $z\in K_\theta$, write $\widehat L_x(z)-\widehat L_x(\theta)=Z_n(z)+\sum_i\alpha_i\E\{\psi_z^\varepsilon(O_i)-\psi_\theta^\varepsilon(O_i)\mid\mathcal F_n\}$. By Proposition~\ref{prop:metric_bias}, each conditional mean differs from the ideal contrast by at most $C_1\rho_\varepsilon(X_i)d(z,\theta)$. Summing the preceding lower bound with weights $\alpha_i$ therefore gives $\widehat L_x(z)-\widehat L_x(\theta)\ge d^2(z,\theta)/2-d(z,\theta)(b_n+C_1B_n)+Z_n(z)$. Taking $z=\widehat\theta$ and using $\widehat L_x(\widehat\theta)-\widehat L_x(\theta)\le\zeta_n$ from~\eqref{eq:metric_approx} yields
\begin{equation}\label{eq:metric_basic}\frac12d(\widehat\theta,\theta)^2\le d(\widehat\theta,\theta)(b_n+C_1B_n)+|Z_n(\widehat\theta)|+\zeta_n.\end{equation}
Put $A_n=b_n+C_1B_n$ and $R=K\{A_n+v s_nJ+\sqrt{\zeta_n}\}$ for a sufficiently large fixed $K$. If $R\ge D_\theta$ the result is immediate. Otherwise, consider a shell $2^jR<d(\widehat\theta,\theta)\le2^{j+1}R$. There $d(\widehat\theta,\theta)>R$, so $d(\widehat\theta,\theta)A_n\le d(\widehat\theta,\theta)^2/K$ and $\zeta_n\le d(\widehat\theta,\theta)^2/K^2$. For $K\ge8$, the basic inequality therefore forces
\[\sup_{d(z,\theta)\le2^{j+1}R}|Z_n(z)|\ge c\,2^{2j}R^2.\]
Only radii at most $2D_\theta$ are needed. By~\eqref{eq:metric_local_expect} and Markov's inequality, the probability of this shell is at most $C4^{-j}s_n^2J^2/R^2$. Summing over $j\ge0$ bounds the conditional probability of $d(\widehat\theta,\theta)>R$ by $Cs_n^2J^2/R^2\le C/(K^2v^2)$, because $R\ge Kvs_nJ$. This proves~\eqref{eq:metric_rate} with $C_2=K\max(1,C_1)$. This also gives the asserted stochastic order and consistency.

\end{proof}

Theorem~\ref{thm:metric_rate} holds conditionally on the forest weights. The neighborhood bias and sampling scale vanish when the leaves containing $x$ shrink while retaining a growing number of estimation observations. Trees grown by the same randomized procedure are identically distributed given the samples, so it suffices to consider the leaf $B_1(x)$ of one tree.

\begin{corollary}\label{cor:forest_consistency}
Suppose that the conditions of Theorem~\ref{thm:metric_rate} hold with a fixed constant $J$, that $\theta$ is continuous at $x$, and that $\sup_{x'}\rho_\varepsilon(x')$ and $\zeta_n$ tend to zero in probability. If every leaf containing $x$ has at least $k_n$ estimation observations, where $k_n\to\infty$, and the Euclidean diameter of $B_1(x)$ tends to zero in probability, then $d\{\widehat\theta(x),\theta(x)\}\to0$ in probability.
\end{corollary}

\begin{proof}[Proof of Corollary~\ref{cor:forest_consistency}]
Each leaf containing $x$ has at least $k_n$ estimation observations, so no tree uses an ancestor in~\eqref{eq:weights}, and $\alpha_i(x)\le T^{-1}\sum_tN_t(x)^{-1}\le k_n^{-1}$ for every $i$. Hence $s_n^2=\sum_i\alpha_i(x)^2\le\max_i\alpha_i(x)\le k_n^{-1}$. By Assumption~\ref{ass:metric_transport}, the aligned effect at every covariate value lies in $K_\theta$, so
\[\omega(r)=\sup\{d\{\theta(x'),\theta(x)\}:\norm{x'-x}\le r\}\]
is at most $D_\theta$, and it tends to zero as $r\to0$ because $\theta$ is continuous at $x$. Every estimation observation receiving weight from tree $t$ lies in $B_t(x)$, which contains $x$, so $b_n\le T^{-1}\sum_t\omega[\operatorname{diam}\{B_t(x)\}]$. The trees are identically distributed given the samples, so $\E(b_n)\le\E(\omega[\operatorname{diam}\{B_1(x)\}])$. This expectation tends to zero by bounded convergence, because the diameter tends to zero in probability; Markov's inequality then gives $b_n\to0$ in probability. Moreover, $B_n\le\sup_{x'}\rho_\varepsilon(x')\to0$ in probability. For $\eta>0$ and $v\ge1$, Theorem~\ref{thm:metric_rate} gives
\[P\{d(\widehat\theta,\theta)>\eta\}\le C_3v^{-2}+P\left\{C_2\left(b_n+B_n+vk_n^{-1/2}J+\sqrt{\zeta_n}\right)>\eta\right\}.\]
The second term tends to zero for each fixed $v$, and letting $v\to\infty$ completes the proof.
\end{proof}

The leaf conditions in Corollary~\ref{cor:forest_consistency} concern only the tree construction and the estimation covariates. They hold for regular random-split trees \citep{wager2018causal} under the following conditions. Suppose the covariates are uniform on $[0,1]^p$, that every coordinate is chosen for a split with probability at least $\nu/p$ for a fixed $0<\nu\le1$, and that each split leaves at least a fixed fraction of the estimation observations of its node in each child, with between $k_n$ and $2k_n-1$ of them in every leaf. If $k_n\to\infty$ and $k_n/n\to0$, Lemma~2 of \citet{wager2018causal} shows that the diameters of the leaves tend to zero in probability at a polynomial rate in $n/k_n$, so that, for H\"older continuous effects, the neighborhood bias $b_n$ decreases at a polynomial rate in $n/k_n$ whose exponent deteriorates with $p$. A polynomial rate for the estimation error requires, in addition, polynomial growth of $k_n$, which controls the sampling term $k_n^{-1/2}J$ in the proof above, and polynomial rates for $B_n$ and $\zeta_n$; Corollary~\ref{cor:forest_consistency} requires only that these terms vanish. Using only the estimation covariates, these constraints preserve honesty, as in the double-sample trees of \citet{wager2018causal}; the implementation in Section~\ref{subsec:forest_estimation} imposes its leaf sizes on the structure sample instead, and its random splits give every covariate a positive splitting probability.

Neither Corollary~\ref{cor:forest_consistency} nor Theorem~\ref{thm:localized_split} shows that the trees concentrate their neighborhoods in the covariates that govern the effect, which would give a rate depending only on the number of such covariates. Suppose that $\theta$ depends only on the coordinates in a set $\mathcal I\subset\{1,\ldots,p\}$ of size $p_0$ and is $\beta$-H\"older continuous in them with constant $L_X$, where $0<\beta\le1$, and that, with probability tending to one, $\norm{X_{i,\mathcal I}-x_{\mathcal I}}\le h$ whenever $\alpha_i(x)>0$ and $\sum_i\alpha_i(x)^2\le C/(nh^{p_0})$. Then $b_n\le L_Xh^\beta$ and $s_n^2\le C/(nh^{p_0})$, so, under the other conditions of Theorem~\ref{thm:metric_rate} with a fixed $J$, $d\{\widehat\theta(x),\theta(x)\}=O_p\{h^\beta+(nh^{p_0})^{-1/2}+B_n+\sqrt{\zeta_n}\}$. Balancing neighborhood bias and sampling error with $h\asymp n^{-1/(2\beta+p_0)}$ gives the usual nonparametric rate $n^{-\beta/(2\beta+p_0)}$, provided $B_n+\sqrt{\zeta_n}$ is no larger. This rate is conditional on the assumed localization, and all covariates needed for causal adjustment still enter the nuisance estimates.

For the feasible estimator~\eqref{eq:feasible_estimator}, the true effect $\theta(x)$ need not belong to $\Gamma\{\widehat b(x)\}$ because this set uses the estimated untreated mean. The rate proof instead compares the fitted criterion with its value at $F_o\{\widehat b(x),m_1(x)\}$. When $\widehat b(x)$ lies within $\eta_\theta$ of $m_0(x)$, this effect belongs to $K_\theta$ by Assumption~\ref{ass:metric_transport} and reconstructs to $m_1(x)$ by~\eqref{eq:metric_inverse}, so it is feasible at $\widehat b(x)$; transport stability bounds its distance from $\theta(x)$ by the baseline estimation error.

\begin{assumption}\label{ass:feasible_estimation}
The baseline estimate $\widehat b(x)\in K_0$ is fitted on the nuisance sample, and the reported estimate satisfies
\begin{equation}\label{eq:feasible_approx}\widehat L_x(\widehat\theta_{\mathrm F})\le\inf_{z\in\Gamma\{\widehat b(x)\}}\widehat L_x(z)+\zeta_n,\qquad \widehat\theta_{\mathrm F}\in\Gamma\{\widehat b(x)\},\end{equation}
where the tolerance $\zeta_n\ge0$ is chosen without the estimation treatments and outcomes, and $\Gamma$ is the feasible set in~\eqref{eq:feasible_domain}.
\end{assumption}

Continuity of $G_o$ and closedness of $\M$ make $\Gamma\{\widehat b(x)\}$ compact, and it contains the no-change effect $o$ (Section~\ref{subsec:feasible_estimation}). The continuous fitted criterion therefore attains a minimum on this set.

\begin{theorem}\label{thm:feasible_rate}
Under the assumptions of Theorem~\ref{thm:metric_rate}, with Assumption~\ref{ass:feasible_estimation} replacing~\eqref{eq:metric_approx}, put $\delta_n=d\{\widehat b(x),m_0(x)\}$. For constants $C_2,C_3$ depending only on the fixed regularity bounds and every $v\ge1$,
\begin{equation}\label{eq:feasible_rate}P\left\{d(\widehat\theta_{\mathrm F},\theta)>C_2\left[L_F\delta_n+b_n+B_n+v s_nJ+\sqrt{\zeta_n}\right]\,\middle|\,\mathcal F_n\right\}\le C_3v^{-2}.\end{equation}

In particular, if the baseline error, neighborhood bias, nuisance remainder, sampling scale, and optimization tolerance converge to zero in probability, the feasible effect is consistent and, under Assumption~\ref{ass:geodesic}, so is its reconstructed path.
\end{theorem}

\begin{proof}[Proof of Theorem~\ref{thm:feasible_rate}]
Condition on $\mathcal F_n$ and write $\theta=\theta(x)$. Both $\widehat\theta_{\mathrm F}$ and $\theta$ belong to $K_\theta$, so if $\delta_n>\eta_\theta$, the bound holds trivially once $C_2\ge D_\theta/(L_F\eta_\theta)$. Otherwise, the point $e_n=F_o\{\widehat b(x),m_1(x)\}$ belongs to $K_\theta$ by Assumption~\ref{ass:metric_transport} and is feasible because $G_o(\widehat b,e_n)=m_1(x)$. It is fixed given $\mathcal F_n$ and satisfies $a=d(e_n,\theta)\le L_F\delta_n$. Use the centered loss-increment process $Z_n$ in~\eqref{eq:local_centered_process}, and put $A_n=b_n+C_1B_n$. The lower bound for the conditional expected loss used in~\eqref{eq:metric_basic} holds at every $z\in K_\theta$. The triangle inequality also gives
\[\frac12\{d^2(e_n,\theta(X_i))-d^2(\theta,\theta(X_i))\}\le\frac12a^2+a\,d\{\theta(X_i),\theta\}.\]
Together with the bias increment in Proposition~\ref{prop:metric_bias}, this bounds the conditional mean of $\widehat L_x(e_n)-\widehat L_x(\theta)$ by $a^2/2+aA_n$. Comparing the feasible approximate minimizer with $e_n$ in~\eqref{eq:feasible_approx} yields, for $r=d(\widehat\theta_{\mathrm F},\theta)$,
\begin{equation}\label{eq:feasible_basic}\frac12r^2\le\frac12a^2+(r+a)A_n+|Z_n(\widehat\theta_{\mathrm F})|+|Z_n(e_n)|+\zeta_n.\end{equation}
Choose $R=C\{L_F\delta_n+A_n+v s_nJ+\sqrt{\zeta_n}\}$ with $C$ sufficiently large. On a nonempty shell $2^jR<r\le2^{j+1}R$, both $\widehat\theta_{\mathrm F}$ and $e_n$ belong to the ball of radius $2^{j+1}R$ around $\theta$, and~\eqref{eq:feasible_basic} requires its centered-process supremum to exceed $c2^{2j}R^2$. The second-moment bound~\eqref{eq:metric_local_expect} bounds this shell probability by $C4^{-j}s_n^2J^2/R^2$. Summation proves~\eqref{eq:feasible_rate}.

The reconstruction bound~\eqref{eq:metric_stability} gives $d\{G_o(\widehat b,\widehat\theta_{\mathrm F}),m_1(x)\}\le L_G\{\delta_n+d(\widehat\theta_{\mathrm F},\theta)\}$, and the endpoints are in $\M$. Continuity of its geodesics gives path consistency.
\end{proof}

\section{Confidence regions for conditional geodesic effects}\label{sec:inference}
Without a linear structure, central limit theorems for the estimator are not directly available, so, as in \citet{kurisu2024gate}, we base inference on a real-valued objective, the corrected loss. A confidence region contains the grid points whose corrected losses are not significantly larger than those of competing candidates. It uses only loss differences and therefore applies to smooth and nonsmooth effect sets satisfying the mixed-increment bound in Assumption~\ref{ass:metric_confidence}.

Fix a covariate value $x$ and write $\theta=\theta(x)$. We condition on $\mathcal F_n$ from Section~\ref{subsec:metric_rates}, so that the honest weights $\alpha_i=\alpha_i(x)$ and the nuisance estimates are fixed, use the criterion $\widehat L_x$ in~\eqref{eq:estimator} and the sampling scale $s_n$ in~\eqref{eq:metric_local}, and abbreviate the corrected loss by $\psi_z(O_i)=\psi_z^\varepsilon(O_i)$. The target of the criterion is the weighted risk $R_n^0(z)=\sum_i\alpha_i d^2\{z,\theta(X_i)\}/2$. Candidates are compared on a finite $\eta_n$-net $\mathcal G_n\subset K_\theta$ with $N_n$ points. The net, its mesh $\eta_n$, and the bounds below are chosen without the estimation treatments and outcomes.
\begin{assumption}\label{ass:metric_confidence}
For a known constant $1\le\Lambda_Z<\infty$, the half squared distance satisfies the mixed-increment bound
\begin{equation}\label{eq:effect_quadruple}|\ell_z(v)-\ell_u(v)-\ell_z(w)+\ell_u(w)|\le\Lambda_Zd(z,u)d(v,w),\qquad z,u,v,w\in K_\theta.\end{equation}
Known finite constants $L_n,\bar B_n,\bar b_n\ge0$ satisfy, conditionally on $\mathcal F_n$ and for all $z,u\in\mathcal G_n$, $|\psi_z(O_i)-\psi_u(O_i)|\le L_nd(z,u)$ almost surely for every estimation observation, and
\begin{equation}\label{eq:confidence_bias}\left|\E\{\widehat L_x(z)-\widehat L_x(u)\mid\mathcal F_n\}-\{R_n^0(z)-R_n^0(u)\}\right|\le \bar B_n d(z,u),\qquad \sum_i\alpha_i d\{\theta(X_i),\theta\}\le\bar b_n.\end{equation}
\end{assumption}

The mixed-increment bound~\eqref{eq:effect_quadruple} controls how the loss difference between two candidates changes when the true effect changes. A compact convex subset of a Hadamard space satisfies it with $\Lambda_Z=1$ \citep{schotz2019quadruple}, and Proposition~\ref{prop:effect_curvature_domains} gives $\Lambda_Z$ for spherical effect sets of radius less than $\pi/2$. The bounds $L_n$, $\bar B_n$, and $\bar b_n$ control sampling variation, nuisance bias, and neighborhood bias in each loss comparison. A known almost-sure bound on $\|\psi^\varepsilon(O_i)\|_{\mathbb Z}$, such as $C_0$ when $A_Q(Y)\le M_Q$ almost surely, can serve as $L_n$, and a known upper bound for $C_1B_n$ gives $\bar B_n$ by Proposition~\ref{prop:metric_bias}. If the effect is $\beta$-H\"older with known constant $L_X$ and the neighborhood radius in the relevant coordinates is $h$, one can take $\bar b_n=L_Xh^\beta$. Multiplication by $d(z,u)$ makes the allowance smaller when comparing nearby candidates.

For conditional failure probability $0<\delta<1$, the allowance for sampling variation, nuisance error, and neighborhood bias is
\begin{equation}\label{eq:confidence_allowance}a_{n,\delta}=L_ns_n\sqrt{2\log(2N_n^2/\delta)}+\bar B_n+\Lambda_Z\bar b_n.\end{equation}
A grid point $z$ is included if its fitted loss is within $(a_{n,\delta}+\eta_n)d(z,u)$ of that of every competitor $u$:
\begin{equation}\label{eq:confidence_grid}\mathcal C_n^{\mathrm{grid}}=\left\{z\in\mathcal G_n:\widehat L_x(z)-\widehat L_x(u)\le(a_{n,\delta}+\eta_n)d(z,u)\text{ for every }u\in\mathcal G_n\right\}.\end{equation}
The mesh radius $\eta_n$ accounts for effects between grid points. Enlarging the selected grid points gives the confidence region
\begin{equation}\label{eq:confidence_region}\mathcal C_n=K_\theta\cap\bigcup_{z\in\mathcal C_n^{\mathrm{grid}}}\overline B(z,\eta_n).\end{equation}
This region is always nonempty: every minimizer of $\widehat L_x$ over $\mathcal G_n$ belongs to $\mathcal C_n^{\mathrm{grid}}$.

\begin{theorem}\label{thm:metric_confidence}
Under Assumptions~\ref{ass:causal}, \ref{ass:metric_transport}, and~\ref{ass:metric_confidence}, conditional on $\mathcal F_n$, an event of probability at least $1-\delta$ satisfies
\begin{equation}\label{eq:confidence_coverage}\theta\in\mathcal C_n,\qquad \sup_{v\in\mathcal C_n}d(v,\theta)\le4(a_{n,\delta}+\eta_n),\qquad \operatorname{diam}(\mathcal C_n)\le8(a_{n,\delta}+\eta_n).\end{equation}
\end{theorem}

The confidence radius is of order $L_ns_n\sqrt{\log(2N_n)}+\bar B_n+\Lambda_Z\bar b_n+\eta_n$. Its terms have the same sources as the estimation bound, with an additional logarithmic factor for comparing the grid points and the mesh error $\eta_n$. The coverage guarantee holds in finite samples and does not rely on a distributional limit for $\widehat\theta$, but it requires valid bounds on the nuisance and neighborhood biases. Because $L_n$ bounds every loss comparison almost surely and the allowance covers all $N_n^2$ comparisons, the region is conservative at moderate effective sample sizes.

\begin{proof}[Proof of Theorem~\ref{thm:metric_confidence}]
Condition on $\mathcal F_n$. Use the fitted criterion $\widehat L_x$ from~\eqref{eq:estimator} and the allowance $a_{n,\delta}$ from~\eqref{eq:confidence_allowance}. For fixed $z,u\in\mathcal G_n$, the independent variables $\alpha_i\{\psi_z(O_i)-\psi_u(O_i)\}$ have ranges of length at most $2\alpha_iL_nd(z,u)$. Hoeffding's inequality \citep[Chapter~2]{boucheron2013concentration} and a union bound over at most $N_n^2$ ordered pairs give an event $\mathcal E_n$ with conditional probability at least $1-\delta$ on which
\begin{equation}\label{eq:confidence_hoeffding}\left|\widehat L_x(z)-\widehat L_x(u)-\E\{\widehat L_x(z)-\widehat L_x(u)\mid\mathcal F_n\}\right|\le L_ns_n\sqrt{2\log(2N_n^2/\delta)}\,d(z,u)\end{equation}
simultaneously for all grid pairs.

The mixed-increment bound~\eqref{eq:effect_quadruple} gives
\[\left|\frac{d^2(z,v)-d^2(u,v)-d^2(z,w)+d^2(u,w)}2\right|\le\Lambda_Zd(z,u)d(v,w).\]
Applying it with $v=\theta(X_i)$ and $w=\theta$, summing with weights $\alpha_i$, and using~\eqref{eq:confidence_bias} and~\eqref{eq:confidence_hoeffding}, we obtain on $\mathcal E_n$
\begin{equation}\label{eq:confidence_pair_event}\left|\widehat L_x(z)-\widehat L_x(u)-\frac{d^2(z,\theta)-d^2(u,\theta)}2\right|\le a_{n,\delta}d(z,u)\end{equation}
for every grid pair.

Choose $q\in\mathcal G_n$ with $e=d(q,\theta)\le\eta_n$. The reverse triangle inequality gives, for every $u\in\mathcal G_n$,
\[\frac{d^2(q,\theta)-d^2(u,\theta)}2\le e\,d(q,u)-\frac12d^2(q,u)\le\eta_n d(q,u).\]
Together with~\eqref{eq:confidence_pair_event}, this proves $q\in\mathcal C_n^{\mathrm{grid}}$, and hence $\theta\in\mathcal C_n$ by the enlargement in~\eqref{eq:confidence_region}.

For any $z\in\mathcal C_n^{\mathrm{grid}}$, compare $z$ with $q$ in~\eqref{eq:confidence_grid} and~\eqref{eq:confidence_pair_event}. With $t=d(z,\theta)$, the triangle inequality yields
\[\frac{t^2-e^2}{2}\le(2a_{n,\delta}+\eta_n)d(z,q)\le(2a_{n,\delta}+\eta_n)(t+e).\]
If $t+e>0$, division gives $t\le4a_{n,\delta}+3\eta_n$; the same inequality is immediate when $t+e=0$. Every $v\in\mathcal C_n$ is within $\eta_n$ of such a point $z$, proving the radius bound in~\eqref{eq:confidence_coverage}. The diameter bound follows from the triangle inequality.
\end{proof}

\section{Simulation details and additional results}\label{app:simulations}
\subsection{Properties of the outcome laws}\label{app:simulation_laws}
For the outcome law~\eqref{eq:simulation_outcomes} in both settings of Section~\ref{subsec:simulation_models}, the noise matrix $Z_a$ takes finitely many equally likely values, each with $\norm{Z_a}_F=1$, and $\E(Z_a)=0$. Under the affine-invariant logarithm, $\Log_{m_a(x)}Y(a)=\sigma m_a(x)^{1/2}Z_am_a(x)^{1/2}$, whose conditional expectation is zero. The expected squared-distance risk therefore has zero gradient at $m_a(x)$; strict geodesic convexity makes this its unique minimizer. Congruence invariance gives $d\{Y(a),m_a(x)\}=\sigma$ exactly, and $d\{I,m_0(x)\}=\norm{\Omega(x)}_F\le0.4$ because $|u_k(x)|\le1$ for $k=1,2,3$.

The coefficients in~\eqref{eq:simulation_directional} balance the local contributions of magnitude and orientation to the affine-invariant distance. Write $\theta(\lambda,\phi)$ for the noncommuting effect in~\eqref{eq:simulation_directional} as a function of $\lambda=\lambda(x)$ and $\phi=\alpha x_2$. Its squared line element is $\mathrm{d}s^2=2(\mathrm{d}\lambda)^2+8\sinh^2(\lambda)(\mathrm{d}\phi)^2$. Since $\mathrm{d}\lambda=\delta\,\mathrm{d}x_1$ and $\mathrm{d}\phi=\alpha\,\mathrm{d}x_2$, the expected coefficients of $(\mathrm{d}x_1)^2$ and $(\mathrm{d}x_2)^2$ are $2\delta^2$ and $8\alpha^2\E[\sinh^2\{1.3+\delta X_1\}]$. Equating them gives $\alpha=\delta/\{2\sqrt{\E[\sinh^2\{1.3+\delta X_1\}]}\}=0.1547684850$ for $\delta=0.6$.

For the commuting control, let $V_\theta=\inf_z\E[d^2\{z,\theta(X)\}]$ be the Fr\'echet variance of the noncommuting effect. The commuting effect has log-diagonal coordinates $\lambda(x)+v(x)$ and $-\lambda(x)+v(x)$, so its Fr\'echet variance is $(2\delta^2+\beta^2)/3$, and choosing $\beta=(3V_\theta-2\delta^2)^{1/2}$ matches it to $V_\theta$. Deterministic integration over $(X_1,X_2)$, independent of the Monte Carlo samples, gives $V_\theta\approx0.421633$ and $\beta\approx0.738172$.

The noncommuting effects have magnitude $\sqrt2\,\lambda(x)\le\sqrt2(1.3+0.6)$, and the commuting effects have magnitude at most $\{2(1.3+0.6)^2+\beta^2\}^{1/2}$. Adding the baseline and noise radii shows that all outcome support points in both settings lie within distance 5 of $I$. We use $K_\theta=\overline B(I,10)$. Positive-weight fitted and contaminated Fr\'echet means stay in $\overline B(I,5)$, and the length identity bounds the distance of their aligned endpoints from $I$ by 10. Reconstruction yields an SPD matrix for every admitted baseline and effect.

\subsection{Fitting, evaluation, and additional results}\label{app:simulation_protocol}
The Monte Carlo study uses 200 independent replications for each outcome configuration and total sample size. Within a replication, the estimators share the generated observations, evaluation covariates, and applicable sample-role assignments. Each arm-specific regression forest for an entry of $\log Y$ has 100 trees and minimum terminal-node size 10, and the nonnegative leaf weights are averaged across entries and trees. Since $\log Y=\Log_I(Y)$, these forests measure dispersion by the log-Euclidean distance $\norm{\log Y-\log Y'}_F$, which equals the affine-invariant distance only for commuting matrices. They are used because they are fast. Any nonnegative weights satisfy~\eqref{eq:profilefit}, their errors affect the corrected loss only at second order (Proposition~\ref{prop:metric_bias}), and forests built from the metric alone could be used instead. Fr\'echet means and contaminated means are computed from the resulting probability weights using the original affine-invariant metric. In the commuting control, all outcome and effect minimizations are restricted to the diagonal subspace.

The GCF has 200 trees grown on 80\% subsamples of the structure observations. All six covariates are eligible for splitting. Candidate cuts are at the empirical 0.1, 0.25, 0.5, 0.75, and 0.9 quantiles of each covariate. Each child must contain at least four observations from each treatment arm and at least $k_S=\lceil\sqrt{N}\rceil$ observations in total, where $N=|\calD_S|$ is the structure-sample size, so that every leaf contains a number of observations that grows with $N$. With probability 0.05, a split is selected uniformly from the admissible candidates; otherwise the corrected-loss gain determines the choice. Trees have no additional depth limit.

The corrected loss uses $\varepsilon=\min(0.005,n_{\mathrm{tot}}^{-1})$ and re-minimizes each contaminated probability law. The GCF minimizes the continuous criterion on $K_\theta$ within each node and in the final fit. The common effect uses the same final criterion with $\alpha_i(x)=1/|\calD_E|$ for every $i\in\calD_E$. The uncorrected forest uses the fitted effect loss $\mathcal T_z^{\widetilde Q}(X_i)$ in place of $\psi_z^\varepsilon(O_i)$ in the same node and final minimizations, with the same subsamples, candidate cuts, and random-split draws as the GCF. The plug-in refits the arm-specific nuisance forests, with the same tuning, to all $n_{\mathrm{tot}}$ observations and aligns their fitted conditional Fr\'echet means at $I$. All estimators were fitted successfully in all 200 replications of each configuration.

As for the integrated squared geodesic error~\eqref{eq:simulation_ISE} in Table~\ref{tab:simulation_primary}, we summarize the errors in effect magnitude and in the reconstructed treated mean by their means and standard deviations across the 200 replications. These errors are
\[\begin{aligned}
L_{\mathrm{mag}}(\widehat\theta)&=\E_X\!\left[\big\{d(I,\widehat\theta(X))-d(I,\theta(X))\big\}^2\right],\\
L_{\mathrm{rec}}(\widehat\theta)&=\E_X\!\left[d^2\!\left\{G_I(\widehat b(X),\widehat\theta(X)),m_1(X)\right\}\right],
\end{aligned}\]
approximated on the same 400 evaluation covariates using the same nuisance-sample baseline estimate $\widehat b$ for all estimators.

Table~\ref{tab:simulation_secondary} reports the mean squared errors in effect magnitude and in the reconstructed treated mean. For both quantities, the GCF error decreases as the sample size increases and is much smaller than those of the uncorrected forest and the common effect in both settings and than that of the plug-in in the noncommuting setting. In the commuting control, the plug-in and the GCF have similar errors.

\begin{table}[!htbp]
\centering
\begingroup
\small
\setlength{\tabcolsep}{4pt}
\renewcommand{\arraystretch}{1.15}
\begin{tabular}{@{}lr*{4}{r@{\hspace{0.3em}}l}@{}}
\toprule
Setting & $n_{\mathrm{tot}}$ & \multicolumn{2}{c}{GCF} & \multicolumn{2}{c}{Uncorrected} & \multicolumn{2}{c}{Plug-in} & \multicolumn{2}{c}{Common effect} \\
\midrule
\multicolumn{10}{@{}l}{Effect magnitude} \\
\addlinespace[2pt]
Noncommuting & 600 & \textbf{3.147} & (0.873) & 10.828 & (1.298) & 5.822 & (0.606) & 24.820 & (1.330) \\
 & 1{,}200 & \textbf{1.951} & (0.521) & 7.837 & (0.877) & 4.343 & (0.454) & 24.913 & (1.343) \\
 & 2{,}400 & \textbf{1.350} & (0.265) & 5.799 & (0.577) & 3.237 & (0.303) & 24.636 & (1.238) \\
\addlinespace[5pt]
Commuting & 600 & 2.709 & (0.763) & 7.478 & (1.313) & \textbf{2.682} & (0.453) & 23.352 & (1.214) \\
 & 1{,}200 & \textbf{1.647} & (0.368) & 4.568 & (0.763) & 1.666 & (0.258) & 23.392 & (1.193) \\
 & 2{,}400 & 1.169 & (0.202) & 2.965 & (0.433) & \textbf{1.044} & (0.147) & 23.140 & (1.126) \\
\midrule
\multicolumn{10}{@{}l}{Reconstructed treated mean} \\
\addlinespace[2pt]
Noncommuting & 600 & \textbf{9.519} & (1.422) & 23.731 & (1.647) & 16.179 & (1.207) & 44.975 & (1.864) \\
 & 1{,}200 & \textbf{6.276} & (0.938) & 18.218 & (1.282) & 13.048 & (1.016) & 44.283 & (1.844) \\
 & 2{,}400 & \textbf{4.484} & (0.510) & 14.491 & (1.004) & 10.681 & (0.786) & 43.503 & (1.732) \\
\addlinespace[5pt]
Commuting & 600 & \textbf{6.857} & (0.976) & 16.093 & (1.390) & 7.132 & (0.745) & 43.946 & (1.781) \\
 & 1{,}200 & \textbf{4.301} & (0.537) & 10.019 & (0.815) & 4.717 & (0.431) & 43.488 & (1.664) \\
 & 2{,}400 & \textbf{3.135} & (0.320) & 6.737 & (0.531) & 3.203 & (0.280) & 42.988 & (1.648) \\
\bottomrule
\end{tabular}

\endgroup
\caption{Mean squared errors in effect magnitude and the reconstructed treated mean, as defined in Section~\ref{app:simulation_protocol}. Each cell reports the mean over 200 replications and the standard deviation across replications in parentheses, both in units of $10^{-2}$. All estimators use the same estimated untreated baseline for reconstruction. Boldface indicates the smallest mean error in each row.}\label{tab:simulation_secondary}
\end{table}

Table~\ref{tab:simulation_ablation} separates the two uses of the correction. The forest whose splits are chosen with the fitted effect loss but whose final estimate minimizes the corrected loss has 8\% to 14\% larger error than the GCF, whereas corrected splits followed by an uncorrected final estimate perform like the uncorrected forest. The correction therefore matters most in the final estimate, and corrected splitting gives a further, smaller reduction. The plug-in computed from the nuisance-sample fits that all forests use has errors between those of the uncorrected forest and the full-sample plug-in in Table~\ref{tab:simulation_primary}.

To vary the accuracy of the nuisance fits, we refit the arm-specific nuisance forests with minimum terminal-node size 40 instead of 10, in the noncommuting setting with $n_{\mathrm{tot}}=1{,}200$, and regrow all forests on the same samples. The coarser nuisance fits increase the error of the uncorrected forest from $0.165$ to $0.258$ and that of the nuisance-sample plug-in from $0.149$ to $0.247$, whereas the error of the GCF increases only from $0.043$ to $0.048$, in agreement with the second-order nuisance remainder in Proposition~\ref{prop:metric_bias}. The error of the forest with uncorrected splits and corrected final estimation increases from $0.046$ to $0.068$, so corrected splitting also makes the forest neighborhoods less sensitive to the accuracy of the nuisance fits.

\begin{table}[!htbp]
\centering
\begingroup
\footnotesize
\setlength{\tabcolsep}{3pt}
\renewcommand{\arraystretch}{1.15}
\begin{tabular}{@{}lr*{5}{r@{\hspace{0.3em}}l}@{}}
\toprule
Setting & $n_{\mathrm{tot}}$ & \multicolumn{2}{c}{GCF} & \multicolumn{2}{c}{\begin{tabular}[b]{@{}c@{}}Corrected\\final only\end{tabular}} & \multicolumn{2}{c}{\begin{tabular}[b]{@{}c@{}}Corrected\\splits only\end{tabular}} & \multicolumn{2}{c}{Uncorrected} & \multicolumn{2}{c}{\begin{tabular}[b]{@{}c@{}}Plug-in,\\nuisance sample\end{tabular}} \\
\midrule
\multicolumn{12}{@{}l}{Nuisance forests with minimum node size 10} \\
\addlinespace[2pt]
Noncommuting & 600 & \textbf{6.799} & (1.247) & 7.625 & (1.307) & 22.096 & (1.652) & 21.553 & (1.632) & 19.077 & (1.643) \\
 & 1{,}200 & \textbf{4.274} & (0.791) & 4.607 & (0.804) & 17.021 & (1.247) & 16.543 & (1.248) & 14.911 & (1.192) \\
 & 2{,}400 & \textbf{3.011} & (0.410) & 3.332 & (0.442) & 13.393 & (0.970) & 13.182 & (1.004) & 11.788 & (0.922) \\
\addlinespace[5pt]
Commuting & 600 & \textbf{5.493} & (0.927) & 6.200 & (0.789) & 15.304 & (1.477) & 14.912 & (1.376) & 11.883 & (1.351) \\
 & 1{,}200 & \textbf{3.312} & (0.470) & 3.656 & (0.496) & 9.454 & (0.854) & 9.136 & (0.823) & 7.228 & (0.827) \\
 & 2{,}400 & \textbf{2.386} & (0.284) & 2.729 & (0.308) & 6.025 & (0.534) & 6.025 & (0.523) & 4.390 & (0.494) \\
\midrule
\multicolumn{12}{@{}l}{Nuisance forests with minimum node size 40} \\
\addlinespace[2pt]
Noncommuting & 1{,}200 & \textbf{4.768} & (0.961) & 6.784 & (1.245) & 26.052 & (1.594) & 25.806 & (1.601) & 24.692 & (1.605) \\
\bottomrule
\end{tabular}

\endgroup
\caption{Integrated squared geodesic error when the correction is used only for final estimation or only for splitting, and for the plug-in computed from the nuisance-sample fits. The upper panel uses the nuisance fits of Table~\ref{tab:simulation_primary}; the lower panel refits the nuisance forests with minimum terminal-node size 40 and regrows all forests. Each cell reports the mean over 200 replications and the standard deviation across replications in parentheses, both in units of $10^{-2}$. Boldface indicates the smallest mean error in each row.}\label{tab:simulation_ablation}
\end{table}

\section{Details of the BRAC livelihood analysis}\label{app:brac_analysis}
\subsection{Data construction and paired comparisons}\label{app:brac_paired}
We use version 1.1 of the public replication data for \citet{balboni2022poor}.\footnote{Harvard Dataverse, \url{https://doi.org/10.7910/DVN/PPMEIR}.} The baseline-eligible group is identified by \texttt{sclass=1} in 2007 and linked by household identifier to 2011. The released subdistrict codes recover twenty pairs, each containing one treated and one control branch. The original design stratifies within subdistricts except in Kishoreganj, where the pair was selected at the district level \citep{bandiera2017labor}. Pair identifiers, assignment, and branch membership agree across waves. The analysis contains 3,456 women with baseline agricultural or domestic wage work and 3,276 without it.

Annual hours use the released livestock, agricultural day-labor, domestic-service, and poultry variables. Total recorded hours equal \texttt{total\_hours\_work} plus \texttt{poultry\_hours\_tot}, because the released aggregate excludes poultry. Other hours are the nonnegative remainder after subtracting the four named activities. These five categories and a nonparticipation coordinate determine the composition $S_i$ in~\eqref{eq:livelihood_composition}. Productive assets use the pretransfer value \texttt{pAssets}. The binary productive-assets indicator records whether this value is positive; the continuous predictor is its $\log(1+x)$ transformation. Baseline hours receive the same transformation. Predictors also include age, adult-equivalent household size, female headship, and the first five baseline composition coordinates, with the sixth determined by their sum. The 153 missing baseline ages are imputed using the nuisance-sample median and accompanied by a missingness indicator.

For a scalar outcome or composition coordinate $W$, let $\overline W_{pat}$ be its mean in the branch assigned treatment status $a$ within pair $p$, at wave $t$. The conventional estimate averages the twenty pair-specific differences in changes, $D_p=(\overline W_{p1,2011}-\overline W_{p1,2007})-(\overline W_{p0,2011}-\overline W_{p0,2007})$. Its standard error is the sample standard deviation of the twenty $D_p$ divided by $\sqrt{20}$, and pointwise intervals use $t_{19}$. Subgroup contrasts use branch means within each subgroup. Each branch receives equal weight.

Table~\ref{tab:brac_main_activity} reports paired componentwise contrasts of the livelihood composition within the groups of Figure~\ref{fig:brac_livelihood}, defined by the main 2007 activity. Each contrast averages over the pairs in which the group is present in both branches, and its interval uses a $t$ reference with one fewer degree of freedom than the number of such pairs. These contrasts concern arithmetic mean shares, whereas Figure~\ref{fig:brac_livelihood} reports changes in conditional Fr\'echet means.

\begin{table}[!htbp]
\centering
\footnotesize
\setlength{\tabcolsep}{4pt}
\begin{adjustbox}{max width=\textwidth}
\begin{tabular}{@{}lrcccccc@{}}
\toprule
Main 2007 activity & Pairs & Nonpart. & Livestock & Agric. & Domestic & Poultry & Other \\
\midrule
Nonparticipation & 20 & $-7.4$ & $34.6$ & $-7.6$ & $-12.8$ & $-0.9$ & $-5.9$ \\
 & & $[-12.3,-2.5]$ & $[28.0,41.2]$ & $[-14.4,-0.8]$ & $[-19.0,-6.7]$ & $[-7.8,6.0]$ & $[-11.8,0.0]$ \\
\addlinespace[2pt]
Livestock & 19 & $-5.2$ & $24.6$ & $-3.2$ & $-8.9$ & $-2.4$ & $-4.9$ \\
 & & $[-8.7,-1.7]$ & $[12.8,36.5]$ & $[-7.8,1.4]$ & $[-13.8,-4.1]$ & $[-10.6,5.8]$ & $[-10.1,0.3]$ \\
\addlinespace[2pt]
Agricultural wage & 15 & $-2.4$ & $35.1$ & $-14.7$ & $-14.6$ & $2.6$ & $-5.9$ \\
 & & $[-5.6,0.8]$ & $[27.0,43.1]$ & $[-24.9,-4.5]$ & $[-26.8,-2.5]$ & $[-5.6,10.9]$ & $[-13.9,2.1]$ \\
\addlinespace[2pt]
Domestic wage & 20 & $-1.3$ & $28.9$ & $-3.6$ & $-21.3$ & $1.6$ & $-4.2$ \\
 & & $[-5.8,3.3]$ & $[24.9,32.9]$ & $[-8.6,1.3]$ & $[-30.1,-12.5]$ & $[-2.8,5.9]$ & $[-10.1,1.6]$ \\
\addlinespace[2pt]
Poultry & 19 & $-7.0$ & $31.6$ & $-1.4$ & $-8.8$ & $-10.1$ & $-4.4$ \\
 & & $[-12.2,-1.7]$ & $[24.4,38.9]$ & $[-5.7,3.0]$ & $[-14.0,-3.7]$ & $[-20.3,0.1]$ & $[-9.4,0.6]$ \\
\addlinespace[2pt]
Other & 20 & $-2.2$ & $23.9$ & $-3.0$ & $-8.0$ & $1.0$ & $-11.6$ \\
 & & $[-6.9,2.4]$ & $[18.4,29.3]$ & $[-6.9,0.9]$ & $[-14.1,-1.8]$ & $[-4.2,6.1]$ & $[-19.4,-3.8]$ \\
\bottomrule
\end{tabular}

\end{adjustbox}
\caption{Paired componentwise contrasts by main 2007 activity. Estimates are treated-minus-control differences in the change in arithmetic mean shares from 2007 to 2011, in percentage points, averaged equally over the randomized pairs in which the group is present in both branches. Brackets contain pointwise 95\% confidence intervals.}\label{tab:brac_main_activity}
\end{table}

\subsection{Estimation and held-out validation}\label{app:brac_validation}
The reference is the uniform composition $o=(1,\ldots,1)/\sqrt6$, and the aligned-effect set is the hemisphere $K_\theta=\overline B(o,\pi/2)$, which contains every aligned effect because $d\{o,F_o(u,v)\}=d(u,v)\le\pi/2$ for compositions $u$ and $v$. With this effect set, Assumption~\ref{ass:spherical_domains} requires that no untreated and treated means have disjoint supports. All fitted means have six positive shares; over the estimation and held-out women in all partitions, the largest distance between a fitted untreated mean and a fitted treated mean is 1.41, below $\pi/2$, and the largest estimated effect magnitude is 0.70. Arm-specific forests regress the coordinates of $\Log_o(Y)$ in a fixed orthonormal basis on the pretreatment covariates, using the spherical logarithm in Section~\ref{app:spherical_profiles}. Their averaged leaf weights define the fitted conditional Fr\'echet risks. The propensity score is $1/2$ by randomization and the contamination size is $\varepsilon=1/6732$. Each branch has unit total base weight, divided equally among its eligible women; conditional weights restrict and renormalize these weights within the fitted neighborhoods.

For each of three fixed partitions, four outer folds hold out five complete branch pairs at a time. The remaining fifteen pairs are allocated to nuisance, structure, and estimation roles, each containing five pairs, with three cyclic rotations. Imputation and nuisance fitting use only the nuisance role. The GCF has 100 trees, depth at most two, minimum leaf size 80, and at least 25 observations and two branches from each treatment arm in each admissible child. Each tree is grown on four randomly chosen structure pairs, and with probability 0.1 a split is selected uniformly from the admissible candidates. As in Section~\ref{app:simulations}, the weights in~\eqref{eq:profilefit} come from arm-specific regression forests, here for the five coordinates of $\Log_o(Y)$ in an orthonormal basis of the tangent space at $o$, with 100 trees, minimum node size 40, and case weights giving each branch equal total weight; the leaf weights are averaged across coordinates and trees. These coordinates are used only for speed, and all Fr\'echet means are computed in the spherical metric. The split criterion uses an effect grid with 32 random directions, the positive and negative coordinate axes, and radii $0.05,0.1,0.2,0.4,0.8,1.2,\pi/2$, together with the reference. The effect grid is used for splitting; final estimation minimizes the average of the three role-specific criteria over the continuous effect set. The untreated reconstruction baseline is the equally weighted Fr\'echet mean of the three nuisance-role untreated predictions.

The common effect uses the same nuisance fits and role rotations as the GCF, with each branch in the estimation sample receiving equal total weight. The four group effects are fitted in the same way within each prespecified group. Proposition~\ref{prop:spherical_feasibility} specifies the feasible set $\Gamma\{\widehat b(x)\}$ in~\eqref{eq:feasible_domain} at each woman's predicted untreated mean. The GCF minimizes the corrected loss over this set, the common effect is fitted over the intersection of these sets for all women in the validation fold, and each group effect over the intersection for the women of that group in the fold. The constraints depend only on the women's pretreatment covariates and on untreated means predicted from the fifteen pairs outside the validation fold; held-out assignments and outcomes enter only the validation score. All 60,588 predictions from the GCF, the common effect, and the group effects across the three partitions yield valid compositions.

Let $g=1,\ldots,40$ index branches, $I_g$ contain the observation indices in branch $g$, $n_g=|I_g|$, and $k(g)$ denote its outer validation fold. Write $\psi_z^{\varepsilon,j,-k}(O_i)$ for the corrected loss~\eqref{eq:metric_psi} using nuisance fit $j$ estimated from the pairs outside fold $k$. For a predictor $\widehat f_{-k}$ fitted without that fold, the validation score is
\begin{equation}\label{eq:brac_validation}\widehat{\mathcal V}(\widehat f)=\frac1{40}\sum_{g=1}^{40}\frac1{n_g}\sum_{i\in I_g}\frac13\sum_{j=1}^3\psi_{\widehat f_{-k(g)}(X_i)}^{\varepsilon,j,-k(g)}(O_i).\end{equation}
The score in~\eqref{eq:brac_validation} averages the three nuisance-fit losses within each woman, then averages women within each branch and gives the forty branches equal weight. All predictors are evaluated with the same losses, so the common reference anchor cancels in their score differences.

For each randomized pair, the pair-level difference between two predictors averages the difference in their branch scores over the two branches of the pair and the three partitions; the mean of the 20 pair-level differences is the average difference reported in Section~\ref{subsec:brac_responses}. These differences are not independent: each pair is held out only from its own predictions and helps fit the nuisance functions and predictors that score pairs in other folds. We therefore report their mean, their signs, and their variation across partitions as descriptive summaries, without confidence intervals. In the three partitions, the forest reduces the mean held-out loss relative to the common effect by $0.0087$, $0.0093$, and $0.0115$, and the prespecified groups reduce it by less than the forest in every partition.

\end{document}